\documentclass[11pt,A4paper,oneside]{book}

\usepackage{amsbsy}
\usepackage{amssymb}
\usepackage{amsmath}
\usepackage{geometry}
\usepackage[british]{babel}
\usepackage{graphicx}
\usepackage{setspace}
\usepackage{hyperref}           
\usepackage{mathrsfs}
\usepackage[cmtip,arrow]{xy}	
\usepackage{pb-diagram,pb-xy}   
\usepackage{fancyhdr}           
\usepackage{sectsty}

\makeatletter
\def\@makechapterhead#1{\thispagestyle{plain}%
  \vspace*{50\p@}%
  {\parindent \z@ \raggedright \normalfont
    \ifnum \c@secnumdepth >\m@ne
      \if@mainmatter
       	\Huge\centerline{\bf\thechapter}
        \par\nobreak
        \vskip 20\p@
      \fi
    \fi
    \interlinepenalty\@M
    \Huge  \centerline{#1}\par\nobreak
    \vskip 50\p@
    \cfoot{}
	\rfoot{\thepage}
    }}
\makeatother

	

\makeatletter
	\renewcommand{\section}
	{\@startsection{section}{1}{0mm}
	{-\baselineskip}{0.5\baselineskip}
	{\large\bfseries}} 
\makeatother

\makeatletter
	\renewcommand{\subsection}
	{\@startsection{subsection}{2}{0mm}
	{-\baselineskip}{0.5\baselineskip}
	{\normalfont\bfseries}} 
\makeatother

\makeatletter
	\def\tableofcontents{\chapter{Contents}\vskip1cm\@starttoc{toc}}
\makeatother

\makeatletter
	\def\listoffigures{\chapter{List of figures}\@starttoc{lof}}
\makeatother

\newenvironment{dedication}
{
   \cleardoublepage
   \thispagestyle{empty}
   \vspace*{\stretch{1}}
   \hfill\begin{minipage}[t]{0.66\textwidth}
   \raggedright
}%
{
   \end{minipage}
   \vspace*{\stretch{3}}
   \clearpage
}

	\newcommand{\ncd}{\newcommand}

	\ncd{\mrm}    {\mathrm}
	\ncd{\beq} {\begin{equation}}
	\ncd{\eeq} {\end{equation}}
	\ncd{\glo}[2]{\glossary{name = {#1}, description = {#2}}}

	\def\q{{\rm q}}

	\def\n{{\rm n}}
	\def\s{{\rm s}}

	\def\i{{\rm i}}
	\def\d{{\rm d}}
	
	\def\cv{c_{\mathrm{v}}}
	\ncd{\nns}{n_{\n\s}^2}
	\ncd{\wns}{w_a^{\n\s}}
	\ncd{\us}{u_a^\s}
	\ncd{\bq}{{\bf q}}
	\ncd{\starq}{\star{\bf q}}
	\ncd{\omeganat}{\Omega^\natural}

	\ncd{\thetastar}{\theta^*}
	\ncd{\sstar}{s^*}
	
	\ncd{\un}{u_a^\n}
	\ncd{\Bn}{\mathcal{B}^\n}
	\ncd{\Bs}{\mathcal{B}^\s}
	\ncd{\Ans}{\mathcal{A}^{\n\s}}
	\ncd{\Ann}{\mathcal{A}^{\n\n}}
	\ncd{\Ass}{\mathcal{A}^{\s\s}}
	\ncd{\Xns}{\chi^{\n\s}}
	\ncd{\Asn}{\mathcal{A}^{\s\n}}
	\ncd{\Xsn}{\chi^{\s\n}}
	\ncd{\He}{\rm He}	
	\ncd{\nn}{\nonumber}

\begin{document}

		\frontmatter

\begin{titlepage}
    \begin{center}
      			{\bf UNIVERSITY OF SOUTHAMPTON}
      			\vskip.15cm
      			{\scshape Faculty of Social \& Human Sciences}\\[.1cm]{School of Mathematics}
                \vskip 1cm
                {\Huge Covariant Thermodynamics \&}
                \vskip0.25cm
	       		{\Huge Relativity}
         
         \vskip0.35cm
         		{by}
        \vskip.35cm
                {\Large C\'esar Sim\'on L\'opez-Monsalvo}
         \vskip12.5cm  
           		{Thesis  for the degree of Doctor of Philosophy}
         \vskip.1cm
 				{March 2011}
	\end{center}
\end{titlepage}


\chapter[Abstract]{}
   \vspace{-5cm}
   \begin{center}
   		{\bf UNIVERSITY OF SOUTHAMPTON}
   \vspace{.35cm}
   
   {\underline{ABSTRACT}}
	\vskip.25cm
	
   \textsc{ FACULTY OF SOCIAL AND HUMAN SCIENCES\\SCHOOL OF MATHEMATICS}
   \vspace{.35cm}
   
   {\underline{\textrm{Doctor of Philosophy}}}
   \vspace{.35cm}
   
   {\bfseries\scshape COVARIANT THERMODYNAMICS AND RELATIVITY}
	\vskip0.15cm
	
   {by}
	\vskip0.15cm
	
   {C\'esar Sim\'on L\'opez-Monsalvo}
   \vspace{.35cm}
   
   \end{center}

\begin{singlespace}
\noindent This thesis deals with the dynamics of irreversible processes within the context of  the general theory of relativity. In particular, we address the problem of the `infinite' speed of propagation of thermal disturbances in a dissipative fluid. Although this problem is not new, its best known solution - the Israel and Stewart second order expansion - has an effective, rather than fundamental, character.

\noindent \ \ The present work builds on the multi-fluid variational approach to relativistic dissipation, pioneered by Carter, and provides a dynamical theory of heat conduction. The novel property of such approach is the thermodynamic interpretation associated with a two-fluid system whose constituents  are matter and entropy. The dynamics of this model leads to a relativistic generalisation of the Cattaneo equation; the constitutive relation for causal heat transport. A comparison with the Israel and Stewart model is presented and its equivalence is shown. This discussion provides new insights into the not-well understood definition of a non-equilibrium temperature.     

\noindent \ \ A crucial feature of the multi-fluid approach is the interaction between its constituents. It is a well known fact that when two, or more, fluids interact, instabilities may occur. Within this work, the two-stream instability analysis is extended to the relativistic domain. As far as the author is aware, such extension has not been discussed in the literature. The analysis allows to assess the stability and causality of relativistic models of matter and their linear deviations from thermodynamic equilibrium directly from their equations of state or, equivalently, their Lagrangian densities. For completeness, a brief digression on a consistent (stable and causal) `first-order' model is also included. 

\noindent \ \ Finally, the road to follow  is laid by posing some physical applications together with some future perspectives and closing remarks.

\noindent \ \ To sum up, the variational approach to heat conduction presented in this thesis constitutes a mathematically  promising formalism  to explore the relativistic evolution towards equilibrium of dissipative fluids in a dynamical manner and to get a deeper conceptual understanding of non-equilibrium thermodynamic quantities. Moreover, it might also be useful to explore the more fundamental issues of the irreversible dynamics of relativity and its connections with the time asymmetry of nature.
\end{singlespace}

			\tableofcontents
	  		\listoffigures
	  		
\chapter[Declaration of authorship]{Declaration of authorship}
	\vskip-.5cm
\begin{singlespace}
	I, {\bf C\'esar Sim\'on L\'opez-Monsalvo}, declare that the thesis entitled:
	 \begin{center}
	 {\bf Covariant Thermodynamics and Relativity},
	\end{center}	  
and the work presented in the thesis are both my own, and have been generated
by me as the result of my own original research. I confirm that:

\begin{itemize}

	\item this work was done wholly or mainly while in candidature for a research
degree at this University;
	\item where any part of this thesis has previously been submitted for a degree or
any other qualification at this University or any other institution, this has been
clearly stated;
	\item where I have consulted the published work of others, this is always clearly
attributed;
	\item where I have quoted from the work of others, the source is always given. With
the exception of such quotations, this thesis is entirely my own work;
	\item I have acknowledged all main sources of help;	
	\item where the thesis is based on work done by myself jointly with others, I have
made clear exactly what was done by others and what I have contributed
myself;
	\item parts of this work have been published as: 
	{\small
		\begin{enumerate}			
			\item C. S. Lopez-Monsalvo and N. Andersson, Thermal Dynamics in General Relativity, {\it Proc. Roy. Soc. A} 467:738–759, March 2011. 	arXiv:1006.2978v1 [gr-qc].
			\item L. Samuelsson, C. S. Lopez-Monsalvo, N. Andersson and G. Comer, Relativistic two-stream instability, {\it Gen. Rel. Grav.} {\bf 42},  413-433, 2010.	arXiv:0906.4002v1 [gr-qc].			
			\item C. S. Lopez-Monsalvo, Heat conduction in relativistic systems: alternatives and perspectives, In {\it IoP Gravitational Physics Group newsletter} December 2010. arXiv:1011.6628 [gr-qc]. 
		\end{enumerate}
	}
\end{itemize}

\end{singlespace}	

\vskip.2cm
{\Large Date:}
\vskip1cm
{\Large Signed:}


\chapter{Acknowledgements}

\vskip1cm

Although immensely rewarding, the path of science - I believe - is hardly ever smooth. In this work, I had the fortune to be assisted by two truly inspirational guides.  I am greatly indebted to Nils Andersson, whose helpful advice always arrived with incisive clarity. He has contributed more than anyone to direct my brief explorations on thermodynamics and relativity. I am also thankful to James Vickers for sharing with me a little bit of his mathematical sophistication. Even his tiniest contribution always showed me a clearer way. To both of them, all my admiration. However, should the reader find  errors and imprecisions in this work, those are the sole consequence of my own misconceptions.      

I would like to thank in advance my examiners: Ian Hawke and David Matravers, for the time and effort spent reading this thesis. Their comments and corrections will certainly be welcomed. 

It has been a pleasure to be a member of the General Relativity Group of the University of Southampton. I am thankful to all, past and present, members that I had the opportunity to meet with, their help during the making of this thesis is greatly appreciated. In particular, I want to express my gratitude to Samuel Lander for his enormous efforts trying to understand the meaning (or the lack of it) of my words and ideas within, and beyond, this work. Also, very special thanks goes to Niels Warburton, Lucy Keer and Michael Hogg, many of our discussions made their way into the following pages.   

Through the years I spent between Southampton and London, I had the fortune to be supported by many individual. Thus, it is inevitable that - should I try to make a list  - I would fail to do justice to most of them. Nevertheless, I do wish to name a few. In chronological order, I would like to acknowledge Juan Carlos Hidalgo, Christine Rooks\footnote{. . . and so I did!}, Anna Kostouki, Jesper Greitz, Aida Cuni, Hatice Ozturk, Lourdes Guzman and Luis Alonso Vasquez. Their help and friendship were  crucial factors in the course of this work. 


For her loving care and unconditional support, I am grateful to Helga Elvira Laszlo. This thesis would not have become a tangible reality without her constant encouragement, love and understanding; especially during the final stages of the writing process. To her, all my love and gratitude.  
 
To  my family, who have supported me in every endeavour I have pursued, I cannot do anything else but to dedicate my work to them. In particular, this one is for Maria Elena Monsalvo: Pa'ti ma, con todo mi cari\~no! 

Finally, I give thanks to Queen Mary College for its hospitality and to CONACYT Mexico (fellowship No. 196971) for financial support through my post-graduate studies.

\vskip2cm
{\flushright\small 
C\'esar S. L\'opez-Monsalvo

Southampton $\cdot$ London

2011

}

			\begin{dedication}
			{\it For those who stood against the fear,\\
			Or those that ran, but held the rear,\\
			Who fought a creeping worm of shame,\\
			Who sought no other fool to blame,\\
			Who read the book and thought it through,\\
			And wanted to believe, but knew,\\
			Most likely, that it wasn't true . . .\\}
			{\flushright Felix Dennis}
			\end{dedication}
			
  		\mainmatter	
  		
			\chapter{Introduction}

\begin{quote}
It has become fashionable to justify one's scientific endeavours  by means of their immediate impact in shaping modern society in a manner that seems to accommodate research into two categories: the  self-evident applicable to technological development kind and the interesting but `why do we care?' type, to which most `curiosity-driven' research belongs. The work presented in this thesis is in the latter group.  Therefore, this chapter is written for those readers whose interest in this work was driven by the mere curiosity of knowing \emph{what is this thesis about?} but whose main concern really is \emph{why they should care?}.
\end{quote}

\emph{Does a moving body appear cold?} This remarkably simple question, raised by the late Professor P. T. Landsberg some forty years ago, highlights a profound missing link in our current understanding of classical thermodynamics and relativity. Some might interpret this as the lack of a relativistic transformation law for temperature measurements (see \cite{landsnature,landsmat,landsphl}). Some others, as we do here, believe that this is a misunderstood concern about the nature of thermal motion in a relativistic settings. Such a view  shifts the attention from a mere exercise in velocity transformations to an argument about thermal equilibrium between moving bodies and the inertial properties of heat \cite{israel, ehlers, carternotes, cartermiss}. Thus, it promotes the discussion about the missing Lorentz transformation for temperature to the more elaborate - and physically meaningful - problem of heat exchange and irreversibility in a relativistic context.

Irreversible processes in general relativity represent an outstanding challenge in our understanding of the  macroscopic dynamics of a wide range of physical systems. Although the problem of relativistic dissipation found a pragmatic solution through the Israel \& Stewart formalism ({\it c.f.} Chapter 4), in recent years there has been a resurgence of interest in the topic.  This thesis is motivated by phenomena whose leading mechanism of dissipation is due to heat flow. Such is  the case of the dynamics of super-fluid neutron stars, where thermal and dissipative effects are expected to impact on observations; or high energy gases, with photons providing the dominant pressure contribution.

\section{The empirical nature of Thermodynamics}
\label{intro.thermo}

Progress in each individual aspect of science is normally preceded by a revolution. The twentieth century is the closest and most vivid example of this, not only because the early years saw the rise of Relativity and Quantum Theory - our two most prominent scientific achievements -  but also due to the urgent need to reconcile the problems they pose in our quest of a unified vision of the universe. This manuscript is concerned with the motion of  systems whose macroscopic dynamics follow the rules prescribed by the general theory of relativity, and where the interactions between their microscopic components  become apparent only through a process of averaging, {\it i.e.} through a thermodynamic description. 

The central concept in thermodynamics is that of \emph{thermal equilibrium}. We will define precisely what  it means for a system to be in such a state. For the time being it will suffice to say that it corresponds to the highly idealised (macroscopic) state in which an observer would not be able to distinguish past from future solely from the dynamics of the system, {\it i.e.} the system's motion is time symmetric. The vast majority of physical situations we experience occur away from thermal equilibrium {\it e.g.}  the Earth's climate, the Sun's internal reactions, black hole formation, the expansion of the universe, life itself... That a system's evolution occurs in such a manner that its final state is thermal equilibrium follows from a set of \emph{empirical laws};  it is not a consequence of the dynamics. In this sense, thermodynamics cannot be considered as a `fundamental' theory, nevertheless its scope is \emph{universal}.

\emph{What is the role  thermodynamics plays in a theory of macroscopic dynamics?} The Laws of Thermodynamics are statements about nature which stem from the observation that certain phenomena - although allowed by the available theory of motion - simply do not occur.  We need to impose these laws `on top' of our more fundamental dynamical basis. Thus, the correct way to understand the role of thermodynamics is by regarding its laws as \emph{auxiliary hypotheses} which rule out entire classes of dynamical processes. 

If we have found a fundamental theory whose elegance and scope satisfy our needs of beauty\footnote{This is indeed a very subjective notion. For the present purpose of discourse it will suffice to consider any theory with a non-empty physical content and a minimal set of hypotheses.}, supplying the auxiliary thermodynamic hypothesis cannot do anything but to make our theory stronger. To illustrate this point, let us suppose that we have a theory which allows us to describe  precisely the motions of the microscopic constituents of a given material such that the total energy and linear and angular momentum are conserved. A collective measure of such motions corresponds to the temperature of the material. \emph{What are the possible dynamical outcomes of bringing  into contact two pieces of the same material at different temperature?} We cannot \emph{decide} the answer to this kind of question solely from the consequences of our fundamental theory, {\it i.e.} its  conservation laws. We need more  hypotheses. For example, if we call heat the energy and momentum transferred  from one body to the other, we can introduce the following observation:
	\vskip1cm
	\begin{quote}
	{\bf (Second  Law of Thermodynamics)} In every natural process involving two bodies at different temperatures heat \emph{never} flows from the colder to the warmer. 
	\end{quote}
The addition of this new statement to our fundamental theory rules out every  dynamical possibility which fails this \emph{empirical} criterion. Indeed, this is a part of what we \emph{observe} as a universal \emph{fact}. However, we cannot yet decide a unique outcome from the set of possibilities allowed by this `restricted' dynamics. A less obvious hypothesis is the one which asserts that:
	\begin{quote}
	{\bf (First Law of Thermodynamics)} There is \emph{no} perpetual motion machine.
	\end{quote}
This new hypothesis prevents our system from transferring heat from one body to the other in a periodic manner, analogous to a harmonic oscillator. This, however, still does not give us conclusive information about any `final' state. A closer look at the question we originally asked shows that we have made a tacit assumption, that we \emph{know} when two bodies are at different temperatures. Thus, in order to clarify such a subtlety, we introduce one more hypothesis
	\begin{quote}
	{\bf(Zero-th Law of Thermodynamics)} If  two bodies are brought into thermal contact and there is \emph{no} heat transfer, they are at the same temperature. 
	\end{quote} 
From these three additional fact-like laws, we can conclude that whenever two bodies at different temperatures are brought into thermal contact the warmer will transfer energy to the colder. If the size of the bodies is `small enough', there may be `thermal bounces' from their boundaries, however those will be damped for otherwise we would be able to construct a perpetual motion machine. Finally, their temperatures will equilibrate when no further heat transfer occurs. The auxiliary hypotheses do no tell us how the process of temperature equilibration  occurs. All the processes involved are driven by the dynamics of our original theory. These auxiliary hypotheses, the \emph{Laws of Thermodynamics}, simply rule out every solution that does not seem to correspond to an observed fact.   

 We may wonder if this process of adding new hypotheses will ever come to an end. Perhaps we have been sloppy while interpreting the predictions of or our theory, or maybe it was just not good enough to reproduce the observed reality. These are serious concerns for which every proposed solution would lie far beyond the scope of this treatise. The best we can do is to keep a critical eye in every conclusion we can reach from the principles of our proposed theories. We are granted by \emph{experience} that the three laws of thermodynamics we have introduced are universal and do not contradict each other. Should we find an internal inconsistency while incorporating these hypotheses into our theoretical construct we cannot do anything else but to withdraw our theory\footnote{In the words of Eddington:
 \begin{quote}
 `If someone points out to you that your pet theory of the universe is in disagreement with Maxwell's equations -- then so much the worse for Maxwell's equations. If it is found to be contradicted by observation -- well these experimentalists do bungle things sometimes. But if your theory is found to be against the second law of thermodynamics I can give you no hope; there is nothing for it but to collapse in deepest humiliation.'
  \end{quote}}. If on the contrary, the addition of these laws to our dynamical principles poses no contradiction, we have made our dynamics  less ambiguous; our theory is more complete.

\subsubsection{What is the empirical basis of this work?}

In this thesis, we use general relativity as our fundamental theory of motion. In the next chapter we will present the principles  upon which the general theory of relativity is based together with its implications as a dynamical theory. Here, we should answer \emph{what do we mean by a fundamental theory?} Throughout the entire manuscript, whenever we claim that a theory is fundamental, it means that there is a variational principle and a \emph{minimal} set of empirical hypothesis from which the dynamics of the theory can be obtained. In this sense, general relativity does qualify as a fundamental theory whose basic hypotheses consist of the principle of equivalence and general covariance. A detailed introduction to the empirical basis of general relativity is presented in the next chapter and the consequences of adding the thermodynamic auxiliary hypotheses to the \emph{material sector} of the theory is the central theme of the present work.

 The  marriage of thermodynamics and general relativity is a very active debate. Here  we address the problem of heat in the context of the relativistic dynamics of a non-perfect fluid. To this end, we will invoke the principle of equivalence to justify the study of the inertial properties of heat. In this sense, the dynamics of heat will be traced down to the variational ingredients used to derive the specific matter model we use. We will explain the precise meaning of this words in due time. First, let us verify that there is indeed an problem of heat and relativity.
 
 \section{Historical problems  of relativistic heat conduction}

    Let us begin our discussion of heat transfer from a non-relativistic point of view.  Consider  some metallic object or any other good thermal conductor. How long will it take to feel the heat propagated if we were holding one of the ends? According to Fourier's theory of heat - the standard transport theory used in most `practical' applications -  the effect  would be felt instantaneously all across the object. This is due to the \emph{parabolic} nature of the partial differential equation describing the heat transport within the body. 

The heat equation is obtained by assuming the validity of the first and second laws of thermodynamics together with Fourier's law relating  heat with temperature gradients in an \emph{un-relaxed} manner. This implies that  when two bodies at different temperatures are put  in thermal contact, heat spontaneously flows from the warmer to the colder without any delay.  Such conclusion cannot be satisfactory. One would expect the speed of thermal disturbances to be bound by the internal structure of the media they travel on. Cattaneo saw a way around this problem by introducing, in a reasonable (but arbitrary) manner, a modification to Fourier's law which takes into account the characteristic time a material  takes to react to thermal stimuli \cite{cattaneo}
	\beq
	\label{fourier}
	{\bf q} = -\kappa \nabla T \quad \rightarrow  \quad \tau\dot{\bf q} + {\bf q} = -\kappa \nabla T.
	\eeq
Here ${\bf q}$ denotes the heat flux, $T$ the temperature, $\kappa$ the thermal conductivity, $\tau$ represents the relaxation time of the medium  and the dot denotes time differentiation. Such an amendment leads to the telegrapher equation for the propagation of heat signals, with a finite bound on the speed they can reach. It is worth mentioning that Fourier's law is the simplest, but not the most general, possible ansatz to ensure that the second law is satisfied. This is done by explicitly making the change in entropy a quadratic function of the heat flow. We should keep this in mind in the forthcoming discussion.

In the relativistic case, the unbounded speed of thermal disturbances implied by the parabolic nature of the heat equation is more than mere inconvenience, it is indeed  a fundamental problem. To give context to our discussion, let us consider the case where  matter's motion is represented by a fluid whose normalised four-velocity is\footnote{ We will elaborate on the choice for such four-velocity  in Chapters 3 and 4.} given by $u^a$. The first attempt of a relativistic extension of the heat equation was introduced by Eckart \cite{eckart03}. The kind of model he proposed has become a stereotype of a class of theories referred as \emph{first-order} relativistic theories of dissipation. In this class, the entropy current of the model, denoted by a vector field $s^a$ which is generally not aligned with the matter four-velocity, may only depend on terms which are linear in deviations from equilibrium, namely the heat flow or the shear viscosity.  For this class of theories, the simplest way to \emph{impose} the second law, which locally takes the form $s^a_{\ ;a}\geq 0$, leads to a relativistic version of  Fourier's law
	\beq
	\label{eckart}
	q^a = - \kappa h^{ab}\left[T_{;b} + T \dot u_b \right], 
	\eeq
where $q^a$ represents the heat flux, $h^{ab}$ is a projector orthogonal to the matter flow and the semi-colon and dot denote covariant and proper time differentiation, respectively. Being essentially identical to \eqref{fourier},  equation \eqref{eckart} inevitably produces  a non-causal theory. Furthermore, as shown by Hiscock and Lindblom \cite{hiscock01}, it also suffers from stability problems, {\it i.e.} some thermal disturbances may grow without bound. However, we also note that Eckart's proposal exhibits an extra piece of information which is missing in the non-relativistic treatment; the acceleration term. This purely relativistic effect can  be interpreted as being due to an effective `mass' per unit entropy given precisely by the temperature (see discussions in \cite{ehlers,carternotes,heatpaper,lrrnils}). It is worth  mentioning  that  this term has recently been suggested to be the origin of the afore mentioned instabilities (see \cite{mexicans1}). However, here we adopt the view that the appearance of the four acceleration in \eqref{eckart} is an inevitable, and physically meaningful, feature of any relativistic theory of dissipation.      

The failure of first order theories to produce a theory of heat conduction compatible with the principle of causality can be tracked down to their definition of the entropy current. In an effort driven by simplicity, it is not permissible to prematurely drop  higher order terms in deviations from equilibrium, since they may give rise to linear terms after the differentiation required by the second law. This point of view is at the heart of the class of \emph{second order theories} of heat conduction whose key contribution is the  widely known Israel \& Stewart model. Here, the entropy current used by Eckart is extended to include \emph{all} the possible second order combinations of dissipative effects \cite{Stewart77,israel,israndstew2}. This strategy, analogous to the Grad's 14-moment method, is firmly ground on kinetic theory and provides a causal and stable account of relativistic dissipation.  It is not our intention here to further explain this particular framework, but to note that,  in spite of its success regarding stability and causality tests, the price to pay is the introduction of a set of second order couplings that, in principle, can be measured but which cannot be obtained within the realm of the theory. A detailed description of the Israel and Stewart formalism is given in Chapter 4.

\subsubsection{Why do we care?}

Owing to an increasing interest in hydrodynamic descriptions of high-energy relativistic plasmas, attention to relativistic theories of dissipation of the Israel \& Stewart type has been renewed. In many such applications, whose main source of dissipation is due to viscosity, heat can effectively be considered as a secondary effect. Indeed, it was this kind  scenario, in which the ratio of viscous to thermal resistivity can be pushed down to order unity, that was the main source of motivation for the development of Israel \& Stewart second order theory.  Our interests, however, follow from  quite a  different motivation. It ranges from recent efforts to model the dynamics of super-fluid neutron stars \cite{nilsgreg, nils2003} to high energy gases with photons providing the dominant pressure contribution \cite{carternotes}.

\section{The multi-fluid approach to thermal dynamics}

In this brief account of a long standing problem in relativity, it is our wish to shed some light upon the more  fundamentally  satisfactory \emph{variational} approach to relativistic heat conduction \cite{carternotes,cartermiss,carter1988}. Historically, due to a simplistic view of the components of the theory, this model, pioneered by Carter, failed the  tests of  stability and causality. However, it was later shown by Priou \cite{priou1} that the predictions of Carter's \emph{complete} theory are equivalent to, and at second order physically indistinguishable from,  those of Israel \& Stewart.  

Perhaps the most attractive feature of a variational construction of relativistic heat conduction is that, once the equation of state (or, equivalently, the Lagrangian density) of the system is known, the theory contains no free parameters. The price for this however is the inclusion of transport quantities in the ``Lagrangian'' density of the system. This is nevertheless inevitable in any circumstance where one's aspiration is to describe, at least to linear accuracy, situations departing from local thermal equilibrium.

In Carter's variational approach to heat conduction, one considers a multi-fluid system whose species are represented by a particle number density current $n^a$ and an entropy flux $s^a$. These two currents, together with the spacetime metric, constitute the fundamental fields of the matter sector for the Einstein-Hilbert action. General covariance requires the Lagrangian to be a proper scalar, therefore, it should depend only on covariant combinations of its fundamental fields. If we consider the metric as a passive field, the Lagrangian density can only depend on combinations of the two fluxes, which includes the relative flow between them. This is precisely what we mean by the inclusion of transport quantities in the Lagrangian density. 

A constrained variation of the matter action, whose Lagrangian density has the characteristics described in the preceding paragraph and which is constrained by the conservation law of the  particle number density flux, allows us to write the local conservation of energy and momentum as a ``force balance'' equation.  It is worth noting that each of the individual `forces' appearing in such balance, takes a form completely analogous to the Lorentz force for electromagnetism in a general relativistic setting ({\it c.f.} Section 5.2). The requirement for the local energy conservation law to follow as a Noether identity of the variational principle,  only needs one of the currents to be strictly conserved, $n^a_{\ ;a}=0$ say.  This allows an ``extra'' freedom to allocate the second law of thermodynamics  by  taking into account the possibility in which the production term $s^a_{\ ;a}$ takes values different from zero. 

One of the central problems  to be faced before giving a \emph{real} thermodynamic interpretation to the variational construction, lies in the correct interpretation of temperature measurements. As stressed in the opening sentence, this is a non-trivial problem in relativity and, therefore, a choice of frame is forced upon us. In this simple case, where we have only one conserved current, the choice is somewhat natural. In the spirit of physical interpretation, we can chose to equip each infinitesimal part of the matter fluid with a thermometer and consider all physical measurements to be taken in, and with respect to, the matter frame. In such a case one can show that the projection of the  canonical conjugate momentum to the entropy flux in the matter frame, which we denote here by $\theta^\parallel$, corresponds to the thermodynamic temperature in the  Gibbs sense\footnote{The rate of change of energy with respect to entropy with all other independent thermodynamic quantities fixed.}. Having clarified this, it is not difficult to show that the equation for the heat flux relative to the matter frame takes the form \cite{heatpaper}
\beq
\label{heat}
\check{\tau}\left[\dot q^a + u^{c;a}q_c\right] + q^a = - \check{\kappa} h^{ab}\left[\theta^\parallel_{;b} + \theta^\parallel \dot u_b \right].
\eeq 
Here all the quantities, with the exception of $\check{\tau}$ and $\check{\kappa}$ which correspond to the \emph{effective} relaxation time and  thermal conductivity, have been  introduced  earlier. This result, which is a relativistic generalization of Cattaneo's equation [the modified Fourier's law \eqref{fourier}], follows directly from the variational principle together with the simplest assumption (in the sense previously discussed) to make the entropy production satisfy the second law of thermodynamics. 

The structure of \eqref{heat} combines all the features present of both \eqref{fourier} and \eqref{eckart}, including the acceleration term which in the variational context arises as a consequence of the equations of motion. The ``check'' marks  indicate that these quantities depend on deviations from thermal equilibrium which are higher than second order.  A comparison with an analogous expression obtained from the Israel \& Stewart theory  is beyond the scope of present means of experimental verification.

\subsubsection{Is this a covariant theory of the {\it caloric}?}

The answer to such a question is of ontological, rather than physical, character. As a reminder, the theory of caloric was the standard explanation of the mechanism of heat conduction until mid-nineteenth century. It posits the existence of a fluid-like substance - the caloric - which is self-repulsive and attracted to other forms of matter.  In this view,  temperature is a measure of the caloric density of a body. Thus, the previous observation that heat flows from warmer to colder bodies is a mere consequence of the natural tendency of a fluid to flow from from  high to low density regions. It is now known that there is no such a substance, that the temperature of a material  corresponds to the kinetic energy of its microscopic constituents and that it is overwhelmingly unlikely that the transfer of this kinetic energy occurs from `colder' to `warmer' bodies. Nevertheless, the caloric theory still provides accurate descriptions of thermal processes, provided it is applied in the right context. This is nothing new in the practice of science, it is simply the regular process of recovering older theories from newer ones in special limits.  For example, in many practical calculations a Newtonian treatment would give quite a satisfactory description of the planets' orbits \emph{as if there was} a gravitational pull from the Sun or the tides of ocean as if the Moon would exert a force on the water. This is better explained by Norton in the following quote:
	\begin{quote}
	...In returning the older theories, the [reduction] relations revive a defunct ontology. More precisely, they do not show that heat is a fluid, or gravity a force; rather they show that in the right domain the world behaves as if they were...\cite{nortonfolk}.
	\end{quote}

While reading through the forthcoming chapters,  it may become quite a tempting  thought that entropy can be considered as  a `real' fluid.  Being an ingredient of a variational construction, it is indeed no less real than the matter density current $n^a$ or the spacetime metric $g_{ab}$. In this sense, it is clear that the multi-fluid approach to heat conduction is an effective model to deal with thermal disturbances under  specific circumstances; those in which entropy behaves \emph{as if it were} a fluid. For some of us, this may seem  a rather un-natural feature of entropy, even though we may be quite happy to use a fluid description for the collective motion of a set of particles. This is because there is a conceptual barrier associated with the `kinetic meaning' of entropy ({\it c.f.} section 3.3) and the properties of a fluid. However, soon we will show that it is, in fact, kinetic theory the one responsible for the fluid interpretation of entropy\footnote{The modern idea of entropy as ``a measure of a system's disorder'' makes it hard to visualise a situation in which such a disorder would have fluid properties. Remarkably, the thermodynamic limit of relativistic kinetic theory includes an entropy current, whose description can be given in terms of an `entropic' four-velocity.  We will address this point towards the end of Chapter 3.} used in our variational formalism. Although the caloric theory was once considered to be equivalent to the kinetic theory of gases, it was eventually overthrown. Kinetic theory became the foundation of modern thermodynamics and the caloric has been regarded as a dead theory ever since. 

 In the multi-fluid approach to heat conduction, which shall be developed in chapter 5, the relative flow between the matter and entropy currents plays a central role in the dynamical setting of a system away from thermal equilibrium. Moreover, it is through the inertial properties of heat that the entropy current can carry momentum from one part of the system to another, driving the dynamics towards equilibrium as dictated by our empirical laws. Therefore, in our view, treating the entropy density current as a `real' fluid in the matter Lagrangian allows us to use the principle of equivalence  to associate some inertial properties to the heat flux. Hence, allowing a hyperbolic description for the propagation of thermal perturbations. 

In conclusion, one could say that, physically, this is not a modern, covariant version of the caloric theory. However, it is one that certainly shares its main ontological properties.

\subsubsection{Past objections and our solution}

Let us emphasize that the main objection to Carter's approach was based on issues about stability and causality. This is a consequence of the simplistic spirit of the original model, ignoring a crucial effect present in almost every multi-fluid system; \emph{entrainment}. This effect tilts the momenta  due to the coupling between their associated currents.

Compared to the standard single-fluid analysis, a multi-fluid system has more degrees of freedom; we need
to account for relative flows between them. When two or more fluids are allowed to interact, instabilities may occur.
Such instabilities are known to exist for a variety of configurations. For example, in shearing motion at an
interface, this corresponds to the well known Kelvin--Helmholtz instability. One of the central results of this work, corresponds to a relativistic generalisation of the two-stream instability analysis. This is a generic phenomenon that
does not require particular fine-tuning to be triggered, nor is limited to any specific physical system. The only
requirements are that there must be a relative (background) flow and a coupling between the fluids.
 
It is worth noting that thermodynamic instabilities are the source of pattern and structure formation and that
our analysis provides a stability criteria directly from the prescribed (interacting) Lagrangian density or equation
of state. The dynamical role of this instability has not been explored  beyond the
linear regime.

\section{The state of the art}

This thesis originally had  a simple goal: to understand heat in a wide range of physical situations. To this end, we needed to provide a satisfactory account of dissipative processes in general relativity. This will remain a challenge with many interesting directions emerging from the work presented here.

So far, we have only scratched the surface of what is in fact a very difficult problem. In the following Chapters, the reader will find a summary of the most significant efforts  to obtain a physically sound theory of heat conduction compatible with the tenets of relativity.  A point to note is that in spite of the recent stir made on first order theories, in our view, physical indicators such as the presence of second sound in materials hint that the correct physics to model departures from thermal equilibrium cannot be less than second order. 

 Although the Israel \& Stewart model is the most prominent and widely used tool to describe dissipative systems, the additional couplings - necessarily introduced in their expansion - give the theory an effective, rather than fundamental, character. In this sense, the variational formalism not only provides us with an alternative to the Israel \& Stewart model, it  centres the attention in the dynamical actors of any canonical theory; the canonical conjugate momenta. This observation makes the multi-fluid construction a very powerful tool which may help us tackle deeper problems in non-equilibrium thermodynamics of relativistic systems.

		  	\chapter{Relativity and Thermodynamics}

The question we address in this thesis is whether we can find a formulation of non-equilibrium thermodynamics compatible with the tenets of general relativity from a fundamental point of view. In particular, we want to study the case of heat conduction for  matter models  described  by fluid dynamics. As highlighted in the introduction, this has been a long standing problem, partially solved by Israel and Stewart some thirty years ago \cite{israel}. It is one of our aims to explain the philosophy and achievements of their theory as a standard to compare our developments from a variational principle, continuing the work started by  Carter \cite{cartermiss,carternotes}.

We start this chapter with a primer of general relativity. Although the reader may find these derivations in most of the references cited in this work, we try to make explicit many assumptions and calculations often omitted in a general reading on the matter. This will also serve to introduce the conventions and notation used in the forthcoming chapters. The second part of this chapter is devoted to irreversible processes. It is a brief exposition of some of the developments in non-equilibrium thermodynamics relevant to this work, therefore many of the ideas will not be presented with the detail they  deserve. The interested reader may wish to have a look at the following references \cite{mulleit,joueit,mullentr}.  

\section{Spacetime and matter}

The aim of this section is to obtain the dynamical equations which govern the interplay between matter and spacetime. Although the derivation presented here is now standard knowledge, the construction and principles upon which relativity is based and the precise meaning of them is still an active debate among relativists \cite{norton}. Therefore, it would be negligent to leave aside a discussion of these matters.

\subsection{The principles of relativity}

The elegance and power of any theory, not necessarily physical, can be judged objectively by the simplicity\footnote{Simplicity is a highly non-trivial concept within the philosophy of science. Here we follow the convention by Popper \cite{popper} which states that a the degree of simplicity of a theory is in correspondence with its susceptibility to be `easily' falsified.} of its assumptions and the scope of its validity\footnote{Its universality.}. In this way, Einstein's theory of relativity is, to the author's taste, the most remarkable achievement in our efforts to understand the most familiar state of nature: \emph{motion}.

Up to the present date, general relativity has stood the most rigorous experimental tests. An excellent account of this was given by Clifford Will in \cite{cliffwill}, and a more recent account can be found in \cite{grexperimental}. The elegance of general relativity, however, can be questioned on the grounds of its principles. It is not an easy task to write a version of them which is not contentious. Strong arguments and objections have been put forward for over almost a century regarding their true  physical content. It is not the intention here to participate in such debate, but to make a clear exposition of the tenets of relativity in a manner relevant to our attempts to reconcile thermodynamics with them.

Einstein built his special theory of relativity on the empirical evidence that the speed of light in vacuum has the same constant value for all inertial observers - those moving with uniform relative velocity between each other - and the mathematical principle of \emph{Lorentz covariance}, which states that any physical law should attain the same mathematical form in every frame related to another by a \emph{Lorentz transformation} of coordinates, that is, independent of the \emph{inertial} state of motion of the observer. This is known as the \emph{principle of relativity}. 

It is tempting to enlarge the group of transformations which keep the covariance of physical laws to include a larger class of possible observers, such as those moving in an uniformly accelerated manner. This is the programme which gave rise to Einstein's theory of gravitation. The connection between accelerated motion and gravity is highly non-trivial and represents the fundamental physical assumption of the general theory of relativity. Just like the invariance of the speed of light for all observers, the equivalence between \emph{inertial} and \emph{gravitational} mass is an empirical fact which was promoted to a general principle, \emph{the principle of equivalence}. 

The information encoded in the equivalence principle is purely physical. However, it has a far reaching consequence in Einstein's programme: it suggests a possible connection between the covariance of the physical laws for accelerated observers and gravity. 

The extension of the relativity principle requires  physically meaningful statements to be independent of the \emph{general} state of motion of the observer, not just inertial as in the case of special relativity. That is, the laws of physics should preserve their form under a general coordinate transformation. This is a formulation of the principle of \emph{general covariance}. It is a statement of mathematical nature whose connection with gravity through the equivalence principle form the core of the general theory of relativity. 

It has been argued  that every physical theory\footnote{We consider a theory as physical if it satisfies the empirical criteria prescribed by the underlying empirical laws. If in addition it can be derived from a variational principle we regard such a theory as fundamental.} can be made to satisfy the principle of general covariance, and therefore it has been objected that its physical content is empty \cite{norton}. One should be aware of how this conclusion was reached - the fact that intrinsically there is no physics in it. It is together with the equivalence principle that we can learn what general covariance is for relativity: 
	\begin{quote}
	... it is the language for describing a world without distinction between the spacetime entities and the dynamical entities...\cite{rovelliQG}
	\end{quote}

\subsubsection{Dynamics and geometry}

To  understand better the interplay between the mathematical language of general covariance and the physical principle of equivalence, it is worth discussing the simpler case of the electromagnetic field and Lorentz transformations. This exercise also makes explicit the underlying physical path to the general theory.

The road to build the special theory of relativity combines the  experimental results on the non-existence of an absolute reference frame\footnote{The rigid aether assumption is a consequence of the finite value of the speed of light and the preferred Newtonian frame it singles out. This introduced the notion of conformal structure which is defined, even in Newtonian theory, by a field of light cones. See section 1.3 in \cite{ehlers}.} (which motivated the formulation of the relativity principle), together with the mathematical fact that Maxwell's equations do not change their form under Lorentz transformations, those which leave the Minkowski metric invariant. 

The combination of  Lorentz covariance with the relativity principle leads to the physical notion of \emph{spacetime}. In special relativity spacetime is fixed and plays no role in the dynamics of physical processes. This is a consequence of the fact that Lorentz transformations only relate inertial observers, in this sense we consider them as defining the observer \emph{kinematics} of relativity. Thus, Einstein's programme is that of finding the observer \emph{dynamics} which, through the equivalence principle, is coupled with the dynamics of spacetime.

\subsection{The symmetry of general relativity}

Now we proceed to understand how the symmetry implied by general covariance merges with the principle of equivalence to produce the dynamics of matter and geometry.  The construction presented here follows that of \cite{plebanski, nakahara, deFelice,hawkingellis}. 

The mathematical model for space-time we use consists of  a four-dimensional pseudo-Riemannian manifold $\mathcal{M}$\glossary{name = {$\mathcal{M}$} , description = {Spacetime manifold}}, whose metric ${\bf g}$\glossary{name={${\bf g}$},description={Spacetime metric}} has the signature $\{-,+,+,+\}$. The aim should be then to obtain a set of equations for the components of the metric, $g_{ab}$, which in the non-relativistic limit reproduce the Newtonian theory of gravity as described by the Poisson equation, a linear second order partial differential equation for the Newtonian potential.

For the gravitational theory, Hilbert proposed the following axioms:
     \begin{enumerate}
          \item The field equations should follow from a variational principle. The independent variables in the action integral should be the components of the metric tensor.
          \item The action functional should be a scalar,
          \item The Euler-Lagrange equations that  follow should be of second order in the space-time metric $g_{ab}$.
     \end{enumerate}

Before introducing Hilbert's form of the action for general relativity, a reminder is in order. We  are looking for a condition in which all dynamical entities are in equilibrium with the background geometry. In terms of a variational principle this entails  considering a single action for matter and geometry, which is invariant under the local group transformations imposed by general covariance, and whose equations of motion fully reflect the equivalence principle. To this end, we consider the following action 
     \beq
     \label{gf.00}
     S(\Omega) = S_{g}(\Omega) + S_{m}(\Omega) = \int_\Omega (\mathscr{L}_g + \mathscr{L}_m)\d \Omega.
     \eeq
where the Lagrangian densities for geometry and matter, $\mathscr{L}_g$ and $\mathscr{L}_m$ respectively, are to be integrated over an open set $\Omega$\glossary{name = {$\Omega$}, description = {Open subset of spacetime with compact closure}} of the spacetime $\mathcal{M}$ with compact closure. Now, we need to ensure that all the quantities involved in \eqref{gf.00} are well defined. 

 We consider first the part of the action describing the dynamics of the geometry. First we note that since the volume element of space-time $\d \Omega$\glossary{name = {$\d \Omega$}, description = {spacetime volume element}} is a scalar density of weight $+1$, it follows from axiom 2 that $\mathscr{L}_g$ should be a scalar density\footnote{A tensor density differs from the corresponding tensor in that when transformed from one coordinate system to another, it gets multiplied by a certain power $w$ of the Jacobian transformation, i.e. $\sqrt{-g}^w$. The exponent of the Jacobian is called the weight of the density. A proper scalar should have density 0. The weight is additive under tensor product, that is: given two scalar densities of weight $w$ and $w'$ respectively, their product is a scalar density of weight $w+w'$ (See Chapter 3 in \cite{plebanski} for a full discussion on general tensor densities).} of weight $-1$. Also, from the third axiom above, $\mathscr{L}_g$ cannot be just $\sqrt{-g}$,  it should also contain the derivatives of the metric. Thus, we could write the spacetime Lagrangian density as 
 	\beq
 	\label{gf.LD}
 	\mathscr{L}_g=\sqrt{-g} \Lambda_g,
 	\eeq
where $\Lambda_g$ is a proper scalar constructed from covariant combinations of the metric and its derivatives only.

 The order of the Euler-Lagrange equations is twice the order of the highest derivative in the action integral, however it is not possible to construct a proper scalar out of combinations of the metric and its first partial derivatives since the Christoffel symbols do not transform as tensors. Thus we need to include second order derivatives of the metric in $\Lambda_g$. These terms will not contribute to the equations of motion if they can be collected into a divergence of an expression which vanishes on the boundary $\partial \Omega$, as it is shown below. The geometry part of the action is postulated to be 
     \beq
     \label{fr.HE}
     S_g(\Omega)= \int_\Omega \sqrt{-g} R \d \Omega,
     \eeq
where $R$ is the curvature scalar.

In a similar way, the Lagrangian density for the matter $\mathscr{L}_m$ can be written as a product of $\sqrt{-g}$ with a proper scalar built out of covariant combinations of all the matter fields we observe. Notice that the metric itself may play a role in these combinations by producing the appropriate Lorentz scalars. In this sense, the metric enters the matter sector of the action in an algebraic way only. Thus, the matter part of  the action is written as
	\beq
	\label{gf.matterL}
	S_{m}(\Omega) = \int_\Omega \sqrt{-g}\Lambda_m(\Psi,\Psi_{;a} ; {\bf g}) \d \Omega
	\eeq
where we denote collectively all the matter fields by the letter $\Psi$ and the semi-colon denotes metric compatible covariant differentiation.

\subsubsection{General covariance and spacetime coincidences}
\label{coincidences}

The equations of motion are obtained in the usual way by requiring the action $S_m(\Omega)$ to be stationary under variations of the fields in the interior of $\Omega$. The following argument has been taken from Hawking and Ellis \cite{hawkingellis}. It is known as `the hole argument' for reasons that will shortly become apparent, and it plays a central role while discussing the meaning of general covariance ({\it c.f.} discussions in \cite{nortonhole,rovelliQG}).

 A variation of the fields, including the metric, is realised by a one-parameter family of diffeomorphisms from spacetime into itself, $\varphi_s \in {\rm Diff}(\mathcal{M})$\glossary{name = {${\rm Diff}(\mathcal{M})$}, description = {Group of diffeomorphisms from spacetime into itself}}, which induces the maps 
 	\beq
 	\label{gf.onepf}
 	\varphi^*_s\Psi = \tilde\Psi(s,x) \quad {\rm and} \quad \varphi^*_s {\bf g} = \tilde{\bf g}(s,x),
 	\eeq
where the parameter $s$ takes its values in a sufficiently small open interval of the real numbers centred at the origin, and $x$ is a point in $\mathcal{M}$, not necessarily inside $\Omega$ such that
	\begin{enumerate}
		\item $\tilde\Psi(0,x) = \Psi(x)$, $\tilde{\bf g}(0,x)={\bf g}(x)$, i.e. $\varphi_{s=0} = {\rm id}_{\mathcal{M}}$\glossary{name = {${\rm id}_{\mathcal{M}}$}, description = {Spacetime identity map}}
		\item $\tilde\Psi(s,x) = \Psi(x)$, $\tilde{\bf g}(s,x) = {\bf g}(x)$ when $x$ is not in the interior of $\Omega$. 
	\end{enumerate}
Thus, the fields variations are denoted by
	\beq
	\label{gf.varia}
	\delta \Psi = \frac{\d}{\d s}\varphi^*_s\Psi\Big|_{s=0} \quad {\rm and} \quad \delta {\bf g} = \frac{\d}{\d s}\varphi^*_s {\bf g}\Big|_{s=0}.
	\eeq

The term `hole' now becomes evident. We are looking at variations which leave the value of the fields unchanged at each point of $\mathcal{M}$, except inside an open region - the hole - $\Omega$. This implies, in particular, that the variations of the fields and the metric vanish on the boundary, $\partial \Omega$\glossary{name = {$\partial \Omega$}, description = {Boundary of $\Omega$}}. 

General covariance is equivalent to the requirement of the invariance of the action under a general transformation of the spacetime manifold
	\beq
	\label{gf.diff}
	S_m(\Omega;\Psi,{\bf g}) = S_m(\Omega;\varphi^*_s\Psi,\varphi^*_s{\bf g}).
	\eeq  
A crucial point to be noted is that a transformation $\varphi_s$ \emph{pulls back} both, the fields and the metric. This removes any physical significance from spacetime points. This fact is a statement in modern language of what Einstein referred as spacetime \emph{coincidences}.

\subsection{Equations of motion}

\subsubsection{Matter equations}

The requirement of invariance of the action under diffeomorphisms of spacetime imposed by general covariance, equation \eqref{gf.diff}, makes evident what the matter dynamics is. From the definitions \eqref{gf.varia}, the variation of the action is
	\beq
	\label{gf.varsm}
	\delta S_m =  \left[\frac{\delta^\flat S_m}{\delta^\flat \Psi}\frac{\d\varphi^*_s \Psi}{\d s} + \frac{\delta^\sharp S_m}{\delta^\sharp {\bf g}}\frac{\d \varphi^*_s {\bf g}}{\d s}\right]_{s=0} =\frac{\d}{\d s} S_m(\Omega)\Big|_{s=0} = 0.
	\eeq
	
This should hold for arbitrary variations of the fields and of the metric. Therefore, each term of the variation \eqref{gf.varsm} should vanish independently.

The first term inside the brackets in \eqref{gf.varsm} corresponds to variations of the action which keep the metric fixed\glossary{name = {$\delta^\flat$}, description = {Variations of the action which keep the metric fixed}}, whilst the second term corresponds to variations in the metric which leave the values of the fields unchanged\glossary{name = {$\delta^\sharp$}, description = {Variations in the metric which leave the values of the fields unchanged}}. We denote with a `flat' superscript the first kind and with a `sharp' the second. Thus, the flat variation of the action is 
	\beq
	\label{gf.mfield}
	\delta^\flat S_m  =  \int_\Omega \left(\frac{\partial \Lambda_m}{\partial \Psi} \delta^\flat \Psi + \frac{\partial \Lambda_m}{\partial \Psi_{;a}} \delta^\flat \Psi_{;a}\right) \sqrt{-g} \d\Omega.
	\eeq
Since the parameter $s$ is completely independent of spacetime, it follows from \eqref{gf.onepf} that
	\beq
	(\delta \Psi)_{;a} = \left(\frac{\d}{\d s}\varphi^*_s \Psi\right)_{;a} = \frac{\d}{\d s}\varphi^*_s \Psi_{;a} = \delta \Psi_{;a},
	\eeq
we can integrate by parts the second term in \eqref{gf.mfield}
	\beq
	\label{gf.intpart}
	\int_\Omega \frac{\partial \Lambda_m}{\partial \Psi_{;a}}\delta^\flat \Psi_{;a} \sqrt{-g}\d\Omega = \int_{\Omega} \left[ \left( \frac{\partial\Lambda_m}{\partial \Psi_{;a}}\delta^\flat \Psi\right)_{;a} - \left(\frac{\partial \Lambda_m}{\partial \Psi_{;a}} \right)_{;a}\delta^\flat \Psi \right]\sqrt{-g}\d\Omega,
	\eeq
and write the `flat' variation of the action for matter as
	\beq
	\label{gf.e-l}
	\delta^\flat S_m = \int_{\Omega}\left[\frac{\partial \Lambda_m}{\partial \Psi} - \left(\frac{\partial \Lambda_m}{\partial \Psi_{;a}} \right)_{;a} \right]\delta^\flat \Psi \sqrt{-g}\d\Omega + \int_{\Omega} \left(\frac{\partial \Lambda_m}{\partial \Psi_{;a}}\delta^\flat \Psi \right)_{;a}\sqrt{-g}\d\Omega.
	\eeq
	
We can use Stokes' theorem in the last term to transform the total divergence into a surface integral over the boundary  $\partial\Omega$ where the field variations are identically zero, as previously argued. Therefore, the change in the action $S_m$  inside $\Omega$ under arbitrary variations of the fields $\Psi$,
	\beq
	\label{gf.e-l2}
	\frac{\delta^\flat S_m}{\delta^\flat \Psi} = \int_{\Omega}\left[\frac{\partial \Lambda_m}{\partial \Psi} - \left(\frac{\partial \Lambda_m}{\partial \Psi_{;a}} \right)_{;a} \right] \sqrt{-g}\d\Omega=0,
	\eeq 
if they satisfy the Euler-Lagrange equations, i.e. if \eqref{gf.e-l2} is identically zero. These are the equations for the matter fields. 

The variation of $S_m$ with respect to the metric is slightly more elaborate since in addition to the field variation we also need to take into account the changes in the density $\sqrt{-g}$ and, possibly, in the covariant derivative of the matter fields induced by the transformations $\varphi_s$. However, since the connection is a coarser geometric object,  its variations can be considered  to be independent of the metric. The variation of the action with respect to the connection will not produce an equation of motion for it since the connection itself is not an observable field.  Instead, it will fix the Levi-Civita connection of the metric ${\bf g}$. We will follow \emph{Palatini's} approach and consider the connection variations as independent so that 
	\beq
	\label{gf.mgvar}
	\delta^\sharp S_m = \int_\Omega \left[ \frac{\partial \Lambda_m}{\partial {\bf g}}\right]\delta{\bf g} \sqrt{-g}\d\Omega + \int_\Omega \Lambda_m \frac{\partial \sqrt{-g}}{\partial {\bf g}} \delta {\bf g}\d\Omega.
	\eeq 
does not include variations of the covariant derivatives of the matter fields\footnote{See chapter 2 in \cite{hawkingellis} for a discussion where these variations are included.} .

Let us start with the second term in \eqref{gf.mgvar}. The change of the density $\sqrt{-g}$ with respect to the metric is given by
	\beq
	\label{gf.sqrt}
	\frac{\partial \sqrt{-g}}{\partial g_{ab}} = -\frac{1}{2\sqrt{-g}} \frac{\partial g}{\partial g_{ab}}= -\frac{1}{2\sqrt{-g}}  g g^{ab} = -\frac{1}{2} g^{ab} \sqrt{-g}, 
	\eeq
so  the  `sharp' variation \eqref{gf.mgvar} becomes
	\beq
	\label{gf.sharpem}
	\delta^\sharp S_m = \frac{1}{2} \int_{\Omega} T^{ab} \delta^\sharp g_{ab} \sqrt{-g} \d \Omega,
	\eeq
where 
	\beq
	\label{gf.emtensor}
	T^{ab} = 2 \frac{\partial \Lambda_m}{\partial g^{ab}} - \Lambda_m g^{ab}
	\eeq
is a symmetric tensor which represents the energy and momentum distribution of the matter fields with respect to the undetermined background geometry.

Finally, let us assume that the matter fields satisfy the Euler-Lagrange equations so that \eqref{gf.e-l2} vanishes. By the fundamental theorem of differential equations, the integral curves of $\varphi_s$ correspond to the flow generated by some vector field $\xi$, therefore, the variation of the metric, \eqref{gf.varia} can be written as
	\beq
	\label{gf.metvar}
	\delta {\bf g} = \frac{\partial \varphi^* {\bf g}}{\partial s}\Big|_{s=0} = \mathcal{L}_{\xi} g_{ab} = \xi_{(a;b)},
	\eeq
where $\mathcal{L}_{\xi}$\glossary{name = {$\mathcal{L}_{\xi}$}, description = {Lie-differentiation with respect to the vector field $\xi$}} denotes Lie-differentiation with respect to the vector field $\xi$. 
Substituting this into \eqref{gf.sharpem} yields
	\beq
	\label{gf.emintp}
	\delta^\sharp S_m = \frac{1}{2}\int_\Omega T^{ab} \xi_{(a;b)} \sqrt{-g}\d\Omega,
	\eeq
and using the symmetry of the energy momentum tensor, we can integrate \eqref{gf.emintp} by parts  to obtain
	\beq
	\int_\Omega T^{ab} \xi_{(a;b)} \sqrt{-g}\d\Omega = \int_\Omega T^{ab}_{\ \ ;b} \xi_a \sqrt{-g} \d\Omega - \int_\Omega \left(T^{ab} \xi_a \right)_{;b} \sqrt{-g} \d \Omega
	\eeq
where we can use Stokes' theorem again to transform the last term into a surface integral over the boundary of the hole - where the vector field $\xi$ vanishes. Thus, the variation of \eqref{gf.sharpem} becomes
	\beq
	\label{gf.Tcons1}
	\delta^\sharp S_m = \int_\Omega  T^{ab}_{\ \ ;b} \xi_a \sqrt{-g}\d\Omega.
	\eeq
The equations of motion follow from the vanishing of \eqref{gf.Tcons1} for the arbitrary vector field $\xi$ inducing the metric variation \eqref{gf.metvar}. These are equivalent to the \emph{local} conservation law
	\beq
	\label{gf.Tcons2}
	T^{ab}_{\ \ ;b} = 0.
	\eeq

Before we continue with the derivation of the equations of motion for general relativity, let us summarize the consequences of Hilbert's variational approach for matter. The requirement of general covariance in the formulation of physical laws is equivalent to the invariance of the action $S_m$ under diffeomorphisms of spacetime, equation \eqref{gf.diff}. This symmetry implies the dynamics of matter, given by \eqref{gf.e-l2} and \eqref{gf.Tcons2} as \emph{Noether identities} and, therefore, they do not need to be postulated independently.  However, this is nothing particular about general relativity, as discussed in the previous section. So far we only have re-written classical dynamics in a generally covariant manner. The next step is to incorporate the principle of equivalence to identify the \emph{full} dynamics of spacetime and matter.

\subsubsection{Geometry equations}

 The last stage of the variation formalism is to obtain the dynamics of the geometry implied by general covariance. The procedure is straightforward. As previously stated, we follow Palatini's formalism by considering the connection as an independent field. We stress again that no equations of motion for the connection would arise from its variations, those will only fix the Levi-Civita connection\footnote{The proof of this can be found in Section 5.3.1 of \cite{kriele}.}. 

Following the previous convention, we denote with a sharp super-script a variation of the metric which leaves the other fields (the connection) fixed, thus, the geometry equations of motion are those for which 
	\beq
	\label{gf.varsg}
	\frac{\d}{\d s} S_g(\Omega)\Big|_{s=0} = \delta^\sharp S_g  = \int_{\Omega} \left[\sqrt{-g}\delta^\sharp R + R \delta^\sharp{\sqrt{-g}} \right]\d\Omega = 0.
	\eeq

We have already calculated the variation of the density $\sqrt{-g}$  in \eqref{gf.sqrt}. Hence, we only need to compute the variation of the curvature scalar. We do this in two steps. First, we consider the variation
	\beq
	\label{gf.varRi}
	\delta^\sharp R = R_{ab} \delta^\sharp g^{ab} + g^{ab} \delta^\sharp R_{ab}.
	\eeq
We notice that the Ricci tensor depends only on the connection coefficients, and therefore its `sharp' variations are identically zero. Thus we only need to compute the covariant components of the metric variations
     \beq
     \label{gf.shcov}
     \delta^\sharp g^{ab}  = -g^{ac}g^{db} \delta^\sharp g_{cd}.
     \eeq
     
Substituting the derivative of $\sqrt{-g}$ given by \eqref{gf.sqrt}, the variations of the curvature scalar \eqref{gf.varRi} and the covariant components of the metric \eqref{gf.shcov} into the variation of $S_g$, equation \eqref{gf.varsg} becomes
	\beq
	\label{gf.fcuk}
	\delta^\sharp S_g = - \int_{\Omega} G^{ab}\delta^\sharp g_{ab} \sqrt{-g}\d\Omega,
	\eeq
where 
	\beq
	\label{gf.einstensor}
	G^{ab} =   R^{ab} - \frac{1}{2}Rg^{ab}
	\eeq
is called the \emph{Einstein tensor}. Following arguments analogous to those leading to equation \eqref{gf.Tcons2} for the stress-energy tensor, \eqref{gf.einstensor} satisfies the Bianchi identities\footnote{This was the fact that originally motivated Einstein to postulate the tensor ${\bf G}$ for the right hand side in equations \eqref{gf.eins}.}
	\beq
	G^{ab}_{\ \ ;b} = 0.
	\eeq

This completes the variational program. It just remains to read the dynamics encoded in the symmetry of general relativity.

\subsubsection{Einstein equations and the equivalence principle}
 
The complete dynamics of general relativity follows solely from the invariance of the Einstein-Hilbert action \eqref{gf.00} under the group of diffeomorphisms of spacetime. Assuming that the matter fields satisfy the Euler-Lagrange equations, general covariance implies
	\beq
	\frac{\d}{\d s} S(\Omega)\Big|_{s=0} = \int_{\Omega} \left[G^{ab} -\frac{1}{2} T^{ab}\right] \delta^\sharp g_{ab} \sqrt{-g}\d\Omega = 0,
	\eeq 
therefore, from the arbitrariness of the diffeomorphisms $\varphi_s$, the equations of motion for general relativity  are 
	\beq
	\label{gf.eins}
	R_{ab} - \frac{1}{2} g_{ab} R = 8 \pi T_{ab}.
	\eeq
These  are ten, coupled, non-linear, second order partial differential equations for the metric coefficients as required by Hilbert's axioms and obtained as a direct consequence of the general covariance of physical laws. Encoded in them is the principle of equivalence. Matter and geometry appear on an equal footing. Their interaction does not distinguish any specific set of physical fields encoded in the stress-energy tensor provided they have the same energy-momentum distribution \cite{deFelice}.  This is the most dramatic insight of general relativity; spacetime curvature is gravity and energy is its source.   

The class of solutions admitted by the system of Einstein's equations is very large. In principle, any metric which is differentiable up to second order in a given region of spacetime could be used to define an energy-momentum tensor. This would leave the task of finding which kind of `matter dynamics' is defined by the geometry prescribed by the chosen metric. The matter models of physical interest, however, are not that arbitrary. There is a set of conditions which must be satisfied by the energy-momentum tensor  in order to consider a theory as physical. In practice, one proposes matter models which satisfy a physically relevant condition and proceeds to find the geometry associated with them.




\section{Irreversible processes}

It may seem that all there is left for classical theories of physics is to write down a Lagrangian density or an energy-momentum tensor which satisfies some energy condition and then solve Einstein's equations for the particular matter model of interest. Here is a sharp distinction between the interpretation of a fundamental theory of nature and the observable phenomena it was abstracted from.

General relativity is the product of the mathematical language of general covariance and the physical principle of equivalence whose dynamics is obtained from the Einstein-Hilbert action. As a dynamical theory it is complete. Then, the question is if dynamics is a faithful description of all natural motions, whether every process described by dynamics can be found in nature. 

We understand a natural process as a consecutive set of motions changing  the properties describing a physical system from one configuration to another. Implicit in this reasoning is the notion of an evolution which makes a clear distinction between the directions of the past and the future, {\it i.e.} it presupposes an   arrow of time. In the vast majority of mechanical contexts, such a distinction is  merely conventional. All the dynamical laws are symmetric with respect to a parameter which we identify with \emph{time}. Thus, we normally follow the convention that a  system naturally evolves in the positive direction of this time parameter.  If the goal of any dynamical theory is to give an account of the natural evolution of physical situations we encounter in the universe, we are faced with the empirical fact that most processes we experience do not possess such a time symmetry - they are \emph{irreversible}.

The aim of this section is to highlight the questions raised by incorporating the Laws of Thermodynamics ({\it c.f.} Section \ref{intro.thermo}) to the empirical basis of a dynamical theory.

\subsection{Equilibrium thermodynamics}

The detailed dynamics of a system composed of a large number of interacting particles in an arbitrary manner is prohibitively complicated. Nevertheless we can still extract a significant amount of physically relevant information if certain restrictions on the collective motion of the individual components are imposed. In this manner, we can get some insight into the most complex scenarios by considering progressively more elaborate steps.

We have argued that thermal equilibrium corresponds to a highly idealised state in which the dynamics of a system are time symmetrical. In this section we will clarify this notion from a macroscopic point of view, leaving the micro-physics details for the following chapter.  Therefore, for any macroscopic system, we say: 
	\begin{quote}
	{\bf Thermal equilibrium} is that state in which the \emph{total} energy is conserved and  its \emph{internal}\footnote{Here,  we understand total energy as a global quantity, while for internal  energy we mean  that possessed by the system's sub-parts.} energy is maximally distributed.
	\end{quote}
	
Notice that this definition does not make any assumption about the material nature of the system. As far as thermal equilibrium is concerned, it makes no difference if we are dealing with atoms, molecules, photons or any other exotic type of matter. This is what gives thermodynamics its universal character. 

The assumption of thermal equilibrium allows us to reduce the very large number of degrees of freedom of a system to a handful of the relevant variables that describe its state. In general, not all such variables are independent, being connected through an \emph{equation of state}. Such equations are not part, nor can be deduced from, the thermodynamic hypotheses. They contain the material information of the system we try to analyse. We shall see that not every relation between the state variables of a given thermodynamic system corresponds to a \emph{real} matter model, for they can lead to dynamical inconsistencies within the underlying fundamental theory. We will explore these matters in further detail in Chapter 6.

As pointed out in \cite{lrrnils}, the number of independent variables can also be reduced if the system has an overall additivity property. To illustrate this, let us  consider the case of a fluid consisting of a single species of particles whose  total energy of  $E$  is related to its entropy $S$, volume $V$ and number of particles $N$ through and equation of state $E=E(S,N,V)$. In this case, we can express the change in the total energy of the system as
	\beq
	\label{thermo.gb01}
	\d E = T \d S + \mu \d N - p\d V
	\eeq
where, by definition, the temperature $T$, chemical potential $\mu$ and pressure $p$ are
	\beq
	\label{thermo.intensive}
	T \equiv \frac{\partial E}{\partial S}, \quad \mu \equiv \frac{\partial E}{\partial N} \quad \text{and} \quad p = - \frac{\partial E}{\partial V}.
	\eeq
In this example, the variables $E$, $S$, $N$ and $V$ satisfy the following scaling property. Let us denote with a tilde the scaled variables
	\beq
	\tilde S = \lambda S, \quad \tilde N = \lambda N  \quad \text{and} \quad \tilde V = \lambda V, \quad  \text{where} \quad \lambda \in \mathbb{R}. 
	\eeq
Then, the total energy of the system becomes	
	\beq
	\label{thermo.additivity}
	\tilde E(\tilde S, \tilde N, \tilde V) = \lambda E(S,N,V).
	\eeq
whilst the variables \eqref{thermo.intensive} remain unchanged. The local form \eqref{thermo.gb01} of the scaled energy takes the form
	\beq
	\d \tilde E = \lambda \d E + E \d\lambda = T \d \tilde S + \mu \d \tilde N - p \d \tilde V.
	\eeq 
Comparing the last two equalities it easy to check that the total energy is
	\beq
	\label{thermo.euler1}
	E = TS + \mu N - pV.
	\eeq 
This result is known as the fundamental \emph{Euler relation} of equilibrium thermodynamics. If instead of working with absolute quantities we use densities, we can rewrite \eqref{thermo.euler1} in the form
	\beq
	\rho + p = T s + \mu n,
	\eeq
where $\rho=E/V$, $s= S/V$ and $n=N/V$ are the energy, entropy and particle number densities, respectively. Now, if we let the scale factor to be  $\lambda = 1/V$, the energy \eqref{thermo.additivity} becomes
	\beq
	\tilde E = (\tilde S, \tilde N, \tilde V) = \tilde E (s, n , 1) = \rho(n,s),
	\eeq 
and the differential form \eqref{thermo.gb01} reduces to
	\beq
	\label{thermo.gibbs}
	\d \rho = T \d s + \mu \d n.
	\eeq
Therefore, whenever a system admits an additivity property of the form \eqref{thermo.additivity}, we can effectively reduce the number of independent variables by one. In our case, equation \eqref{thermo.gibbs} will henceforth be referred as the equilibrium \emph{Gibbs' relation}. It will serve as our thermodynamic reference point in the relativistic discussions of the forthcoming chapters.  

Finally, let us note that thermal equilibrium implies that the rate of change of the total energy with respect to time vanishes identically. Thus, in the case of a single species fluid, the entropy and particle productions should be in balance. In particular, if the particle number is conserved, the entropy, as required by the definition, should be maximal. We will give a clearer explanation of the meaning of maximally distributed energy in the next chapter, when we introduce the microscopic motivation for the concept of entropy. For the time being, let us note that the Gibbs' relation \eqref{thermo.gibbs} completely characterises the state of thermal equilibrium for a fluid\footnote{The extension to the case with more than one species is straightforward.}.  Therefore, for a fluid with constant particle number density, the above definition of thermal equilibrium is equivalent to the weaker version:
	\begin{quote}
	{\bf Thermal Equilibrium (isentropic)} is that state for which the entropy density of the system is a conserved quantity, {\it i.e.}
		\beq
		\d s = 0.
		\eeq
	\end{quote}
	
Notice that this definition of thermal equilibrium is indeed a time symmetric one since the requirement of maximal energy distribution implies that there are no hotter or colder regions; there is no net energy transport from one part of the fluid to the other. In this sense, equilibrium thermodynamics gives an account of collective phenomena in a time reversible fashion.

The above isentropic definition of thermal equilibrium differs from the general one in a crucial point. Suppose we have a thermodynamic system in thermal equilibrium. There are essentially two different ways we can perturb the system away from equilibrium: reversibly and irreversibly. The difference between the two is that, in the former, the total energy of the system is modified without  changing its entropy while, in the latter, entropy is  not a conserved quantity, even when the energy itself may be conserved. 

Here, we have been implicitly assuming that time indeed `flows' in one direction, {\it i.e.} towards the future. Thus, when we speak about reversibility we mean that, from the present state of a system, we have enough information to \emph{retrodict} the system's previous state. This is obviously  the case for thermal equilibrium, although it may come as a surprise that it is also the case of some systems away from it. In such a case, the only relevant auxiliary hypothesis are the zero-th and first laws of thermodynamics discussed in the introduction ({\it c.f.} Section \ref{intro.thermo}); the `time-symmetric'  laws. It is  the second law that  describes systems which, internally\footnote{At the level of description, the distinction between past and future is purely a convention on the direction of the future, not a true feature of the dynamics.},  can distinguish past from future; we gradually lose our ability to retrodict the past while the final state becomes progressively more predictable. Moreover, if the final state of thermal equilibrium is reached, we have lost all previous information about the systems evolution; from the system's point of view, it has never departed from it\footnote{Determinism is lost! One could argue that a system takes an infinitely long time to completely relax to equilibrium and, therefore, its evolution is deterministic for all times.  Surprisingly, we can find purely mechanical examples for which determinism is also lost ({\it c.f.} Norton's slippery dome \cite{nortonfolk}). The solution to this apparent loss of predictability can be traced down to the differentiability of the system's constraints (in the case of Norton's dome the surface is analytic everywhere except at the summit).}. We can give a more precise statement of the second law for a system which has been left to evolve without any external influence:
	\begin{quote}
	{\bf Second law of thermodynamics} The entropy of a system, which internally can distinguish the orientation of time, always increases towards the future, {\it i.e.}
	\beq
	\label{thermo.2ndlaw}
	\frac{\d S}{\d t}  \geq 0.
	\eeq 
	\end{quote}

Let us denote  the rate of change of entropy \eqref{thermo.2ndlaw} by $\Gamma_\s$ and call this quantity the \emph{entropy production}. To give account of irreversible processes we need a detailed study of situations when the entropy production does not vanish. We cannot  be satisfied with the simple inequality stated by the second law. It is our task to relate entropy production to a well-defined physical processes. 

\subsection{Linear Irreversible Thermodynamics}

This subsection is devoted to the first step in the study of irreversible processes. The physics of systems far away from thermal equilibrium remains a topical problem. Thus, the natural approach towards a theory of irreversible processes would seem to require an expansion in deviations from the state of equilibrium. Here, we give an account of the Linear 
Irreversible Thermodynamics (LIT) programme pioneered by Onsager and later developed by Prigogine \cite{prignetex,prigvar}. The central result of this subsection are the Onsager reciprocity relations, of which Fourier's law is a consequence. We will argue that, although it has been experimentally verified to a good level of accuracy, LIT will necessarily fail  when the relaxation time scale of thermal phenomena is \emph{long} compared with the dynamical one - {\it e.g.} in superfluid helium - even when the deviations from equilibrium are not large.  Therefore, although LIT will not play a significant role in this thesis, this brief introduction will serve as a motivation for the Extended Irreversible Thermodynamics (EIT) programme which permeates the multi-fluid approach to heat conduction presented in Chapter 5.

A simple evaluation of the entropy production becomes possible if we assume that entropy away from equilibrium depends on the same variables  as it does while in equilibrium. We will introduce no restriction as to the nature of the system except that we shall assume that, although the system as a whole will not be in equilibrium, there exists at every point a state of \emph{local equilibrium} for which the entropy is given by the classical Gibbs formula \eqref{thermo.gibbs}. This restricts us to systems whose evolution may be described in terms of macroscopic thermodynamics and hydrodynamics without explicit introduction of molecular concepts.

The central quantity is the entropy production per unit time \eqref{thermo.2ndlaw}. We should notice that this will become an issue when dealing with relativistic systems, when we will have to refer to a particular frame to have a definite interpretation. For the time being we shall consider the Newtonian framework where thermodynamic systems  are Galilean invariant, and hence  have a universal notion of time. In this case, the entropy production per unit time may be expressed as the volume integral
     \beq
     \label{fr.ep}
     \frac{\d S}{\d t} = \int \sigma \d V \geq 0,
     \eeq
where $\sigma$ denotes the entropy production per unit time and volume. 

It is very rare that thermodynamic processes occur in complete isolation. It was noted by Onsager that when two or more irreversible processes take place simultaneously in a thermodynamic system, they may interfere with one another\cite{ons1}; leading  to the discovery of the celebrated reciprocity relations. The central assumption in Onsager's reasoning was that of \emph{microscopic reversibility}, which guarantees an independent balancing of different classes of molecular processes maintaining statistical equilibrium. This balance is a crucial point if we want to extend the LIT programme to physical regimes where non-inertial effects may not preserve the statistical equilibrium of the system.

The LIT programme posits that the entropy production  $\sigma$ is a bilinear form of the generalised forces $X_i$ (affinities, gradients of temperature, gradients of chemical potentials, etc.), and fluxes $J_i$ representing the rates of the various irreversible processes taking place (chemical reactions, heat flow, diffusion, etc.)
     \beq
     \label{fr.entrprod}
     \sigma = \sum_i J^i X_i \geq 0.
     \eeq 
     
It should be noted that the validity of the Gibbs relation \eqref{thermo.gb01} is assumed in \eqref{fr.ep}. The Gibbs relation, however, can only be established in some neighbourhood of equilibrium which defines the region of local equilibrium. At thermal equilibrium the fluxes and forces vanish identically, {\it i.e.}
     \beq
     J^i = 0, \qquad X_i=0
     \eeq
for all irreversible processes simultaneously. This is the justifying fact that allows us to assume linear homogeneous relations between the forces and the fluxes for conditions near equilibrium. 

In this representation, the fluxes are the unknown quantities, and the forces are given functions of the state variables or their gradients. Thus, we can then write
     \beq
     J^i = J^i(X_1, X_2,..,X_n).
     \eeq 
Therefore, in the neighbourhood of equilibrium, we can make the expansion
     \beq
     J^i = \sum_j \left(\frac{\partial J^i}{\partial X_j} \right) X_j + \mathcal{O}(X_j X_k), 
     \eeq
and neglect second order terms\footnote{Microscopic reversibility assumption.} to obtain the fluxes as linear transformations of the forces
     \beq
     \label{fr.onsa01}
     J^i = \sum_j L^{ij}X_j.
     \eeq
Here the coefficients $L^{ij}$ are measurable constants which are defined in terms of correlation functions \cite{mpptext}.

Substituting the \emph{linearised} fluxes \eqref{fr.onsa01} into the entropy production \eqref{fr.entrprod} gives us the form 
     \beq
     \label{fr.cons01}
     \sigma = \sum_{i,j} L^{ij} X_i X_j.
     \eeq
The positivity of entropy production  constrains  the coefficients $L_{ij}$ to be a \emph{quadratic form}. Thus, the simplest constitutive equations which guarantee the semi-positiveness of the rate of entropy production are given by Onsager's reciprocity relations
     \beq
     \label{fr.recip}
     L^{ij} = L^{ji}.
     \eeq
The Onsager relations are the central result in the linear thermodynamics of irreversible processes programme. Their strength  resides in their generality; they showed for the first time that non-equilibrium thermodynamics leads to general results independent of the molecular model used \cite{prignetex}. 

Onsager's reasoning  was inspired by William Thomson's idea of a variational principle for dissipative systems, namely, a \emph{principle of least dissipation of energy}. There is no evidence of such a principle for a generic non-equilibrium system. However, for relatively small deviations from thermal equilibrium one can argue in favour of a \emph{principle of least entropy production} \cite{prigvar}. To illustrate this point, let us consider the case of heat conduction alone. In this case we are left only with the single pair
     \begin{subequations}
     \begin{align}
     \label{irr.fourier}
     J^q & =-\kappa \frac{\partial T}{\partial x},\\
     X_q   & =\frac{\partial}{\partial x}\left(\frac{1}{T} \right),
     \end{align}
     \end{subequations}
where $J^q$ is the one-dimensional heat flow, $T$ is the absolute temperature and $\kappa$ is the thermal resistivity. The volume integral \eqref{fr.ep} becomes
     \beq
     \frac{\d S}{\d t} = \int  \frac{\kappa}{T^2}\left(\frac{\partial T}{\partial x} \right)^2 \d V.
      \eeq
If in the whole system the temperature deviates locally from some average $T_0$ 
     \beq
     T=T_0 + \delta T,
     \eeq
neglecting higher order terms, the perturbed entropy production is
     \begin{align}
     \label{fr.minentr}
     \delta \frac{\d S}{\d t} & = \int  \frac{\kappa}{T_0 + \delta T}\left(\frac{\partial \delta T}{\partial x} \right)^2 \d V\nn\\
                              & = \int \ \frac{\kappa}{T_0\left(1 + \delta T/T_0\right)}\left(\frac{\partial \delta T}{\partial x}\right)^2\d V.
     \end{align}
Since the temperature is a function of the position only, we write the temperature perturbation in terms of the displacement $\delta x$ as
          \beq
          \delta T= \frac{\partial T}{\partial x} \delta x.    
          \eeq
Hence, since $\delta T/T_0\ll 1$, the variation of entropy production \eqref{fr.minentr} becomes
     \beq
     \label{thermo.minent}
     \delta \frac{\d S}{\d t} = \frac{1}{T_0}\int  \kappa \left(\frac{\partial^2 T}{\partial x^2} \right)^2 \delta x\d V.
     \eeq
     
Note that the \emph{minimum} of \eqref{thermo.minent} corresponds to the stationary state away from equilibrium
     \beq
     \kappa \frac{\partial^2 T}{\partial x^2} = 0.
     \eeq
Prigogine referred to this result as an example of a kind of inertial property of non-equilibrium systems
	\begin{quote}
	\ldots When given boundary conditions prevent the system from reaching thermodynamic equilibrium (that is, zero entropy production) the system settles down to the state of least dissipation \ldots \cite{prignetex}
	\end{quote}

It should be stressed that the theorem of minimum entropy production is formulated under the assumption of small deviations from equilibrium. It can easily be shown that systems far from thermodynamic equilibrium could behave in quite a different manner, in fact, directly opposite from the one expected from the minimum entropy production theorem.

It was also noticed in \cite{joueit} that this approach has some other limitations, the most important for us being that of the parabolic nature of Fourier's law. This implies that thermal disturbances propagate information with boundless speed. Cattaneo proposed a solution to this problem by taking into account the microscopic time-scale in which such disturbances would propagate information in a material medium, thus deducing a hyperbolic equation for heat.


\subsubsection{Hyperbolic heat conduction: a road to Extended Thermodynamics}

 The one dimensional version of Fourier's law \eqref{irr.fourier} together with an energy balance of the form
 	\beq
 	\frac{\partial}{\partial t} T(t,x) = - \frac{\partial}{\partial x} J^q(t,x),
 	\eeq 
leads to the heat equation
     \beq
      \label{cvf.01}
     \frac{\partial}{\partial t}  T(t,x) - \kappa \frac{\partial^2}{\partial x^2} T(t,x) = 0.
     \eeq  

We may write the solution of \eqref{cvf.01} for an initial value problem  in an unbounded region of space in the form
     \beq
     T(t,x) =\frac{1}{(4\pi \kappa t)^{3/2}}\int_{-\infty}^{\infty} T(0,y)\exp\left[-\frac{(y-x)^2}{4\kappa t} \right] \d y.
     \eeq

This solution implies that $T(t,x)$ is non-zero zero for all $x$ and $t>0$, even when $T(0,x)$ may have support on a finite interval only. 

The above result is far from  satisfactory. One would expect that temperature disturbances have a bound on their propagation speeds given by the material internal structure. In this interpretation, heat conduction can be understood as the energy exchange during collisions of the molecules forming the material. The chance of being hit depends on the length a molecule travels before hitting another one. This length is the mean free path. Cattaneo proposed a modification to Fourier's law to incorporate this view
	\beq
	\label{dis.insf}
	\tilde J^q = - \kappa \left[\frac{\partial T}{\partial x} - \tau \frac{\partial}{\partial t}\left(\frac{\partial T}{\partial x} \right) \right],
	\eeq
where $\tau$  corresponds to a real parameter with units of time. He called \eqref{dis.insf} `non-stationary' Fourier's law. This leads to
	\beq
	\frac{\partial}{\partial t} T(t,x) = \kappa \left[\frac{\partial^2}{\partial x^2}T(x,t) - \tilde\tau \frac{\partial}{\partial t}\left(\frac{\partial^2}{\partial x^2} T(t,x) \right)\right].
	\eeq
Still, this is not a hyperbolic differential equation, this attempt did not remedied the problem. Thus, Cattaneo proposed a further modification of \eqref{dis.insf}. He assumed that the the operator $\tau(\partial/\partial t)$ was `sufficiently' small so the difference between $J^q$ and $\tilde J^q$ is negligible. Thus, he could  approximate  the non-stationary Fourier's law by 
	\beq
	\label{dis.cattaneo}
	\tilde J^q(t,x) + \tilde \tau \frac{\partial }{\partial t}\tilde J^q(t,x) = -\kappa \frac{\partial}{\partial x} T(t,x),
	\eeq
where $\tilde \tau=\tau/\kappa$. This result is called the \emph{Cattaneo equation}. This will be the reference point in the coming chapters, when we deal with hyperbolic theories of dissipation. The reason for this is that it leads to a telegraph equation for thermal disturbances, {\it i.e.} 	
	\beq
	\label{thermo.telegraph}
	\tilde \tau \frac{\partial^2}{\partial t^2} T(t,x) + \frac{\partial}{\partial t} T(t,x)  -\kappa \frac{\partial^2}{\partial x^2} T(t,x) = 0.
	\eeq

\begin{figure}
	\begin{center}
		\includegraphics[width=.45\columnwidth]{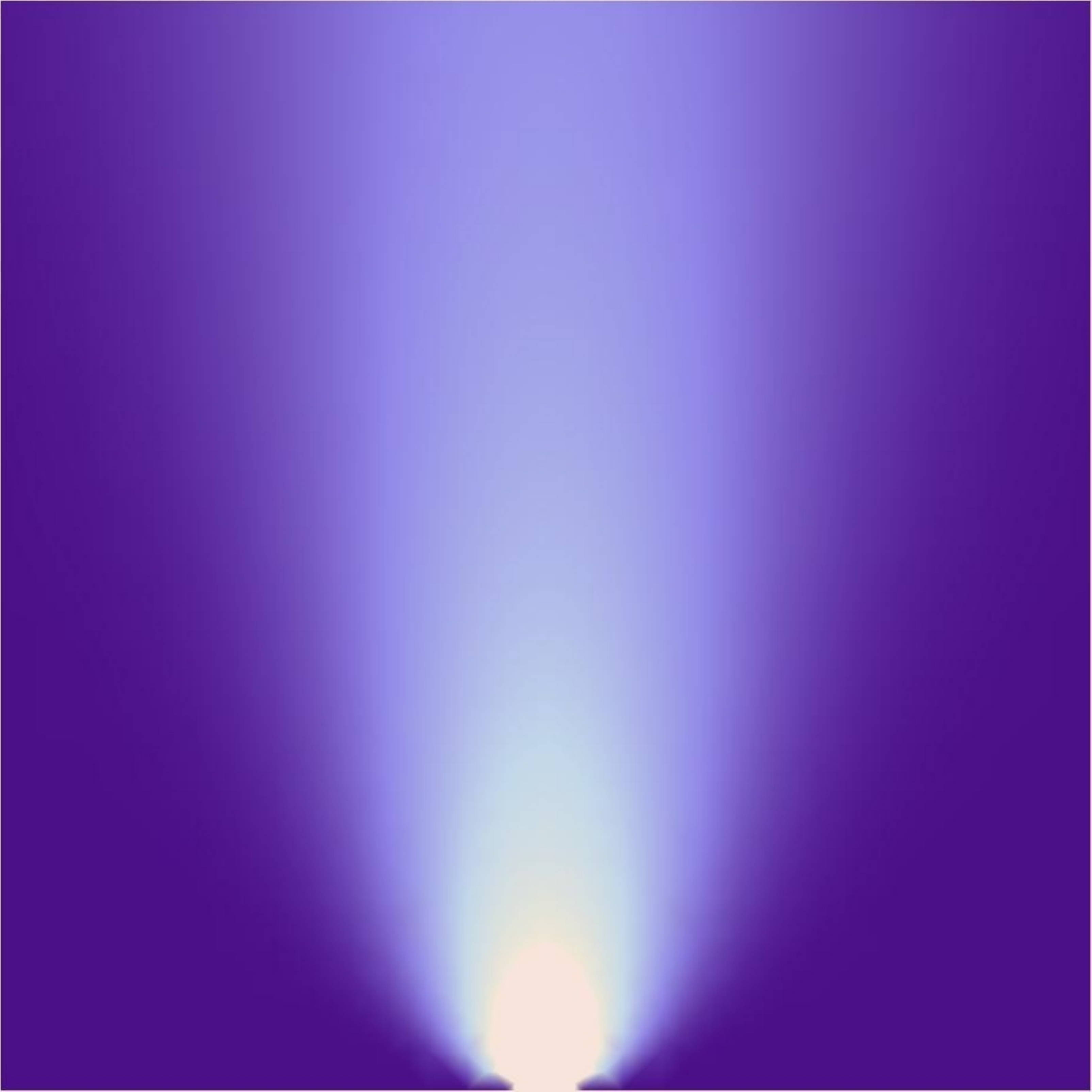} \includegraphics[width=.45\columnwidth]{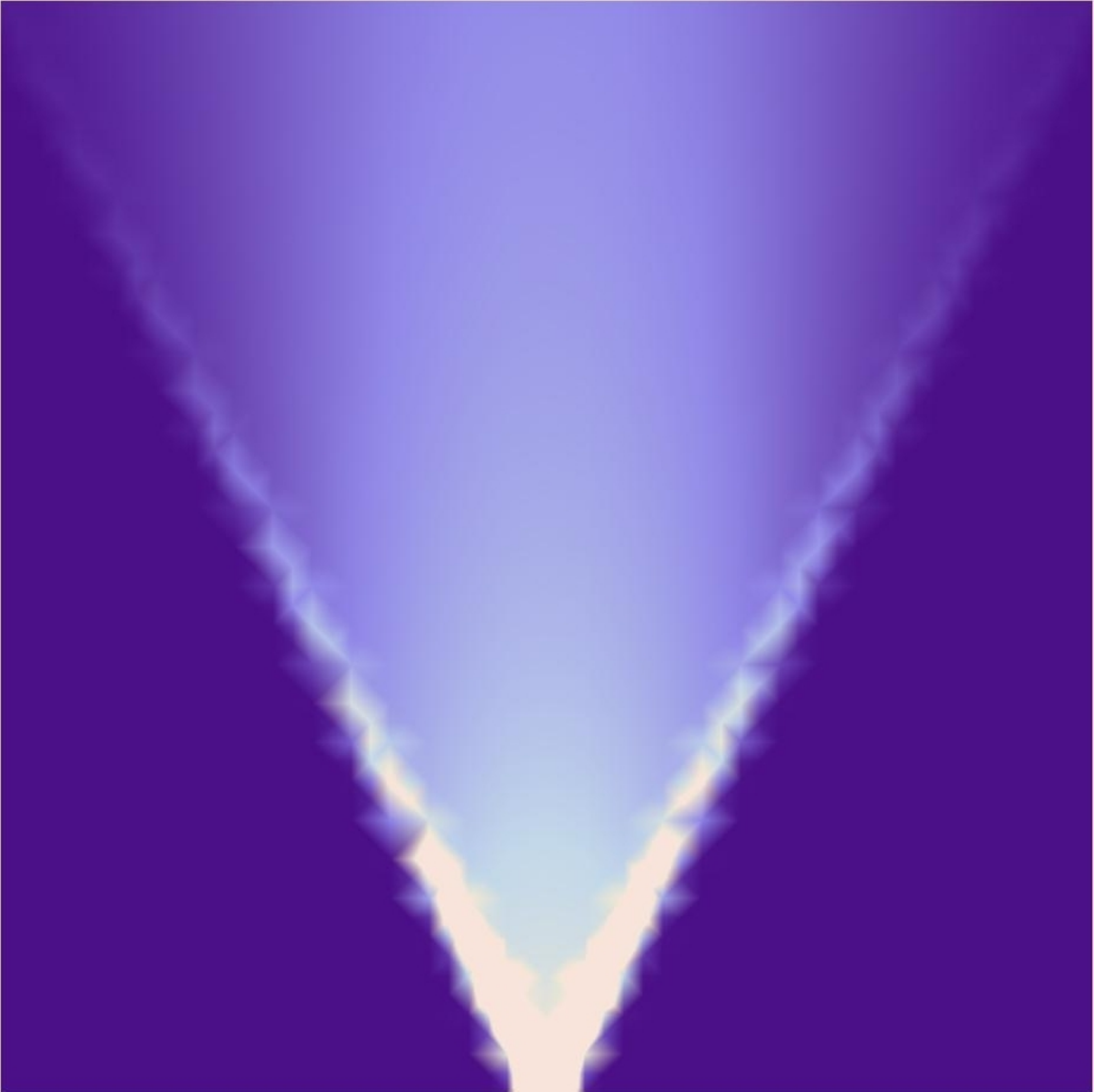}\\[.1cm]
		\includegraphics[width=.455\columnwidth]{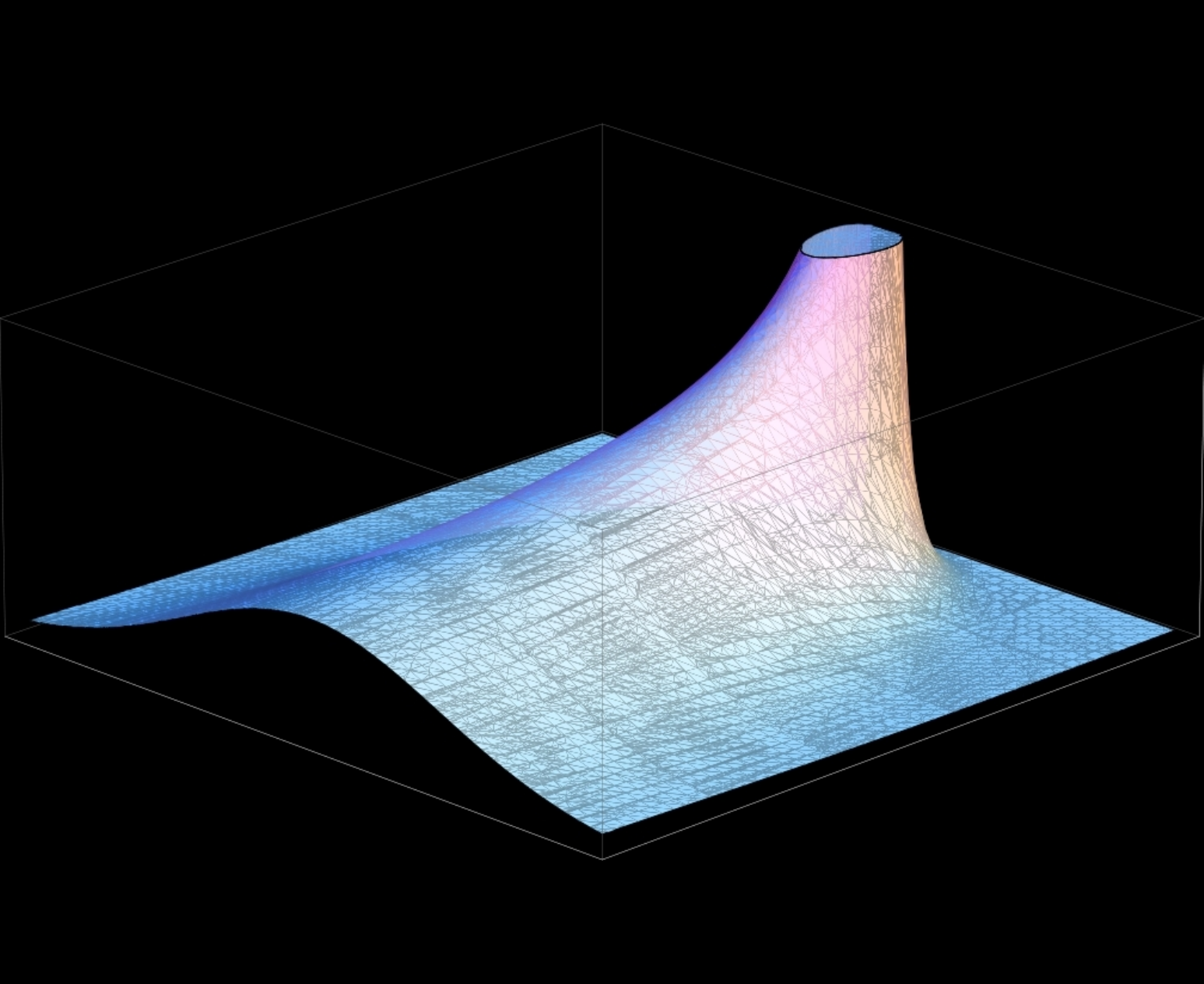}\includegraphics[width=.455\columnwidth]{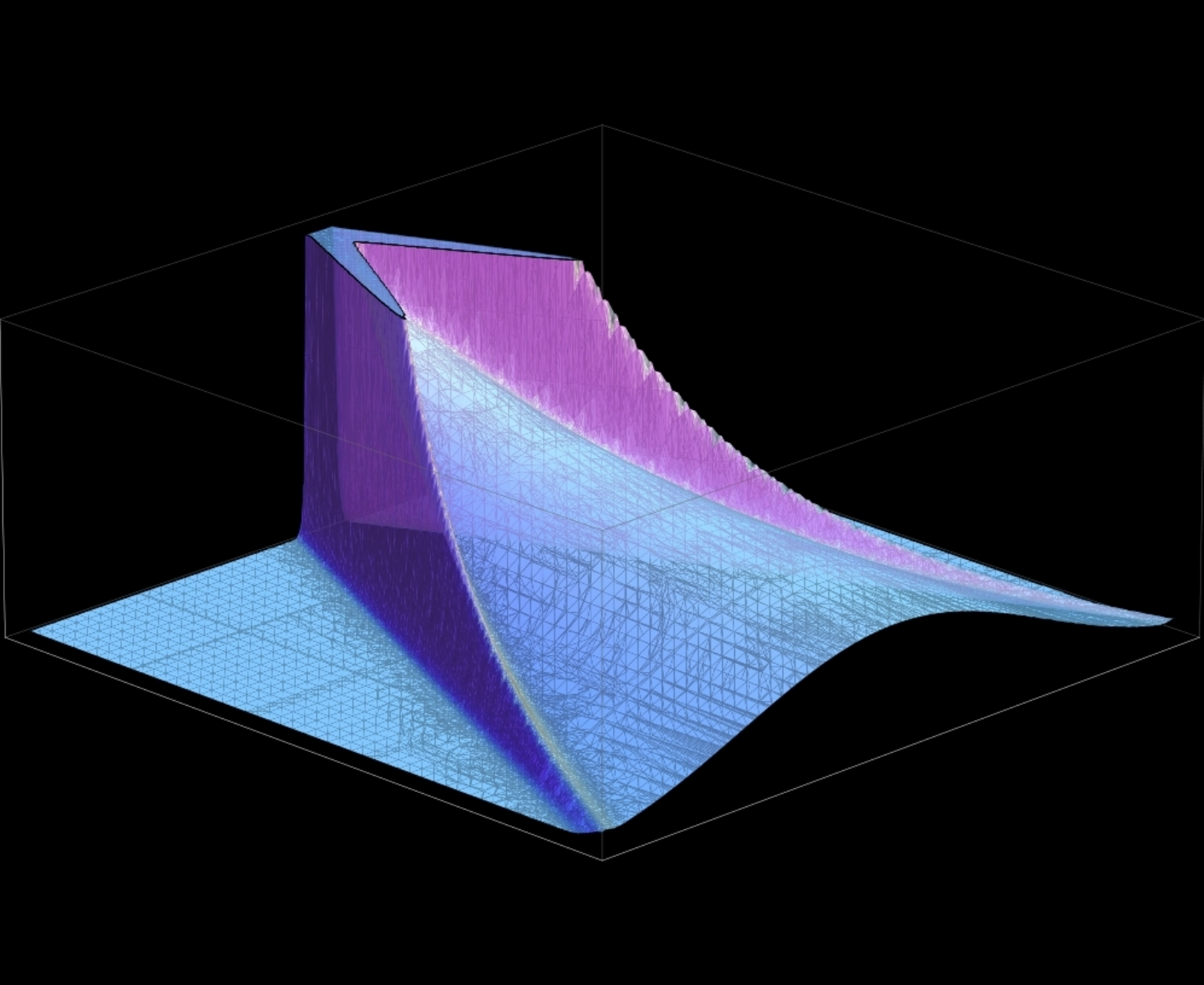}
	\end{center}
	\caption{This figure shows a comparison of the evolution of a thermal disturbance modelled with the heat (left) and telegrapher (right) equations. The vertical axis represents time, while the horizontal corresponds to one-dimensional space. The initial condition is a delta function at the origin of the spatial direction. The evolution occurs towards the top of the page. We can see that, for the heat equation, the pulse spreads over the entire region without any lag. On the contrary, the evolution provided by the telegraph equation propagates a thermal signal  which dissipates away to merge the long term solution of the heat equation (the transition between `fast' and `slow' heat flux). The bottom graph shows a how the two distinct evolutions become indistinguishable in the `long' run. Here, the `z'-axis corresponds to temperature, while the spatial and temporal directions should be obvious from the description of the top images.} 
	\label{graf.01}
\end{figure}

Equation \eqref{thermo.telegraph} is hyperbolic if  the parameter $\tilde \tau >0$. In such a case, it predicts the propagation of heat pulses at the finite  speed
	\beq
	v= \pm \sqrt{\frac{\kappa}{\tilde \tau}}.
	\eeq
Thus, we see that for arbitrary small values of the `relaxation' time $\tilde \tau$ the characteristic speeds become unbounded. Figure \ref{graf.01} shows a comparison of the evolution of a pulse by considering the heat and the telegraph equations, respectively.

Note that in order to obtain the hyperbolic heat equation \eqref{thermo.telegraph} two assumptions were made. The first is that the `modified' heat flux $\tilde J^q$ is no longer related to a `force gradient' as in the ordinary Fourier's law.  The second is the assumption that the quantity
	\beq
	\tau\frac{\partial}{\partial t}\left(\frac{\partial T}{\partial x}\right)
	\eeq
is small, thus setting a dissipation time-scale which is long compared to the system's dynamical evolution. This kind of behaviour is better seen in liquids, as in the case of the second sound in superfluid helium or the dissipation to solidification time scales  of molten polymeric fluids\footnote{An example of a polymeric fluid is `gel-wax', a fluid typically used in  candle making. One can try the following experiment: put some molten gel wax in a Petri dish to form a layer of about $1$mm. Heat from below in a uniform manner trying to preserve a steep temperature gradient between the bottom and the surface of the fluid. With enough luck hexagonal convection patterns (Benard cells) will form. Remove the Petri dish from the heat source. If the cooling down is sufficiently uniform, the wax will stiffen before the patterns are completely dissipated.}. In these cases, heat acts as a `slow' variable of the process. Thus, one could treat the heat flux as if it were a state variable. This constitutes the motivation to introduce Extended Irreversible Thermodynamics.

\subsection{Extended Irreversible Thermodynamics}

From the discussion in the previous section, the reader may have some feeling of discomfort about the applicability of the heat equation. For many `practical' applications, Fourier's law is a very good approximation. However, when  one is interested in sufficiently rapid phenomena, the  LIT description is not good enough. Moreover, when the underlying principles of the dynamical theory\footnote{The underlying Lorentzian geometry encoding the `principle of causality' through general covariance.} prevent signalling faster than a critical speed, then the traditional approach is simply unacceptable. Therefore, there is a clear conflict between  Linear Irreversible Thermodynamics  and relativity.

The general philosophy behind a thermodynamic description of macroscopic phenomena is that of averaging. In such a process, one reduces the (very large) number of degrees of freedom, to a manageable set of `thermodynamic variables' which are directly related to conserved quantities, {\it i.e.} mass, energy and momentum. All other variables decay very rapidly in the averaging process and thus they are not conserved, {\it e.g.} heat. If, however, the time scale of certain phenomena is `fast enough', the decay of some of these variables may become slow compared to the total duration of the phenomenon. In this case, the otherwise not conserved fast variable, can be treated as a slow, conserved one. The Extended Irreversible Thermodynamics programme (EIT) was designed to systematically approach thermodynamic phenomena whose set of conserved slow variables is enlarged by a set of `dissipative' yet slow ones which complement the thermodynamic description of a system evolving away from thermal equilibrium. The interested reader is referred to the work by Jou \cite{joueit}.

Continuing our previous fluid example ({\it c.f.} Section 2.2.1), in EIT one considers the energy of the system to be a function of the equilibrium (conserved) quantities and the dissipative (slow) fluxes. In our case, we have
	\beq
	E=E(N,S,V,\tilde J^q),
	\eeq
or, equivalently, the energy density is written as
	\beq
	\rho = \rho(n,s,q).
	\eeq
Thus, we can write an `extended' Gibbs relation of the form
	\beq
	\d \rho = \mu \d n + T \d s + \mu_q \d q,
	\eeq
where $\mu_q$ stands for the `heat chemical potential' given by
	\beq
	\mu_q \equiv \frac{\partial \rho}{\partial q}.
	\eeq
	
It is one of the aims of this work to show how an extended Gibbs relation leads naturally to a hyperbolic description of heat conduction in a relativistic setting. The details of this will be given in Chapter 5.

\section{Irreversible processes in general relativity}

In this chapter, we have introduced the two main components of this thesis. On the one hand, we consider the general theory of relativity as our fundamental dynamical theory ({\it c.f.} Section 2.1, footnote 3). The purpose of presenting the full variational construction from the Einstein-Hilbert action was two-fold: 
	\begin{enumerate}
	\item it allowed us to spell out unambiguously the meaning of general covariance in the context of this work, and
	\item to highlight that the connection of the mathematical formulation of a dynamical theory with the empirical world is manifest through the Lagrangian function and its symmetries.
	\end{enumerate}  
It is hard to overestimate the importance if these two points, specially to correctly interpret the outcome of chapters 5 and 6 - the core of this manuscript.

On the other hand, we acknowledge the fact that many dynamical processes allowed by general relativity, have never been observed in nature\footnote{This kind of induction does not imply that such processes will never be observed, but we may have to wait a `very' long time.} and therefore, it may seem the they are not allowed by some \emph{physical law} as a consequence of its `time-symmetric' nature\footnote{The choice of the direction of time is a matter of convention.}. As discussed in chapter 1, here we adopt the view that thermodynamics is a collection of self-consistent additional hypothesis - the laws of thermodynamics - that  we must  add to our dynamical theory to obtain a \emph{stronger} description about nature\footnote{The addition of the laws of thermodynamics to any dynamical theory  `forbids' a larger class of processes to occur than the dynamics alone would. In this sense it is `easier', at least in principle, to falsify. Thus, we obtain a stronger description of nature \cite{popper}.}.

If we agree that the principles of general relativity and the laws of thermodynamics form the empirical basis of our macroscopic description of nature, how do we assess our individual models of physical processes? In the forthcoming discussion we will show that in essence, the physical information contained in the Lagrangian function is equivalent to the one given by an equation of state. Thus, in a more complete view of natural processes,  we can describe a system's thermodynamics in a dynamical way or, equivalently, we can give a thermodynamic description of its dynamics. 
 
With this, we close our discussion about the framework of this thesis.

			\chapter[Relativistic kinetic theory]{Relativistic kinetic theory}

Irreversible processes in general relativity represent an outstanding challenge for our efforts to understand the
dynamics of macroscopic systems. The equations governing such dynamics were laid down in the previous chapter.  In this chapter we deal with the matter content of those equations. The aim is to give a sound microscopic motivation for our later developments of relativistic non-equilibrium thermodynamics.

Although most of the material in chapter is not new, we have tried to carefully spell out subtleties often omitted in more specialised literature. In particular, on arguments typically invoked in `proofs' of the fundamental theorem of irreversible processes; Boltzmann's $\mathcal{H}$-theorem. The following is an account of the results of non-equilibrium relativistic kinetic theory which are relevant to keep this work self-contained. Central discussions relating to the Einstein-Boltzmann or Einstein-Vlasov systems, have been entirely omitted.  The interested reader is referred to the classic work of Synge \cite{synge}, Tauber \& Weinberg \cite{tauber}, Stewart \cite{KTStewart},  Ehlers \cite{ehlers}, Treciokas \& Ellis \cite{ellistreciokas1}, Ellis, Matravers \& Treciokas \cite{ellistreciokas2} and references therein for a range of different aspects of relativistic kinetic theory. 

The spirit of the following discussion, as a preamble to our treatment of relativistic dissipative fluids,  is the one highlighted by Treciokas \& Ellis \cite{ellistreciokas1}. Therefore, we will use kinetic theory  to assist us  to obtain the form of the macroscopic transport equations. Hence, rather than use it to examine the nature of the fluid theory, it will serve as a motivation for it.

\section[Symplectic structure of phase space]{The symplectic structure of the relativistic phase-space}

This section is the starting point for our exploration of  processes in nature which exhibit a `time-asymmetric' evolution. It was stated in the previous chapter (see footnote 3) that, within this work, we consider a theory as physical only if it can accommodate the features of our observed world, moreover, we say that a theory is fundamental if it can be derived from a variational principle. One can immediately raise an objection  - with  well justified claims - that, under such consideration, we would have to exclude dissipative systems altogether. The reason for this is that, in the strict sense, variational techniques only apply to conservative systems. The view sustained in this thesis is more relaxed. Therefore, the rest of the present work will be devoted to argue that, through the insights stemming from the tenets of relativity, a variational construction for transport phenomena is possible.       

The central aim of this section is to present a covariant description of the underlying microscopic theory of relativistic hydrodynamics. The reason for this exercise is two-fold. On the one hand, it will exhibit the role played by general covariance in the context of classical mechanics when formulated in terms of states on the \emph{relativistic phase-space}. Furthermore, it will remove the notion of `universal time' from the formulation of the second law of thermodynamics.  

The  work  presented in this section was motivated by the relativistic treatment of kinetic theory as given by Synge \cite{synge}, Stewart \cite{KTStewart} and Ehlers \cite{ehlers}. The mathematical account is well known for non-relativistic systems and can be found in the classical text by Arnold \cite{arnold}. A relativistic version of the same symplectic formalism has been given by Rovelli \cite{rovelliQG}. Hence, this brief introduction to relativistic kinetic theory is not new. However, the discussion regarding the covariant formulation of the second law of thermodynamics is presented from a novel point of view, allowing us to introduce the notion of `entropy as a fluid'.

Let us commence our discussion by considering a simple gas, {\it i.e.} a collection of a \emph{large} number $N$ of particles of a single species. As the particles move through spacetime collisions might occur. We will assume for the moment that they are completely elastic so that momentum, mass and number of particles remain conserved. When the particles are not colliding, the principle of equivalence asserts that they move along the \emph{free fall} trajectories of the spacetime metric ${\bf g}$, {\it i.e.} the non-intersecting curves, $\gamma:I\subset\mathbb{R} \longrightarrow \mathcal{M}$, which satisfy  the equations of motion obtained from the Lagrangian function, $\Lambda:T\mathcal{M} \longrightarrow \mathbb{R}$\glo{$\Lambda$}{Lagrangian function (Master function in Chapter 5)},
	\beq
	\label{kin.lagrangian}
	\Lambda(x^c,\dot x^a) = \frac{1}{2}  g_{ab}(x^c)\dot x^a \dot x^b.
	\eeq  
Here $\dot x^a$ denotes the components of an element of the tangent space  $T_x\mathcal{M}$\glo{$T_x\mathcal{M}$}{Tangent space to spacetime at a point $x$} at a point $x$ on the spacetime manifold $\mathcal{M}$. This defines the set of `natural' coordinates - the eight functions $(x^a,\dot x^a)$ - for the tangent bundle of spacetime $T\mathcal{M}$\glo{$T\mathcal{M}$}{Tangent bundle whose base is spacetime}. Through the rest of this work, a dot  represents  differentiation with respect to the affine\footnote{A parameter is called affine if it is related to proper time by an affine transformation: $\tau\rightarrow \lambda = a \tau + b$ where $a$ and $b$ are constants and $\tau$ denotes proper time.} parameter $\lambda$. 

The spacetime metric provides us with an isomorphism between the tangent and the co-tangent bundles of spacetime. Therefore, we can alternatively describe the same free-fall trajectories on $T^*\mathcal{M}$\glo{$T^*\mathcal{M}$}{Co-tangent bundle whose base is spacetime}. Let us define the canonical momentum conjugate to $\dot x^a$ by the spacetime one-form $p\in T_x^*\mathcal{M}$ whose components are given by
	\beq
	p_a = \frac{\partial L}{\partial \dot x^a}.
	\eeq
In this case, the free-fall dynamics of a system of test particles  is encoded in   the free Hamiltonian function, $H:T^*\mathcal{M} \longrightarrow \mathbb{R}$, defined by\glo{$H$}{Hamiltonian function}
	\beq
	\label{freeH}
	 H(x^c,p_a)=\frac{1}{2}g^{ab}(x^c)p_a p_b.
	\eeq
Here, the set of natural  coordinates for $T^*\mathcal{M}$ is given by the eight coordinate functions $(x^a,p_b)$. As the reader may anticipate, the dynamics is independent of such a particular choice of coordinates. However, from an operational point of view, the natural coordinates are indeed `preferred'. In the context of classical mechanics they are referred to as `canonical coordinates'. 

Notice the use of the covariant and contravariant metric representations in the Lagrangian and Hamiltonian descriptions respectively. Their appearance highlights the role of the metric components as potentials of the gravitational field in a non-relativistic situation\footnote{Consider the \emph{weak} field limit of general relativity. In this case we can decompose the spacetime metric into a Minkowski part and a perturbation
	\[
	g_{ab} = \eta_{ab} + h_{ab}, \quad \text{where} \quad \eta_{ab} = \text{diag}(-1,1,1,1) \quad \text{and} \quad |h_{ab}|\ll 1.
	\]
It follows from the definition of the inverse metric 
	\[
	\delta_a^{\ b} = g_{ac}g^{cb} = \left(\eta_{ac} + h_{ac}\right)\left(\eta^{cb} + \tilde h^{cb} \right) = \delta_{a}^{\ b} + \eta_{ac}\tilde h^{cb} + h_{ac}\eta^{cb} + \mathcal{O}(h^2).
	\]
Thus, keeping only the linear terms in $h$, it follows that
	\[
	\eta_{ac}\tilde h^{cb}  + h_{ac}\eta^{cb} = 0, 
	\]
and we have that the inverse metric is simply given by
	\[
	g^{ab} = \eta^{ab} - h^{ab}.
	\]

To obtain the Newtonian limit, we use the equivalence principle, {\it i.e.} from the geodesic equation and the equivalence between `inertial' and `gravitational' masses of test particles, to make the identification
	\[ 
	h_{00} = - 2 \Phi,
	\] 
where $\Phi$ is the Newtonian gravitational potential. Thus, we can compute both the Lagrangian and Hamiltonian [\eqref{kin.lagrangian} and \eqref{freeH} respectively], for a test particle of mass $m$ moving with some velocity $\vec{v}$ to obtain the familiar expressions of classical mechanics (modulo an infinite additive constant!)
	\[
	\Lambda= \frac{1}{2} m v^2 - m \Phi \quad \text{and} \quad H = \frac{1}{2} m v^2 + m\Phi.
	\]
}.

Since the Hamiltonian \eqref{freeH} does not depend explicitly on the parametrisation of the curve, its numerical value is constant on the hypersurfaces   defined by the pseudo-sphere
	\beq
	\label{shell}
	g^{ab}(x) p_a p_b = -m^2.
	\eeq 
If $m>0$,  we call $m$  the mass of a test particle. Two curves passing through the same spacetime event with the same direction differ only in their parametrisation. Thus we see that their momenta define different ``shells''. We say that the particles have different masses and consequently we call the pseudo-sphere \eqref{shell} the \emph{mass shell} for particles with mass $m$.

In non-relativistic classical mechanics, the phase-space of a system of $N$ particles is defined by the $6N$-dimensional manifold whose elements are the collection of spatial locations and momenta of each particle, {\it i.e.} the state of the system. The natural generalisation to a relativistic case would be the $8N$-dimensional manifold with the restriction imposed by the underlying Lorentzian symmetry of spacetime. The relativistic single-particle phase-space is then the collection of points $x^a$ in spacetime together with all the allowed momenta $p_a$ through each point. This implies that $p_a$ should either be timelike or null and future directed. Thus we say that the relativistic single-particle phase-space $\mathcal{P}$\glo{$\mathcal{P}$}{Relativistic single-particle phase-space} is the future non-spacelike part of $T^*\mathcal{M}$ whose boundary $\partial \mathcal{P}$ corresponds to states of massless particles.  Alternatively, we can define the relativistic phase space in terms of the mass of the test particles. The mass shells \eqref{shell} form 7-dimensional subsets of $T^*\mathcal{M}$. Let us denote by $\mathcal{P}_m$\glo{$\mathcal{P}_m$}{Mass shell of mass $m$}  each subset for different values of $m$. We define the single particle phase-space as the collection of all the  individual mass-shells
	\beq
	\label{sym.union}
     \mathcal{P} = \bigcup_m \mathcal{P}_m.
     \eeq  

The dynamics of a system described by a Hamiltonian function  follows from an action principle. However, the role played by the Hamiltonian is different from the one played by the Lagrangian in the variational principle discussed in Chapter 2. This is a direct consequence of the symplectic structure of $\mathcal{P}$.

\subsection{Forms and motion}

It was previously said that the `natural' choice of coordinates for $T^*\mathcal{M}$ was the `pair' $(x^a, p_a)$, where the $x^a$ are the coordinates representing a point $x\in\mathcal{M}$ and the $p_a$ are the components of the one-form
	\beq
	\label{symp.onecan}
	\tilde\alpha = p_a \d x^a, \qquad \text{where} \quad \tilde\alpha \in T_x^*\mathcal{M}.
	\eeq  
By definition, the one-forms that we have encountered so far are elements of the dual to the  tangent space at each point of spacetime, {\it i.e.} they are functions which assign a real number to each tangent vector of spacetime. In particular
	\beq
	\label{symp.onecan1}
	\tilde\alpha:T_x\mathcal{M} \longrightarrow \mathbb{R}.
	\eeq 
The natural projection of the co-tangent bundle, $\pi: T^*\mathcal{M} \longrightarrow \mathcal{M}$, induces the \emph{push-forward}  map $\pi_* : T\mathcal{P}\longrightarrow T\mathcal{M}$. Therefore, given a tangent vector $X\in T_m\mathcal{P}$, we observe that the action of the one-form \eqref{symp.onecan1} on the pushed-forward vector $\pi_* X\in T_x \mathcal{M}$, whose base point is $x = \pi(m)$, defines  the phase-space one-form 
	\beq
	\alpha:T_m\mathcal{P} \longrightarrow \mathbb{R}.
	\eeq
We can express this by 
	\beq
	\label{symp.onecan2}
	\alpha\left[X\right](m) = \tilde\alpha\left[\pi_* X\right](x), \quad \text{where} \quad \alpha\in T_m^*\mathcal{P}.
	\eeq
Equation \eqref{symp.onecan2} is the definition of the \emph{canonical one-form} or \emph{symplectic potential}. Written in this form  it is clear that  $\alpha$ is a one-form on phase-space, not on spacetime.  
The situation is summarised by the commutative diagram;
	\begin{equation}
	\label{symp.diagram}
	\begin{diagram}
		\node{T_m\mathcal{P}} \arrow[11]{e,t}{\pi_*} \arrow[5]{se,r}{\alpha} 
			\node[11]{T_x\mathcal{M}} \arrow[5]{sw,r}{\tilde\alpha} \\[6]
		\node[6]{\hskip.3cm \mathbb{R}}
	\end{diagram}.
	\end{equation}

This rather abstract treatment pays  immediate rewards. It is not difficult to show  that the symplectic structure of the co-tangent bundle (in particular of the phase-space) is simply given by the  closed two-form
	\beq
	\label{kin.symplectic}
	\omega = \d \alpha\overset{!}{=} \d p_a \wedge \d x^a.
	\eeq
Here the coordinate expression follows directly from \eqref{symp.onecan1}. Notice that the warning sign is in place since the coordinate representation is defined from $\d \tilde\alpha$ and, therefore, it belongs to a different space! It is through \eqref{symp.onecan2} that we should understand the last equality in \eqref{kin.symplectic}  without any ambiguity. 

The two-form $\omega$ is obviously skew-symmetric, and it follows from its coordinate expression that its matrix representation is non-singular\footnote{A matrix is non-singular if its determinant is not zero.}. Hence, $\omega$ defines the symplectic structure of phase-space. Since it is a bi-linear non-degenerate form, it provides us with an isomorphism between  $T\mathcal{P}$ and $T^*\mathcal{P}$. In particular, given any smooth function $H$ on $\mathcal{P}$, the symplectic form $\omega$\glo{$\omega$}{Symplectic form} takes the one-form
	\begin{align}
	\label{dh}
	-\d H 	& =  - \frac{\partial H}{\partial p_a} \d p_a - \frac{\partial H}{\partial x^a} \d x^a \nn\\
	      	& = \frac{\partial H}{\partial p_a}\frac{\partial }{\partial x^a} \d p_b \wedge \d x^b - \frac{\partial H}{\partial x_a}\frac{\partial }{\partial p^a} \d p_b \wedge \d x^b  \nn\\	
			&=  \left(\frac{\partial H}{\partial p_a}\frac{\partial }{\partial x^a} - \frac{\partial H}{\partial x^a}\frac{\partial}{\partial p_a}\right) \rfloor (\d p_b \wedge \d x^b) = \omega\left(X_H\right),
	\end{align}
where $\rfloor$ denotes all contractions, into  the  Hamiltonian vector field defined by\glo{$X_H$}{Hamiltonian vector field} 
	\beq
	\label{hmavec}
	X_H \equiv \frac{\partial H}{\partial p_a}\frac{\partial }{\partial x^a} - \frac{\partial H}{\partial x^a}\frac{\partial}{\partial p_a}.
	\eeq

The vector field \eqref{hmavec} generates a one-parameter family of diffeomorphisms of phase-space. The dynamics of a system whose Hamiltonian function is $H$ is given by the integral curves of  $X_H$.

\subsubsection{A fluid example: vorticity and structure}

Let us make a pause in our present argument and note how the previous construction is immediately related to our forthcoming discussion of hydrodynamics  [{\it c.f.} Chapter \ref{ch.canonical}]. 

Consider a single constituent fluid whose elements are described by the conserved particle flux $n^a$. Its   conjugate momentum $\mu_a$ can be related to the chemical potential $\mu$ through the Euler relation
	\beq
	\rho + P = -n^a\mu_a,
	\eeq
where $\rho$ and $P$ are the energy density and pressure of the fluid.

We can express the particle number density current in terms of a family of co-moving observers $n^a = n u^a$ whose normalised four-velocity $u^a u_a = -1$ defines a congruence parametrised by an affine parameter $\lambda$. Let us denote the four velocity as the rate of change of the coordinate functions of spacetime along these curves, $u^a = \dot x^a$, where the dot denotes change with respect to the parameter. The canonical equations of the system are \cite{carterkal}
	\beq
	n^a  = \frac{\d n x^a}{\d \lambda} = \frac{\partial H}{\partial \mu_a} \quad \text{and} \quad \frac{\d \mu_a}{\d \lambda} = \mu_{a;b}n^b = - \frac{\partial H}{\partial x^a},
	\eeq
where $H$ is the `free-fall' Hamiltonian \eqref{freeH}. 

Let us contract these equations for the fluid  with the particle number density current $n^a$
	\begin{subequations}
	\begin{align}
	\label{kin.ucon1}
	g_{ab}n^a n^b &=   -n^2\\
	\label{kin.ucon2}
	\mu_{a;b}n^b n^a =  -\frac{\partial H}{\partial x^a}\frac{\partial H}{\partial \mu_a} & = \frac{\partial H}{\partial \mu_a}\frac{\partial H}{\partial x^a} = - \mu_{a;b}n^b n^a 
	\end{align}
	\end{subequations}  
We see that \eqref{kin.ucon1} simply gives the information about the number density of the fluid particles, while \eqref{kin.ucon2} tells us that the contraction of the momentum equation with the particle flux is identically zero. Thus, it follows that the `forces' acting on the fluid are orthogonal to its four-velocity.  Hence, the dynamics of this simple system is given by the momentum equation	
	\beq
	\label{kin.exeom}
	2\mu_{[a;b]}n^b = -H_{;a}.
	\eeq

The LHS of equation \eqref{kin.exeom} allows us to identify the fluid's vorticity with the symplectic two-form
	\beq
	\label{kin.ident}
	\mu_{a;b} \d x^b \wedge \d x^a = \d \mu_a \wedge \d x^a.
	\eeq

In this example the only way in which the fluid particles  can change their energy is by changing their masses (during collisions say). If we do not allow for particle creation or mass shifts, the dynamics is constrained to one of  the surfaces $\mathcal{P}_m$ of \eqref{sym.union} defined by the equation $H = m$. This is normally known as the Hamiltonian constraint and is a crucial ingredient in Hamilton's variational principle.

Notice that the vanishing in the contraction of \eqref{kin.exeom} with the flux $n^a$ is equivalent to the (trivial) vanishing of $\omega(X_H,X_H)$. We shall return to this point in Chapter \ref{ch.canonical}, when the constraint surface will have more degrees of freedom.

\subsubsection{Motion in phase-space}

The example above gives us some insight into how the classical mechanics of conservative systems is formulated. Without much ado, Hamilton's variational principle can be stated in the following manner\footnote{Up to this point we have only considered states defined on the relativistic single-particle  phase-space. In the case of `larger' systems, the geometric construction is completely analogous. We just need to `count' the correct number of degrees of freedom. In such a setting, the spacetime manifold $\mathcal{M}$ is simply replaced by the space of \emph{kinematic} variables; the configuration manifold $\mathcal{C}$ whose coordinates are typically denoted by $q$. The relativistic phase-space is then the future, non-spacelike part of $T^*\mathcal{C}$ with coordinates $(q^a,p_a)$. Here the indices run over $4N$ values, {\it i.e.} we have $(q^1,q^2,...,q^{4N},p_1,p_2,...,p_{4N})$. }:
	\begin{quote}
	{\bf The single-particle variational principle.} Let $\alpha$ be the canonical one-form of the relativistic single-particle phase-space and let $\gamma\subset\mathcal{P}_m$ be a curve whose restriction $\tilde\gamma \subset\mathcal{M}$ connects two events, $x_0, x_1 \in \mathcal{M}$. We say $\tilde\gamma$ is a physical motion if $\gamma$ extremises the action
		\[
		S[\gamma] = \int_\gamma \alpha.
		\]
	\end{quote}

This principle is an empirical fact and constitutes the \emph{physical} basis of classical mechanics.  Equivalently, we can use Stokes' theorem to state the same principle in terms of the symplectic form in the following way. Consider one of the hyper-surfaces $\mathcal{P}_m$. The restriction of the canonical one-form allows us to define a two-form $\tilde\omega$. Since $\mathcal{P}_m$ is a hyper-surface of $\mathcal{P}$ of co-dimension one, $\tilde\omega$ is necessarily degenerate, {\it i.e.} the bi-linear form $\tilde\omega$ has a \emph{null} eigenvector\footnote{A null eigenvector of a bilinear form is a non-vanishing vector with zero eigenvalue.} $X$. The empirical fact in this case states that the physical motions correspond to the projected orbits of $X$ into spacetime.

 Let us denote by $\mathcal{G}_{H}$\glo{$\mathcal{G}_{H}$}{The one dimensional sub-group of the diffeomorphisms of phase-space generated by $X_H$} the one dimensional sub-group of the diffeomorphisms of phase-space generated by $X_H$. If the mass of every individual test particle is conserved, $X_H$ is tangent to each $\mathcal{P}_m$. We can think of the 6-dimensional quotient manifold $\mathcal{P}_m/\mathcal{G}_H$ as the non-relativistic phase space of classical mechanics for each inertial observer. 	
 
In order to construct a differential operator which encodes the evolution of a dynamical system, we can substitute the definition of the free-fall Hamiltonian \eqref{freeH} into \eqref{hmavec} to obtain\glo{$L$}{Liouville operator}
	\beq
	\label{liouv}
	L \equiv \omega(\cdot,X_H) = \{\cdot,H \}  = g^{ab}p_b\frac{\partial}{\partial x^a} -  g^{bd}g^{ce}\Gamma^a_{bc}p_d p_e \frac{\partial}{\partial p^a}.
	\eeq
Here $\Gamma^a_{bc}$ are the Christoffel symbols associated with the metric $g^{ab}$. We call \eqref{liouv} the Liouville operator. This differential operator acts on functions defined on $\mathcal{P}$. Thus, if $G:\mathcal{P}\rightarrow\mathbb{R}$ and $L[G] = 0$, then $G$ is a constant of the motion. 

Notice that, equivalently, the Liouville operator is simply   a one form on $T^*\mathcal{P}$ defined by the canonical two-form $\omega$ and the Hamiltonian vector field $X_H$. This highlights the leading role played by $\omega$ in the geometric description of a dynamical system. Furthermore, as we will show, the symplectic form acts as a building block for the phase-space volume measure.

\subsection{Volume elements}

In this sub-section, we  try to keep a completely covariant, coordinate-free language. However, we will  make a clear connection with the coordinate language used in most relativistic fluid dynamics work. 

We need to construct a set of measures. One for the base manifold and one for the fibre through each point of the base manifold and also a prescription to transform the volume form as we move along spacetime.


The volume measure in $\mathcal{M}$ is
	\beq
	\label{lio.form1}
	\eta = \sqrt{-g} \d x^0 \wedge \d x^1 \wedge \d x^2 \wedge \d x^3 = \frac{1}{4!}\varepsilon_{abcd} \d x^a \d x^c \d x^c \d x^d
	\eeq

The cotangent space of $\mathcal{M}$ at $x$, $T_x\mathcal{M}$, has its own volume given by
	\beq
	\pi = \left( \sqrt{-g}\right)_{x} \d p^0 \wedge \d p^1 \wedge \d p^2 \wedge \d p^3,
	\eeq
this is the volume element in 4-momentum space at each point.

We can now construct a volume form for $\mathcal{P}$ in terms of products of the two-form $\omega$
	\begin{align}
	\Omega  &= \omega \wedge \omega \wedge \omega \wedge \omega \nn\\
		&= \left[\d p_0 \wedge \d x^0\right] \wedge \cdots \wedge \left[\d p_3 \wedge \d x^3\right] \nn\\
		&= -\left[\d x^0 \wedge \cdots \wedge \d x^3\right] \wedge \left[\d p_0 \wedge \cdots \wedge \d p_3\right]   \nn\\
		&= -\left[\d x^0 \wedge \cdots \wedge \d x^3\right] \wedge \left[\d (g_{0a_0}p^{a_0})\wedge \cdots \d (g_{3a_3}p^{a_3})\right]\nn\\
		&= -g_{0 a_0}g_{1a_1}g_{2a_2}g_{3a_3}\left[\d x^0 \wedge \cdots \wedge \d x^3\right] \wedge \left[\d p^{a_0} \wedge \cdots \d p^{a_3}\right]\nn\\
		&=  \eta \wedge \pi.
	\end{align}

Now we can show a result of central importance in the development of the second law of thermodynamics. We can ask ourselves what happens to the volume form we just defined as it is dragged along the flow of $X_H$. From the Cartan identity for Lie differentiation of r-forms we see 
	\beq
	\label{kin.cartan}
	\mathcal{L}_{X_H}\omega = X_H \rfloor \d \omega + \d \left(X_H \rfloor \omega\right) = 0
	\eeq
and as a consequence
	\beq
	\label{LT}
	\mathcal{L}_{X_H}\Omega = 0
	\eeq
	
Since $8-$forms in $\mathcal{P}$ are a one dimensional vector space and $\mathcal{L}_{X}\Omega$ is an $8-$form for some vector field $X$ (not necessarily Hamiltonian), it must be proportional to $\Omega$. The divergence of $X$ is defined in such manner as the constant of proportionality between the Lie-dragged volume form and itself. Thus we see that any Hamiltonian vector field is divergence-free, that is, the phase flow defined by such field is incompressible! This is precisely what the \emph{Liouville's theorem} states.

Before we move on, we need an extra measure. Consider the $7-$form given by
	\beq
	\label{7form}
	\omeganat = X_H \rfloor \Omega = -4 \d H \wedge \omega \wedge \omega \wedge \omega.
	\eeq
	
The hyper-surfaces of the phase space $\mathcal{P}$ which will be of special interest are those for which \eqref{7form} reduces to
 	\beq
 	\omeganat = p^a \sigma_a \wedge \pi
 	\eeq
where\glo{$\omeganat$}{Volume element for the hypersurfaces of phase space of `orthogonal' to the Hamiltonian flow}
	\beq
	\sigma_a = \frac{1}{3!} \varepsilon_{abcd}\d x^b \d x^c \d x^d.
	\eeq 

 We denote such hyper-surfaces by $\Sigma$\glo{$\Sigma$}{Spacetime hypersurfaces orthogonal to the worldlines of a Hamiltonian observer}. One can verify that the $7-$form \eqref{7form} is closed and it is also dragged along the phase flow
	\beq
	\label{lie.omeganat}
	\mathcal{L}_{X_H} \omeganat = 0.
	\eeq
	
We can consider $\omeganat$ as the volume measure on the space of phase orbits, since Liouville's theorem guarantees that every cross section of a tube of phase orbit have the same volume, thus it can be regarded as a measure of the number of orbits in each tube. Physically the integration of $\omeganat$ over a hyper-surface $\Sigma$ counts the number of states intersecting $\Sigma$ \cite{ehlers}.

\subsection{Individual mass-shell volume measure}


Finally, we will be interested in cases where the mass distribution is discrete, i.e. we will be interested in different mass shells of definite mass $m$, $\mathcal{P}_m$. We observe that another way of interpreting \eqref{LT} is to say that $\Omega$ is an $L-$invariant measure on $\mathcal{P}$ and from that we see that  $h(p^2)\Omega$ is too, for any function $h$. To get rid of the contribution from all the irrelevant mass shells consider the function
	\beq
	h(p^2) = 2He(p) \delta (p^2 + m^2) 
	\eeq
where $He$ is the Heaviside function which is equal to one if $p$ is future directed and zero otherwise. Note that upon integration with respect to $\Omega$ this will only take into account the contributions from the future part of the mass shell $\mathcal{P}_m$. Using coordinates $(x^a,p^i,p^0)$ on $\mathcal{P}$ and `integrating' $h(p^2)\Omega$ over $p^0$  we obtain
	\beq
	\label{ve.single}
	\Omega_m = \eta \wedge \pi_m
	\eeq 
on $\mathcal{P}_m$, where 
	\beq
	\label{ve.single2}
	\pi_m = -\frac{\sqrt{-g}}{p^0}\d p^1 \wedge \d p^2\wedge \d p^3
	\eeq


\section{Boltzmann's equation}

In the rest of this chapter we consider a spacetime region and assume that it is filled with a system of many particles. Following \cite{ehlers}, we suppose that the interaction between the particles can be divided into \emph{long range} and \emph{short range} forces. Then it seems reasonable to account for the long range forces in terms of a \emph{mean field}, in which each particle moves like a test particle except when they undergo a collision, which correspond to the short range interactions. The short range forces can be accounted for in terms of idealised point collisions, whose probability of occurrence is governed by cross-sections taken from special-relativistic scattering theory\footnote{This is the so-called `molecular chaos' hypothesis. It is the only time asymmetric assumption of kinetic theory. A deeper understanding of this point is required to explain the origin of the asymmetry of the second law, in particular the problem of the `arrow of time'.}.  We  also assume here that the collisions only occur  between pairs of particles, thus, we are considering the case of a \emph{dilute gas}.

In this work, the only long range interaction we consider is gravity. If we take the spacetime metric as a mean gravitational field, then the associated phase flow describes all possible free motions of particles between collisions. 

The assumptions of kinetic theory are physically plausible for \emph{dilute gases} in which the mean free path between successive short range interactions is much larger than the range of those interactions. In what follows, we generalise in a covariant manner Boltzmann's programme of a \emph{one-particle phase space}.

A gas state is represented as a collection of phase orbits corresponding to the free fall states occupied by particles between collisions. A collision corresponds to jumps of the participating particles from old to new phase orbits. During a collision,  particle  has no phase orbit assigned to it. The state of a gas in phase space is fully characterised by the occupation number, $N=N[\Sigma]$,  which assigns to a given compact oriented spacetime hyper-surface $\Sigma$ the number of occupied orbits intersecting it. An individual gas state is too complicated to explore in detail. In practice, one uses a \emph{Gibbs-ensemble}, which is a collection of such individual states representing a \emph{macro-state} of the gas, and studies the \emph{average distribution} of states. We assume that the average distribution of occupied states can be described by a continuous density function with respect to the volume measure $\omeganat$ 
	\beq
	\label{be.number}
	\langle N[\Sigma] \rangle = \int_\Sigma f_\Sigma \omeganat,
	\eeq 

We call $f_\Sigma$ the one particle distribution function of the gas. It is a  \emph{Lorentz invariant} representing the density of particles in a spacetime region of 3-volume $\d^3 x$ and momenta in the range $\d^3 p$ with energies in $\d E$.
Throughout the rest of this work we will work with the average distribution of occupied states and will drop the angle brackets. Also, for small enough regions, the distribution function at a point is independent of the hyper-surface we use and, therefore, we will drop the subscript for it\glo{$f_\n$}{One-particle distribution function for the species $\n$}. 

Consider a compact region of phase-space $D$ with boundary $\partial D$. During a collision inside $D$ particles exchange momentum. If we count creations in the positive sense and annihilations in the negative, then the total number of collisions inside $D$ can be expressed by an analogous distribution
	\beq
	N[\partial D] = \int_{D} g \Omega
	\eeq
where $g$ is interpreted as a collision density in $D$.
	
Assuming that $f$ is continuously differentiable, the net number of collisions in $D$ is given by 
	\beq
	N[\partial D] = \int_{\partial D} f \omeganat
	\eeq
and using Stokes' theorem, the product rule for exterior differentiation and the definition of $\omeganat$ we obtain
	\beq
	\label{be.equation}
	N[\partial D] = \int_{\partial D} f \omeganat = \int_D \d(f\omeganat) = \int_D \d f\wedge \omeganat = \int_D \d f \wedge (X_H \rfloor \Omega) = \int_D L(f) \Omega  
	\eeq

Thus, we have obtained the covariant version of Boltzmann's equation 
	\beq
	\label{kin.boltzeq}
	L(f) = g.
	\eeq	
Specific forms of the collision term $g$ depend on the model under consideration. In particular, it is through the mathematical form of this term that one can materialise the time asymmetric molecular chaos and the dilute gas  assumptions of kinetic theory, {\it i.e.} that two un-correlated particles before a collision will be correlated for all times afterwards, even if they never meet again. We will not go further with a discussion relating to the solutions of Boltzmann's equation. We just need to  keep in mind that the `irreversibility hypotheses' for kinetic theory are manifest through \eqref{kin.boltzeq}.

\section{Entropy}

The central notion in any discussion about irreversibility is that of entropy. Unlike most quantities that we have encountered so far, for which we have a clear and definite intuition, entropy will be the most elaborate and abstract concept we will work  with during the course of the present work. It is not the intention here to repeat the philosophical meaning of the concept - beautifully addressed by Penrose \cite{penroseroad} and Price \cite{price} - but to obtain an operational definition to support our future developments on the relativistic dynamics of non-equilibrium processes.  The material presented in this section is mostly referential. The construction follows that of Synge \cite{synge} and Ehlers \cite{ehlers}.

Consider a box divided in $N$ equal cells and a set of $\nu$ marbles. Suppose we accommodate the marbles in the cells in some random manner and then ask  how random our arrangement is. This may sound terribly vague so let us first compare two possible arrangements and ask which one is more likely to be the outcome of a random assignment. To do this, let us denote each arrangement by an string of numbers
	\begin{subequations}
	\label{kin.arrays}
	\begin{align}
	\label{s.01}
	A  & = [\nu_1,\nu_2, \cdots , \nu_n], \\
	\label{s.02}
	A' & = [\nu_1',\nu_2',\cdots, \nu_n'].
	\end{align}
	\end{subequations}
Here $\nu_i$ and $\nu_i'$ represent the number of marbles in the $i$th box for each distribution. Obviously, the total number of marbles does not depend on any particular arrangement 
	\beq
	\label{s.sum}
	\sum_{i=1}^n \nu_i = \sum_{i=1}^n \nu_i' = \nu.
	\eeq

In a truly \emph{random} assignment, each box has an equal probability of $1/n$ of receiving a marble.  It is not difficult to check  that the probability to obtain any of the outcomes  \eqref{kin.arrays} is given by an expression of the form
	\beq
	\label{s.prob0}
	P[A] = \frac{\nu !}{\nu_1!\nu_2!\cdots\nu_n!} \left(\frac{1}{n}\right)^\nu.
	\eeq

If the number of marbles and cells is not very large, the answer follows from the direct computation of \eqref{s.prob0} for each array. Indeed, if the numbers are ``small enough'' one can answer such question by mere inspection. However, as the numbers get larger, equation \eqref{s.prob0} is not very useful to compare the \emph{likelihood} of the outcome $A$ given $A'$. To make this expression  more manageable, we work with its logarithm
	\begin{align}
	\label{s.logs}
	\log P[A] & = \log \nu! - \log \prod_{i=1}^n \nu_i! - \nu \log n\nn\\
		      & = \log \nu! - \sum_{i=1}^n \log \nu_i! -\nu \log n.
	\end{align}
This will be very useful in the case where the number of cells, the number of marbles and the number of marbles in each cell are very large. In such situation, we can approximate the logarithms in \eqref{s.logs} by  Stirling's formula 
	\beq
	\log a! \sim a \log a - a.
	\eeq
Thus, using \eqref{s.sum}, we obtain 
	\beq
	\log P[A] = \nu \log \nu - \sum_{i=1}^n \nu_i \log \nu_i - \nu \log n.
	\eeq 

Now it is straightforward to compare the probabilities to obtain each outcome just by taking the difference of their logarithms
	\beq
	\label{s.logs2}
	\log P[A] - \log P[A'] = \log \frac{P[A]}{P[A']} = \zeta- \zeta',
	\eeq
where we have made the abbreviation
	\beq
	\label{s.zeta}
	\zeta = -\sum_{i=1}^n \nu_i \log \nu_i \quad {\rm and} \quad \zeta'= - \sum_{i=1}^n \nu_i' \log \nu_i'.
	\eeq
Therefore, the desired relation for both probabilities follows from \eqref{s.logs2} and \eqref{s.zeta}. That is
	\begin{align}
	\zeta > \zeta' &\quad \Rightarrow \quad P[A] > P[A'],\nn\\
	\zeta = \zeta' &\quad \Rightarrow \quad P[A] = P[A'],\nn\\
	\zeta < \zeta' &\quad \Rightarrow \quad P[A] < P[A'].\nn
	\end{align}
Thus we see that the quantity $\zeta$ measures the `randomness' of a given array $A$. 

In the example above we were just interested in the `likelihood' of finding  different static configurations of very large samples of balls grouped in cells. We can extend the  analysis to the case where we are also interested in the motions of the balls inside each cell (provided there is a mechanism that governs such motion). In that case, the \emph{distribution} of balls  will not just depend on their positions within the cells, but also on their velocities. In the limit when the marbles and the cells are arbitrarily small, we are essentially looking at a gas in a box\footnote{At this level we are not yet concerned about the particular details of the gas. For our purposes it suffices that each marble/particle is indistinguishable from one another. This corresponds to the ``hard-spheres'' approximation.}. Now, we have seen in the previous section that the total number of particles at a given \emph{time} is given by the occupation functional \eqref{be.number}. Here an instantaneous configuration is only defined with respect to a class of observers; those sharing the same simultaneity hyper-surface. As before, let  $\Sigma$ be a 7-dimensional hyper-surface of phase space\footnote{Note that the one-particle phase space is an 8-dimensional manifold. In this sense $\Sigma$ leaves one degree of freedom. If the projection of $\Sigma$ down to spacetime is a spacelike hyper-surface $G$, then \eqref{number} effectively counts the number of particles at a given time.}. The total number of particles and their respective momentum is given by 
	\begin{subequations}
	\label{kin.nump}
	\begin{align}
	\label{number}
	N[\Sigma] &= \int_\Sigma f_\n \omeganat_\n,\\
	\label{h.momentum}
	P_a[\Sigma] &= \int_\Sigma f_\n p_a \omeganat_\n.
	\end{align}
	\end{subequations}

In analogy with the total number of marbles \eqref{s.sum}, equations \eqref{kin.nump} do not depend on the particular distribution function of the particles; their specific arrangement within the volume $\omeganat_\n$.  Therefore, to be able to compare the degree of randomness among different distributions, we follow \eqref{s.zeta} and  define\glo{$\mathcal{H}$}{Negative of the entropy at a spacelike hypersurface orthogonal to an observer} 
	\beq
	\label{entropy}
	\mathcal{H}[\Sigma,f_\n] = - \int_{\Sigma} (f_\n \log f_\n) \omeganat_\n,
	\eeq
Notice that, unlike the total number of particles and the momentum on a given hypersurface, the quantity  $\mathcal{H}$ does depend on the particular distribution function. Thus, the quantity $\mathcal{H}[\Sigma,f_\n]$ can be regarded as a measure of the ``randomness'' of the distribution $f_\n$ at a specific moment characterised by the hyper-surface $\Sigma$.

Consider a space-like hyper-surface $\Sigma_0$. If we want to know what is the most random distribution of a very large number of particles $N$ with total momentum $P^a$ at a given ``time'' represented by $\Sigma_0$, we just need to extremise \eqref{entropy} with respect to all the possible distributions. A crucial thing to be noticed is that, without any restriction, such question does not make sense. It would be like asking what is the most random distribution of marbles in a box without saying how many marbles there are in the first place! In our case the relevant question is which is the most random distribution over $\Sigma_0$ of an array of $N$ particles \eqref{number} with momentum \eqref{h.momentum}.To answer this question, we introduce the Lagrange multipliers $\alpha_\n$ and $\beta_a$\glo{$\alpha_\n$ and $\beta_a$}{Lagrange multipliers associated with the chemical potential and temperature, respectively, of the $n$ fluid species} and work instead with
	\beq
	\tilde{\mathcal{H}} = \mathcal{H} + \alpha_\n N + \beta^a P_a,
	\eeq
where $\beta^a$ is assumed to be a \emph{timelike} vector field.
Then, we look for the distribution function that makes  $\tilde{\mathcal{H}}$ an extremum   
	\beq
	\delta \tilde{\mathcal{H}} = \int_{\Sigma_0} \left[-(\log f_\n +1) + \alpha_\n + \beta^a p_a\right]\delta f_\n \omeganat_\n= 0.
	\eeq
We can immediately read off from the integrand the most random distribution of $N$ identical particles  with averaged momentum $P^a$  over the hyper-surface $\Sigma_0$ to be
	\beq
	\label{eqdist}
	f_0 = \exp\left[\alpha_\n(x) + \beta^a(x)p_a -1\right].
	\eeq

We say that a system is thermalised at a certain moment characterised by $\Sigma_0$ if it is described by the distribution function \eqref{eqdist}. Note that the choice of hyper-surface is crucial through the whole argument, as it represent  a surface of simultaneity for a family of observers. However, just as in our previous example of marbles in a box,  if each box has an \emph{equal} probability of getting a ball, the result is independent of the particular geometry. Therefore, the next question one could ask is how to compare distributions between different hyper-surfaces, provided there is a manner of \emph{passing} from one to another.

\subsection{$\mathcal{H}$-theorem}
 
Now that we know the distribution function which extremises the randomness functional $\mathcal{H}[\Sigma_0,f_\n]$ of a \emph{gas} described by $f_\n$ at a given \emph{time} $\Sigma_0$, we are interested to know how to relate such measure from one hyper-surface to another. A precise and objective  answer to this problem  is far from trivial. In a rather speculative sense, this is the definition of irreversibility; the fundamental notion of the \emph{arrow of time}. Again, it is not our aim to pursue such quest in any depth, but to obtain a clear characterisation of the irreversibility of relativistic processes.   

Consider the flow generated by the vector field $X$ (not necessarily Hamiltonian). The change of $\mathcal{H}$ along the flow of $X$, parametrised by $\varepsilon$, is 
	\begin{align}
	\label{ent.chg}
	\frac{\d}{\d \varepsilon}\mathcal{H}[\Sigma,f_\n] & = -\frac{\d}{\d \varepsilon}\int_{\Sigma(\varepsilon)} f_\n \log f_\n \omeganat_\n\nn\\
							  & = - \int_{\Sigma} \mathcal{L}_{X}[f_\n \log f_\n \omeganat_\n]\nn\\
							  & = - \int_{\Sigma}[(1 + \log f_\n) \mathcal{L}_{X} f_\n]\omeganat_\n  - \int_{\Sigma}f_\n \log f_\n \mathcal{L}_{X} \omeganat_\n.
	\end{align}

 Our first task is to find the conditions under which  the quantity $\mathcal{H}$ is conserved along the flow of $X$. That is, when two hyper-surfaces have an equally random distribution of particles and momenta. To do this, we need to be careful;   we need to guarantee that we are comparing the right regions of phase space. This fact becomes evident by looking at the second term in the RHS of \eqref{ent.chg};  we need to take into account evolutions of the volume form $\omeganat_\n$ along the flow. To keep the discussion at the simplest level, we only consider those flows which are volume preserving. Thus, from the volume dragging formula, equation \eqref{lie.omeganat}, we see that if the vector field $X$ is  Hamiltonian, the volume element $\omeganat$ is preserved.  In this case, $\mathcal{H}$ is constant along the flow of $X_H$ if the distribution function $f_\n$ is also dragged by the flow, {\it i.e.} if $f_\n$ is constant along the orbits of $X_H$.
 
From the discussion in the previous section we can ask if the  distribution function \eqref{eqdist} is dragged by the phase flow generated by $H$. Consider the Hamiltonian vector field $X_H$ and the distribution function \eqref{eqdist}. From the definition of the Liouville operator \eqref{liouv} we have
	\beq
	\mathcal{L}_{X_H} f_0 = X_H f_0 = L[f_0].
	\eeq
It is straightforward to evaluate this expression to obtain
	\beq
	\label{l.f0}
	L[f_0] = \exp\left(\alpha_\n + \beta^a p_a\right)\left[\alpha_\n^{\ ,a}p_a + \beta^{(a;b)}p_a p_b\right]. 
	\eeq
Substituting \eqref{l.f0} back into \eqref{ent.chg} we obtain 
	\begin{align}
	\label{equ}
	 \left.\frac{\d\mathcal{H}}{\d \varepsilon}\right|_H & = -\int_\Sigma \left\{\left(1 + \log f_0\right) L\left[f_0\right]\right\} \omeganat_n - \int_{\Sigma} f_0 \log f_0 \mathcal{L}_{X_H}\omeganat_n \nn\\
							    & =  \int_\Sigma \left(\alpha_\n + \beta^a p_a\right)\exp\left(\alpha_\n + \beta^a  p_a\right)\left[\alpha_\n^{\ ,a}p_a + \beta^{(a;b)}p_a p_b\right] \omeganat_\n\nn\\
		       					    & = 0,
	\end{align}
where we have made use of  the volume preserving condition \eqref{lie.omeganat}. Therefore, the distribution function $f_0$ is dragged by the phase flow only if the Lagrange multipliers satisfy the conditions
	\beq
	\label{equilibrium}
	\alpha_\n^{\ ,a} = 0 \quad \text{and} \quad \beta^{(a;b)} = 0.
	\eeq	
Note that the choice of a Hamiltonian \emph{observer} is crucial to make the right comparison of the distribution functions that extremise $\mathcal{H}$ over two distinct hyper-surfaces.  To see this, let us recall the marbles in cells problem. A choice of a flow which does not preserve the volume element $\omeganat$, would be equivalent to comparing the distributions of marbles in boxes of the same volume but with different number of cells. 

Equations \eqref{equilibrium} represent one of the central results of relativistic kinetic theory. They are the macroscopic conditions of \emph{global} thermodynamic equilibrium \cite{dixon}. In the next section we will express these relations in terms of fluid variables. For the time being, let us note that in order for the functional $\mathcal{H}$ to be dragged by a Hamiltonian observer, a necessary condition is that the spacetime admits a timelike Killing vector field, {\it i.e.} the spacetime must be \emph{stationary}.	

 We have arrived at the conclusion that the macroscopic conditions \eqref{equilibrium}  are equivalent to the micro-physics condition expressed by the collision-less Boltzmann equation
	\beq
	L[f_0] = 0.
	\eeq
Thus, we see that for any Hamiltonian observer,  a  thermalised system in a given hyper-surface $\Sigma_0$ [described by the distribution function \eqref{eqdist}] will be thermalised in a neighbourhood of $\Sigma_0$ and we say that the system is in \emph{equilibrium}. A distribution function satisfying \eqref{eqdist}, but not \eqref{equilibrium}, is called a \emph{local} equilibrium distribution \cite{isralekinetic}.  Are there any other distribution functions which are dragged by the phase flow? The answer to that rests at the core of the $\mathcal{H}$-theorem.

	\begin{quote} 
	{\bf $\mathcal{H}$-theorem}. Let $f_\n$ be a distribution function describing the \emph{instantaneous} configuration of a system of particles on a hyper-surface $\Sigma$ with  volume measure $\omeganat_\n$ defined in terms of a Hamiltonian vector field $X_H$.  If $f_0$ satisfies \eqref{eqdist} over a hyper-surface $\Sigma_0$ and $f_0 \neq f_\n$, then $\Sigma$ is in the the \emph{past} of $\Sigma_0$.
	\end{quote}

This theorem has been proved a number of times in the literature under two fundamental assumptions: when the system is a \emph{dilute} gas and \emph{molecular chaos} is present. The following argument will show how we need \emph{physical} input to realise a proof of the above mathematical statement.
   
The first thing to note is that, implicit in such a statement, there is a convention. We have chosen the direction of the $future$ to correspond to ``positive'' displacements of the parameter of the flow $\varepsilon$. Now, consider the change of $\mathcal{H}$ along $X_H$ between two neighbouring hyper-surfaces $\Sigma$ and $\Sigma'$ of a system described by the distribution functions $f_\n$ and $f_\n'$ respectively
	\beq
	\label{ht01}
	\mathcal{H}[\Sigma',f'_\n]  = \mathcal{H}[\Sigma,f_n] +  \left.\frac{\d}{\d \varepsilon}\mathcal{H}\right|_H \varepsilon.
	\eeq
From \eqref{ent.chg} we can immediately see that if we can guarantee that the inequality
	\beq
	\label{colless}
	L[f_\n] \geq 0,
	\eeq
is satisfied, then  the last term in the RHS of \eqref{ht01} 
	\beq
	-\int_\Sigma [1+\log f_\n] L[f_\n] \omeganat_\n \leq 0.
	\eeq
Therefore,  it follows that the value of $\mathcal{H}$ decreases in the positive direction of the flow of $X_H$. That is, if $\varepsilon \geq 0$
	\beq
	\label{ht02}
	\mathcal{H}[\Sigma',f'_\n] \leq \mathcal{H}[\Sigma,f_\n].
	\eeq
In particular, the equilibrium distribution function satisfies
	\beq
	\mathcal{H}[\Sigma_0, f_0] \leq \mathcal{H}[\Sigma,f_\n]
	\eeq
for any distribution function $f_\n$. Thus, the equilibrium  value $\mathcal{H}_0 = \mathcal{H}[\Sigma,f_0]$ corresponds to a minimum. The entropy of a system is defined as the negative of the functional $\mathcal{H}$. Hence, we see that \eqref{ht02} implies that entropy \emph{increases} towards the \emph{future} of a given observer. Notice that the conclusion \eqref{ht02} depends on the validity of the inequality \eqref{colless}. This is a particular version of the fundamental problem of the arrow of time. In most cases this result is simply postulated and in the present work we  do as well. However, we should keep in mind through the rest of this work that \eqref{colless} is as far as we can go within the realm of the classical theory of irreversible processes


\section{Macroscopic fluid equations}

To close this chapter, we obtain the macroscopic fluid description of the gas. As noted in \cite{ellistreciokas1}, a fluid model is appropriate if the collision dominance is assumed to imply that the gas is sufficiently close to equilibrium to allow the definition of a unique 4-velocity and use of a conventional thermodynamic formalism.

So far we have been considering the averaged occupation number to describe the number of particles crossing a given hyper-surface. Now we focus on those hyper-surfaces for which we can use the individual mass-shell volume element \eqref{ve.single} where the integration over momentum space is simply \eqref{ve.single2}. In this case the occupation functional \eqref{be.number} can be written as
	\beq
	N[\Sigma] = \int_G \sigma_a \left[\int_{\Sigma^\sharp} f_\n p^a \pi_\n \right],
	\eeq
where $\Sigma^\sharp$ is the part of the mass-shell which is contained in $\Sigma$. Written in this way, the term in brackets represents the number density current for the $\n$ species\glo{$n^a$}{Particle number density current},
	\beq
	\label{fe.number}
	n^a = \int_{\Sigma^\sharp} f_\n p^a \pi_\n,
	\eeq
which integrated over the oriented volume element $\sigma_a$ gives the total number of particles passing through $G$ with momentum in $\Sigma^\sharp$. In what follows, if the region of integration of phase space $\Sigma^\sharp$ is omitted, the integration \eqref{fe.number} should be over the entire mass-shell\footnote{In order to guarantee that all the integrals of the type \eqref{fe.number} exist, we shall always assume that all distribution functions vanish sufficiently fast for large energies at infinity on the mass-shell.}. For any smooth distribution function $f_\n \geq 0$, the particle density current $n^a$ is a timelike, future directed vector field in spacetime.

In a similar manner, the current density associated with the four-momentum distribution of the system defines the \emph{kinetic stress-energy tensor} of particles of the $\n$ species
	\beq
	\label{fe.em}
	T^{ab} = \int f_\n p^a p^b  \pi_\n.
	\eeq 
	
These current densities are special cases of moments of the distribution function in momentum space. 

To obtain conservation laws, we need to compute the covariant divergence of \eqref{fe.number} and \eqref{fe.em}. This is easily done as an application of Liouville's theorem. Thus, we write (see \cite{ehlers})
	\begin{align}
	\label{fe.cont}
	n^a_{\ ;a}      & = \left[\int f_\n \pi_\n \right]_{;a} = \int L[f_\n] \pi_\n \\
	\label{fe.cons}
	T^{ab}_{\ \ ;b} & = \left[\int f_\n p^a p^b \pi_\n \right]_{;b} = \int L[f_\n] p^a \pi_\n.
	\end{align}
	
Notice that for a simple gas, consisting of a single particle species, Einstein's equations together with the Bianchi identities imply that the divergence of the energy-momentum tensor vanishes, and in this case the distribution function $f_\n$ satisfies 
	\beq
	L[f_\n] = 0
	\eeq
and therefore, \eqref{fe.cont} is a continuity equation stating the particle number current $n^a$ is conserved.

For a mixture of particles, we need to consider different distribution functions integrated over different mass-shells. In this case there might not be a conservation of energy for each individual constituent, but again, Bianchi identities impose a balance  law
	\beq
	T^{ab}_{\ \ ;b} = \sum_A T^{ab}_{A\ ;b} = \sum_A \int L_A[f_A] p^a \pi_A = 0
	\eeq	
where the subscript $A$ denotes each individual species.


We can extend the reasoning above and construct an entropy current density from \eqref{entropy} in an analogous way to \eqref{fe.number}
	\beq
	S[\Sigma] = \int_{G} \sigma_a s^a
	\eeq
where again $G$ is a spacelike hyper-surface and \glo{$s^a$}{Entropy density current}
	\beq
	\label{kin.entflux}
	s^a = - \int (f_\n \log f_\n) p^a \pi_\n.
	\eeq
	
This provides us with another route to the second law of thermodynamics, as encoded in the $\mathcal{H}$-theorem stated in the previous section [see \eqref{ht02}]. In this case, if we consider two spacelike hyper-surfaces $G$ and $G'$ as part of the boundary of a `cylindrical' region of spacetime  and using Stokes's theorem we have
	\beq
	S' - S = \int_{G'} s^a \sigma_a  - \int_{G} s^a \sigma_a  = \int_{\partial D} s^a \sigma_a  = \int_D s^a_{\ ;a} \eta 
	\eeq 
where $\eta$ is the spacetime four-volume form \eqref{lio.form1}.

It is clear then, that the total entropy of an adiabatically isolated system will not decrease in time if and only if
	\beq
	\label{kin.entdiv}
	s^a_{\ ;a} \geq 0
	\eeq
	
This requirement is equivalent to the condition on the distribution function \eqref{colless}. There is no direct way to prove this without further assumptions about the dynamics of the gas. The asymmetry in the direction of the arrow of time can thus be reinterpreted as an asymmetry in the boundary conditions.


\subsection{Thermal equilibrium and the Tolman-Ehrenfest effect}

This subsection is entirely based on the work by Ehlers \cite{ehlers} and has been included for completeness and future reference. 

For an observer moving with the normalised four velocity given by the particle flux\glo{$u^a$}{Observer's normalised four-velocity}
	\beq
	u^a= n^a/n,
	\eeq
the equilibrium distribution function \eqref{eqdist} depends only on the projection of the four momentum into its own frame, $E=-u^ap_a$. In this case, one can write the densitised thermodynamic quantities $\rho$, $n$, $s$ and $p$, representing energy, particle number, entropy  densities respectively together with the pressure $p$ as
	\begin{subequations}
	\begin{align}
	\rho &= \int_m^\infty NE \d E,\\
	n    & = \int_m^\infty N \d E,\\
	\label{kin.final01}
	s    &= \beta(\rho + p) - \alpha n,
	\end{align}
	\end{subequations}
where
	\beq
	p = \frac{1}{3}\int_m^\infty N(E^2 - m^2)E^{-1}\d E.
	\eeq

These quantities are defined on a surface of equilibrium states whose coordinates are Lagrange multipliers $(\alpha,\beta)$. One can show that, in addition to \eqref{kin.final01}, they satisfy the equation
	\beq
	\d \rho = \beta^{-1} \d s + \alpha \beta^{-1} \d n.
	\eeq
Thus, comparing with the Gibbs relation \eqref{thermo.gibbs} of section 2.2.1, we can interpret the Lagrange multipliers as the temperature and relativistic chemical potential by
	\begin{subequations}
	\label{kin.identi}
	\begin{align}
	\label{kin.identia}
	T   & = \beta^{-1},\\
	\mu & = \alpha\beta^{-1}.
	\end{align}
	\end{subequations}
	
There is a subtle point to note in the above discussion. The identification \eqref{kin.identi} comes from the identification of the equilibrium thermodynamic variables with the Lagrange multipliers defined to obtain the equilibrium distribution function. These, however, retain their meaning only when the distribution itself is dragged by the vector field defining the evolution of the system, {\it i.e.} $X_H$. What happens if we choose some other vector field which does not commute with $X_H$? Does a moving body look cold? A step towards the answer was found by Tolman. Equilibrium in a region of spacetime demands that the Lagrange multipliers  satisfy the evolution equation $L(f) = 0$, {\it i.e.} that the combination  $\alpha + \beta^a p_a$ should be a constant of the motion. In particular, the second equilibrium condition \eqref{equilibrium} implies the existence of a timelike Killing vector in spacetime, that is, thermal equilibrium can be attained only in stationary spacetimes. This is quite a remarkable constraint for a relativistic description of thermodynamic systems. In particular,  this implies that for relativistic thermodynamics the temperature $T$ is not spatially constant. To see this, we use coordinates adapted to the Killing vector field $\beta^a$ so that
	\beq
	g_{ab;0} = 0,
	\eeq	
and, from \eqref{kin.identia}, the temperature satisfies
	\beq
	g_{ab}\beta^a\beta^b = g_{00} = -\beta^2 = - T^{-2}.
	\eeq
Therefore, we can see that in a stationary gravitational field the equilibrium temperature varies according to the law
	\beq
	T = (-g_{00})^{-1/2}.
	\eeq
This is called the Tolman-Ehrenfest effect. In consequence, a gravitationally red-shifted photon preserves the energy with which it was emitted, that is, there is no energy transfer while freely propagating in a gravitational field. 

We shall come back to this  in more detail in Chapter 5.


\subsection{Deviations from thermal equilibrium}

The above discussion gives us a gist of what is yet to come. The Tolman-Ehrenfest effect is a purely relativistic effect for systems in thermal equilibrium which shows that the Newtonian idea that spatial variations of temperature alone imply a non-vanishing heat flux cannot be extended to the relativistic realm.  Therefore, in relativistic settings,  a non-uniform temperature distribution is not necessarily associated with a non-equilibrium situation.

Within the framework of kinetic theory, we defined thermal equilibrium in terms of a distribution function $f_\n$ and an  observer\footnote{Or a set of \emph{inertial} observers; those defined through a timelike vector field $X_{E}$ which commutes with $X_H$.} $X_H$ by requiring that, in the spacelike hypersurface locally orthogonal to $X_H$, the distribution function should be given by \eqref{eqdist}, where the Lagrange multipliers, $\alpha$ and $\beta^a$, satisfy the conditions \eqref{equilibrium}. Thus, in the macroscopic limit we could associate them with the relativistic chemical potential and the \emph{equilibrium} temperature, respectively. How do we obtain the macroscopic limit of a non-equilibrium situation? or, rather, how do we define entropy (and therefore temperature) in a situation away from equilibrium?

We have a partial answer to the first question, and we face a serious problem with the second. On the one hand, we can consider an expansion order by order in deviations from thermal equilibrium. This is the well known Grad's approach described in detail for the relativistic case by Stewart \cite{Stewart77}. Such an approach leads to the set of constitutive equations of the Israel \& Stewart model of relativistic dissipation discussed in the next chapter. On the other hand, following such a path, we will inevitably face the problem of interpreting the Lagrange multipliers of the equilibrium function $f_0$. This does not represent  an issue as long as the macroscopic model is one for which the local equilibrium assumption holds. That is, in a regime where LIT programme \emph{should} operate. This is a surprising result because the Israel and Stewart approach models systems with finite relaxation times, as one would expect from an EIT point of view.

To stress the physical relevance of the ambiguities arising from any attempt to define a non-equilibrium temperature we consider the following thought experiment proposed by Casas-Vazquez and Jou \cite{casas-jou}. The zeroth-law of thermodynamics, as introduced in Chapter 1, can be used to define an ideal thermometer as the objects which measure the \emph{same} temperature throughout a system which is in thermal equilibrium. However, away from equilibrium this is no longer the case. Imagine a system with \emph{two} components; matter and radiation, and suppose we have two  thermometers; one  with perfectly reflecting walls and one whose walls are black. The first one will be unaware of the radiation and indicate the temperature of the matter only, whilst the latter  will be sensitive to both, radiation and matter. In an equilibrium situation, both thermometers will produce the same reading. However, when we depart from equilibrium, matter and entropy may have  different temperatures, thus the thermometers would provide us with different readings. We will address this particular problem in section 5.3.1 from a dynamical point of view. For the time being, the purpose of this discussion is to make the reader aware that there is, indeed, a physical issue when dealing with systems away from thermal equilibrium.



			\chapter{Relativistic theories of dissipation}

This chapter presents a review of various theories of relativistic dissipation. The aim is to make evident  how the distinct models of relativistic dissipation share the same philosophy in their construction, that is, by providing an energy momentum tensor written in a particular  frame together with  a suitable definition for the entropy flow in order to impose  the second law of thermodynamics. 

We begin this review with  a discussion of the so-called `first-order' theories of dissipation. In particular, we work through the construction of Eckart's model - the first extension of non-equilibrium thermodynamics to a relativistic framework. In this case, the simple definition of the entropy flow leads to  Eckart's hypothesis, which is nothing more than the relativistic analogue of Fourier's law. Therefore, a thermal disturbance will propagate with an unbounded speed, leading to internal inconsistencies in a relativistic setting. For the sake of completeness, we mention the Landau \& Lifshitz model, which is not fundamentally different from Eckart's original proposal. 

Owing to the inconsistencies arising from the first-order theories, a class of `second-order' theories emerged. Here we will introduce two distinct approaches of this kind which differ essentially in their thermodynamic assumptions. Firstly, we will discuss the model proposed by  Israel \& Stewart \cite{israel}.  They  generalise the definition of the entropy flow to include `second order' corrections from thermal equilibrium. This is normally interpreted as a truncated series on deviations from equilibrium.  The second-order terms of such an expansion are given by all the possible  covariant combinations of  the scalars, vectors and tensors available in the theory.  These additional couplings  need to be obtained by external means, {\it i.e.} by indirect  measurement or through a direct kinetic calculation. Interestingly, it has been argued by Geroch and Lindblom \cite{diver} that a direct measurement of these terms may not be possible. However, experiments showing the existence of second-sound in solids or, alternatively, in superfluid helium seem to contradict their result. The imposition of the second law in the second-order theories is made in an analogous manner to  the  Eckart `first-order' model. However,  in this case it leads to a generalised version of the Cattaneo equation [{\it c.f.} equation \eqref{dis.cattaneo}] and, hence, to a relaxed propagation of thermal signals. There is a subtle point to note here. In the Israel and Stewart programme, one assumes the validity of the local equilibrium hypothesis and yet one correctly obtains a finite speed for heat propagation. We shall discuss these matters towards the end of section 4.2.1.

Secondly, we include a  discussion of a different point of view sustained by Carter. Although Carter's original proposal aimed at a simple way of doing `off the peg' calculations,  the discussion presented here will serve two purposes. On the one hand it will shed light onto a highly obscure reference on relativistic thermodynamics. On the other, such approach constitutes the original motivation  for the multifluid approach to relativistic dissipation presented in this thesis. For completeness and accessibility, we include a transcript of Carter's original views in Appendix A.

Let us emphasise that the various theories presented in this chapter does not exhaust all the possibilities for the first and second order theories available. It is simply a collection of the most representative and widely used. The inclusion of a wider selection of models, we believe, would not contribute in a significant manner to our discussion. However, it would be remiss not to mention some other roads that have been taken. Such is the case of the work by Geroch and Lindblom on relativistic theories of dissipation of divergence type \cite{diver}, or the recent re-derivation of a second-order theory from kinetic theory by Koide \cite{koide}. Finally, the interested reader may wish to consult the review article by Herrera and Pavon \cite{herrerap} for a broader discussion about hyperbolic theories of dissipation.

\section{First-order theories}

Let us begin this review  with the simplest generalization to describe dissipative process in a relativistic context.  In 1940, Eckart published a collection of articles entitled `Thermodynamics of Irreversible Processes', the third of which was devoted to irreversible processes in relativity. This publication set the basic strategy followed by later developments in relativistic theories of dissipation. A generic theory of relativistic dissipation consists of at least one fluid, whose particle number density is conserved, together with a local energy balance written in a  manner that allows us to impose the second law of thermodynamics.   This section is based on Eckart's original work \cite{eckart03}.
   
\subsection{Eckart's model}

Let $n^a$ represent the particle number density flux of a single species fluid system. In the previous chapter we saw that we can express the conservation law for $n^a$ by the expression
	\beq
	\label{cons0}
	n^a_{\ ;a}=0.
	\eeq
In Eckart's model, one uses this fact to choose a frame in which to carry out all physical measurements. Thus, the Eckart frame is defined by the family of observers moving with the normalised four-velocity  parallel to the matter fluid
	\beq
	\label{fourv}
	u^a=\frac{n^a}{n}
	\eeq
where $n^2= - g_{ab}n^a n^b$.

 Given an arbitrary vector field $F^a$, we can form the pair 
	\begin{subequations}
	\begin{align}
	f     & = -g_{ab}u^a F^b,\\
	f^a   & =h^a_{\ b} F^b,
	\end{align}
	\end{subequations}
where we have introduced the \emph{orthogonal projector} to the observer's four-velocity\glo{$h^a_{\ b}$}{Orthogonal projector to the observer's four velocity $u^a$}
	\beq
	\label{dis.proj}
	h^a_{\ b}=\delta^a_{\ b} + u^a u_b. 
	\eeq
One can easily verify that, indeed, $h^a_{\ b} u^b = 0$. Therefore, we can decompose any vector field $F^a$ at each point of spacetime into its parallel and orthogonal components to $u^a$
	\beq
	F^a=fu^a+f^a.
	\eeq
This is a very useful decomposition to describe vector quantities from the point of view of an observer moving together with a fluid. In a similar manner, an arbitrary tensor $F^{ab}$ can be decomposed in terms of its components
	\begin{subequations}
	\label{dis.dec}
	\begin{align}
	\label{eck01}
	\phi &=u_a u_b F^{ab},\\
	\label{eck02}
	\phi^a &=h^a_{\ b}u_c F^{bc},\\
	\label{eck03}
	\phi^{ab}&=h^a_{\ c}h^b_{\ d}F^{cd}.
	\end{align}
	\end{subequations}

Instead of proving a variational principle from which the conservation of energy follows as an identity, most  relativistic theories of dissipation postulate that there exists a symmetric energy momentum tensor which satisfies the conservation law
	\beq
	\label{cons01}
	T^{ab}_{\ \ ;b}=0.
	\eeq
Then, using the decomposition \eqref{dis.dec}, they write it in its general form
	\beq
	\label{dec01}
	T^{ab}= \rho u^a u^b +2 q^{(a}u^{b)} + \pi^{ab},
	\eeq
where 
	\beq
	q^au_a = 0 \quad \text{and} \quad \pi^{ab}u_a =0. 
	\eeq
In this sense, any divergence-free symmetric second rank tensor serves to define a matter model if we interpret (with the right units) $\rho$ as the energy density, $q^a$ as the transverse momentum and $\pi^{ab}$  as the anisotropic stress tensor  as measured by an observer moving with the particle flux. With this physical interpretation, one can define the internal energy $\epsilon$ through the relation \cite{hawkingellis}
	\beq
	\label{internal}
	n(\epsilon + a)= \rho
	\eeq
where $a$ is an arbitrary constant.



The dynamical relations contained in our matter model imposed by the constraint \eqref{cons01} are easily obtained by contracting the energy-momentum tensor $T^{ab}$ with  the observer's four velocity $u^a$ to obtain
	\beq
	\label{first01}
	\left(u_a T^{ab}\right)_{;b}- u_{a;b} T^{ab}=0.
	\eeq
Let us begin with the first term in the above expression. It follows from the decomposition \eqref{dec01} that
	\beq
	u_a T^{ab} = -\rho u^b - q^b,
	\eeq
and, from the definition of the internal energy \eqref{internal},  it becomes
	\beq
	\label{dis.first02}
	\left(-u_a T^{ab}\right)_{;b}=\left[n(\epsilon + a)u^b\right]_{;b} + q^b_{\ ;b}.
	\eeq
Finally, using the conservation of the particle number density flux, equation \eqref{cons0}, it follows that 
	\beq
	\left(a n^b\right)_{\ ;b} = 0,
	\eeq
and equation \eqref{dis.first02} reduces to 
	\beq	
	\left(-u_a T^{ab}\right)_{;b}=n{\dot\epsilon} + q^b_{\ ;b}.
	\eeq
The second term in the LHS of \eqref{first01} can be written simply as
	\beq
	u_{a;b}T^{ab}=q^a {\dot u_a} + u_{a;b}\pi^{ab}.
	\eeq
Thus, the local form of the energy balance \eqref{first01} from the fluid observer's point of view is given by
	\beq
	\label{FL}
	n{\dot \epsilon} + q^b_{\ ;b} + q^a {\dot u_a} + u_{a;b}\pi^{ab} = 0.
	\eeq




Before giving the physical significance of \eqref{FL}, a further simplification can be made. Note that  when the energy-momentum tensor takes the form
	\beq
	T^{ab}= (\rho + p)u^a u^b + p g^{ab}
	\eeq
we can define the \emph{hydrostatic pressure} of the fluid through the trace of the stress tensor
	\beq
	\label{preas}
	p \equiv\frac{1}{3}\pi^a_{\ a}.
	\eeq
Therefore, in the general case, this allows us to define the viscous stress tensor 
	\beq
	\label{visc0}
	P^{ab}= -\pi^{ab} + p h^{ab},
	\eeq
which is assumed to be a linear function of $u_{a;b}$, in terms of the hydrostatic pressure. It follows that the contractions of this viscous stress with the observer's four velocity vanish 
	\beq
	\label{ort0}
	u_a P^{ab} = 0, \quad P^a_{\ a}=0,
	\eeq
and we can obtain a general  expression for $P^{ab}$ 
	\begin{align}
	\label{visc}
	P^{ab} &= \lambda \left[2 h^{ac}h^{bd} u_{(c;d)} - \frac{2}{3}h^{ac} h^{cd}u_{c;d} \right]\nn\\
	       &= \lambda \left[u^{(a;b)} + 2\dot u^{(a}u^{b)} - \frac{1}{3} u^c_{\ ;c}h^{ab} \right]\nn\\
	       &= \lambda \left[\sigma^{ab} + 2\dot u^{(a}u^{b)}\right],
	\end{align}
where  $\lambda$ is the  viscosity coefficient and we have introduced  $\sigma^{ab}$ as the traceless shear tensor
	\beq
	\sigma_{ab} = u_{(a;b)} - \frac{1}{3} u^c_{\ ;c} h_{ab}.
	\eeq

Let us define the the \emph{invariant} specific volume  $v$ as the inverse of the particle number density $n$ and use it to express the purely spatial part of $u_{a;b}$ as
	\beq
	\label{aux}
	h^{ab}u_{a;b}= u^b_{\ ;b} = n {\dot v}.
	\eeq
Thus, substituting the above expressions for the viscous stress, equation \eqref{visc0},  and the spatial projection \eqref{aux} back into the local energy balance \eqref{FL} we obtain	\beq
	\label{sc0}
	n({\dot \epsilon} + p{\dot v}) + q^a_{\ ;a} + q^a{\dot u_a} - u_{a;b}P^{ab} = 0.
	\eeq
In this form, equation \eqref{sc0} is completely analogous to the non-relativistic energy balance\footnote{We can write the Newtonian energy balance in as 
	\[
	n\frac{\d \varepsilon}{\d t} + \nabla \cdot {\bf q} - (\mathscr{P} \cdot \nabla) \cdot {\bf v} = 0
	\]
where $\varepsilon$ is the internal energy, ${\bf q}$ is the heat flow, $\mathscr{P}$ is the total stress and ${\bf v}$ is the fluid's three-velocity.}. It is only the term containing the four-acceleration $\dot u^a$  which has no Newtonian counterpart. Formally, it results from the fact that infinitesimal 3-spaces orthogonal to the observer's worldline are not parallel to each other, but relatively tipped because of the curvature of such a line \cite{ehlers}. One can interpret this as a contribution due to the \emph{inertia} of heat.

For a simple fluid,  where the internal energy is only a function of the pressure $p$ and the specific volume $v$, there are always two functions, $\theta$ and $s$ say,  such that \cite{eckart01,eckart03}
	\beq
	\frac{\partial \epsilon}{\partial v} + p = \theta \frac{\partial s}{\partial v}.
	\eeq
Hence, we can write the quantity inside the bracket in the first term of the energy balance \eqref{sc0} in terms of these two functions as
	\beq
	\label{th1}
	\dot\epsilon + p \dot v = \theta \dot s.
	\eeq
This is simply the Gibbs relation we found in section 2.2.1 [{\it c.f.} equation \eqref{thermo.gibbs}]. Thus, the two functions $\theta$ and $s$ correspond to the equilibrium temperature and entropy density. In terms of these variables, we can re-express \eqref{sc0} to get
	\beq
	\label{th2}
	n\dot s + (\theta^{-1}q^a)_{;a} = -\frac{1}{\theta^2}q^a [\theta_{;a} + \theta \dot u_a] + u_{a;b} P^{ab}\theta^{-1}.
	\eeq

Written in this form, we can impose the second law of thermodynamics on the dynamics of the matter model described by the general energy momentum tensor $T^{ab}$ in terms  of the thermodynamic quantities $n$, $\rho$ and $s$ relative to observer's frame in the following manner. Consider a vector field $s^a$ representing the entropy flux within the fluid [{\it c.f.} section 3.4]. From \eqref{th2} one can infer that 
	\beq
	\label{sflow}
	s^a = sn^a + \frac{1}{\theta}q^a.
	\eeq
Note that this is far from being the only manner in which we could have written the entropy density current. It is, however, the simplest for it is `linear' in departures from thermal equilibrium, {\it i.e} it is of `first-order' in the heat flow.
In this case, the second law of thermodynamics takes the local form [{\it c.f.} equation \eqref{kin.entdiv}]
	\beq
	\label{SL}
	s^a_{\ ;a} = n\dot s + (\theta^{-1}q^a)_{;a} \geq 0.
	\eeq
As we have argued in the previous chapter, the inequality has to be imposed from the outset; it cannot be proven within the dynamical content of the matter model.


\subsubsection{Eckart's hypothesis}

In order to make the divergence of the entropy flux positive definite, the simplest assumption is that the two terms in the RHS of \eqref{th2} are independent and non-negative. 

On the one hand, the first term in the RHS of \eqref{th2} lead Eckart to propose the relativistic analogue of Fourier's law, namely
	\beq
	\label{heat01}
	q^a=-\kappa h^{ab}[\theta_{;b} + \theta \dot u_b].
	\eeq
As before, the proportionality scalar $\kappa$  represents the thermal conductivity of the fluid. It is also clear that equation \eqref{heat01}  is orthogonal to the fluid's four-velocity.  The spatial projection of the covariant derivative of the function $\theta$ corresponds to the relativistic temperature gradient. As we have already discussed, there is no Newtonian analogue of the acceleration term $-h^{ab}\theta \dot u_b$. Such a term implies an isothermal flow of heat in accelerated matter in the direction opposite to the acceleration \cite{eckart03}.  We will provide further details on  this point in the next chapter. Soon it will become clear, that every reasonable relativistic generalization of Fourier's law contains this acceleration term as a consequence of the effect of the \emph{thermal inertia}.

On the other hand,  from the definition of the viscous-stress tensor \eqref{visc} and the decomposition of $u_{a;b}$ into a trace, shear, vorticity and acceleration parts
	\beq
	\label{cdecom}
	u_{a;b} = \frac{1}{3}u^c_{\ ;c}h_{ab} + \sigma_{ab} + \omega_{ab} - \dot u_a u_b,
	\eeq	
using a symmetry argument and the orthogonality condition \eqref{ort0}, it follows that
	\beq
	u_{a;b}P^{ab}  = \sigma_{ab} P^{ab} = \lambda\sigma^2.
	\eeq
Hence, we can write the divergence of the entropy flux explicitly as a quadratic function of its sources
	\beq
	\label{ec2l}
	s^a_{\ ;a}= \frac{q^a q_a}{k\theta^2} + \frac{\sigma_{ab}\sigma^{ab}}{\lambda \theta}\geq 0.
	\eeq

This constitutes the first attempt to generalise Fourier's law to the relativistic regime. However, it can be readily seen that this result contains a covariant analogue of the heat equation, without altering its parabolic character. Indeed, it was shown later by Hiscock and Lindblom that the Eckart model not only suffers from non-causal behaviour, but also it possesses  unstable modes for thermal perturbations of ordinary matter models \cite{hiscock01}.



\subsection{Landau \& Lifshitz theory}

The theory of relativistic dissipation proposed  by Landau and Lifshitz as an alternative to Eckart's model shares essentially the same features as in the original proposal. Therefore, there is no need to extend the discussion in this particular direction beyond a few main results to keep the discussion complete and self-contained. The assumptions and calculations are completely analogous to the ones in the previous section and can be found in \cite{landaufm}. 

The main difference this approach has with respect to the one followed by Eckart is the choice of the observer's frame. In the case of Eckart's model we used co-moving observers to describe the dynamics of the fluid whilst Landau \& Lifshitz used a timelike eigenvector of the energy momentum tensor of the matter. In both cases, the equations of motion are determined by the conservation of the particle number density flow and the vanishing of the divergence of the energy-momentum tensor. Thus, the dynamics are completely equivalent and, although the specific form of the local energy balance will take a different form, the physical content of both approaches is necessarily the same. Crucially, they lead to the same ansatz to ensure the positivity of the entropy flux. Therefore, they also share the same pathologies as found by Hiscock and Lindblom.


\section{Second-order theories}

This two-part section contains an account of the well known Israel \& Stewart second-order theory together with the original version of the pioneering work  of Carter known as the `regular model'. One of the aims of this section is to highlight the main ontological difference between the two `second-order' approaches, namely, the local equilibrium assumption for small deviations from thermal equilibrium in the former and, the EIT approach to non-equilibrium thermodynamics of relativistic systems of the latter. There is no claim of originality of the material  presented in this section. However, it has been written in a manner that, we hope, will facilitate an understanding of the details in the original sources. 

\subsection{The Israel \& Stewart second-order theory}

In view of the unsatisfactory results  of Eckart's theory, Israel \cite{israel} and Stewart \cite{Stewart77} developed a new strategy to solve the inconsistencies of the relativistic first-order theories of dissipation. Their approach, firmly grounded in relativistic kinetic theory, was named  the transient relativistic thermodynamics, but it is better known as the second-order theory of relativistic dissipation. This model follows the same dynamical construction as in Eckart's model up to the definition of the entropy density flux, where they included a full set of second-order corrections in the entropy sources. This truncated expansion leads to a relativistic generalization of Cattaneo's equation \eqref{dis.cattaneo}. The following is a detailed  derivation of such an extension.

Let us begin by writing the decomposition of the stress-energy tensor [see \eqref{dec01}] in a slightly different manner 
	\beq
	\label{stress}
	T^{ab} = \rho u^a u^b + (p + \tau)h^{ab} + 2 u^{(a}q^{b)} + \tau^{ab}.
	\eeq
The difference from \eqref{dec01} lies  in the explicit inclusion of a viscous pressure term $\tau$.  Again $q^a$ is the heat flow orthogonal to the matter's four velocity , while $\tau^{ab}$ and $\tau$ are the stresses caused by viscosity in the fluid. Here the tensor $\tau^{ab}$ satisfies the relations
	\beq
	\label{ort1}
	0 = u^a \tau_{ab}= \tau^a_{\ a}=\tau_{[ab]}.
	\eeq
Thus, it relates to Eckart's definition through 
	\beq
	\tau^{ab}= -\pi^{ab} + (p+\tau)h^{ab}.
	\eeq

As before, the equations of motion are given by the conservation laws \eqref{cons0} and \eqref{first01} and the imposition of the second law is completely analogous to the previous section. From the modified form of the energy-momentum tensor \eqref{stress} and  properties \eqref{ort1} of the viscous-stress tensor, the divergence of the entropy density current, equation  \eqref{th2}, can be written as
	\beq
	\label{sflowi}
	s^a_{\ ;a}= -\frac{1}{\theta^2}q^a [\theta_{;a} + \theta \dot u_a] + \langle u_{a;b}\rangle \tau^{ab}\theta^{-1} - \tau u^a_{\ ;a}\theta^{-1}.
	\eeq
Here $s^a$ is the entropy current defined in Eckart's model, equation \eqref{sflow},  the brackets $\langle \rangle$  represent the symmetric traceless part of a given second rank tensor, {\it i.e.} for a given tensor $A^{ab}$ 
	\beq
	\langle A_{ab} \rangle = \frac{1}{2}h^c_{\ a}h^d_{\ b}\left[2A_{(ab)} - \frac{2}{3}h_{cd}h^{ef}A_{ef} \right],	
	\eeq
and, crucially, $\theta$ is the \emph{equilibrium} temperature of the fluid. It is obtained through the  assumption that the local equilibrium hypothesis holds for small deviations from thermal equilibrium, {\it i.e.} that the energy density remains a function of the particle number and entropy densities when the system is driven `slightly' away from thermal equilibrium. This means that, implicitly, the heat flux is still considered to be a \emph{fast} variable in the EIT sense ({\it c.f.} section 2.2.3) and, therefore, it  \emph{cannot} be treated as a state variable.
 
In order to satisfy the requirement of the second law, in addition to Eckart's hypothesis \eqref{heat01}, we require that
	\begin{subequations}
	\begin{align}
	\tau 	  &= -u^a_{\ ;a}\xi,\\
	\tau^{ab} &= -2\eta \langle u^{a;b}\rangle,
	\end{align}	
	\end{subequations}
where $\xi$ and $\eta$ are the bulk and shear viscosity coefficients, respectively. Note that in the case of the Eckart model we only had the coefficient $\lambda$. This is because here we have made an explicit distinction between the trace and shear parts of the congruence.

 We see from \eqref{visc} that $\tau^{ab}$ is proportional to $P^{ab}$. Hence, the quadratic form for the divergence of the entropy density current, equation \eqref{ec2l}, becomes
	\beq
	\label{seclaw}
	s^a_{\ ;a}= \frac{\tau^2}{\xi \theta} + \frac{q^a q_a}{k\theta^2} + \frac{\tau_{ab}\tau^{ab}}{2\eta \theta}\geq 0.
	\eeq

Motivated by the non-causal and unstable behaviour of Eckart's proposal, Israel and Stewart proposed to generalise the definition of the entropy density flux \eqref{sflow} by including a complete set of second order terms, namely
	\beq
	\label{sisrael}
	s^a=sn^a +\frac{q^a}{\theta} - \frac{1}{2}\left(\beta_0 \tau^2 + \beta_1 q^b q_b + \beta_2 \tau_{bc}\tau^{bc} \right)\frac{u^a}{\theta} + \alpha_0 \frac{\tau q^a}{\theta} + \alpha_1\frac{\tau^a_{\ b} q^b}{\theta}.
	\eeq 
Here the  coefficients $\beta_0$, $\beta_1$, $\beta_2$, $\alpha_0$ and $\alpha_1$ correspond to the different couplings for the second-order terms. In particular, these quantities need to be provided by some other means, {\it i.e.} by direct measurement or directly through kinetic theory, but cannot be obtained from an equilibrium equation of state.  

To compute the entropy production $s^a_{\ ;a}\theta$, we can split \eqref{sisrael} into  first and second order pieces. The divergence of the linear part is given by \eqref{sflowi}
	\beq
	s^a_{I\ ;a}\theta = -\frac{1}{\theta}q^a [\theta_{;a} + \theta \dot u_a] + \langle u_{a;b}\rangle \tau^{ab} - \tau u^a_{\ ;a}.
	\eeq
Therefore we only have to compute the divergence of the second-order part
	\begin{align}
	\label{sflowii}
	s^a_{II\ ;a}\theta =&\frac{\theta}{2}\left[ \left(\frac{\beta_0 \tau^2 u^a}{\theta}\right)_{;a} + \left(\frac{\beta_1 q_b q^b u^a}{\theta}\right)_{;a} + \left(\frac{\beta_2 \tau_{bc}\tau^{bc} u^a}{\theta}\right)_{;a}\right]\nn \\
                &+\theta\left[\left(\frac{\alpha_0 \tau q^a}{\theta}\right)_{;a} + \left(\frac{\alpha_1 q_b \tau^{ab}}{\theta}\right)_{;a} \right].
	\end{align} 

Expanding term by term, we have the following factorisations for the first square bracket
	\begin{subequations}
	\begin{align}
	\left(\frac{\beta_0 \tau^2 u^a}{\theta}\right)_{;a} & = \tau \left[ \left(\frac{\beta_0 u^a}{\theta}\right)_{;a}\tau + 2\frac{\tau_{;a}u^a \beta_0}{\theta} \right],\\
	\left(\frac{\beta_1 q_b q^b u^a}{\theta}\right)_{;a}& = q^a \left[\left(\frac{\beta_1 u^b}{\theta}\right)_{;b}q_a + 2 \frac{q_{a;b}u^b \beta_1}{\theta} \right],\\
	\left(\frac{\beta_2 \tau_{bc}\tau^{bc}u^a}{\theta} \right)_{;a}& = \tau^{ab}\left[\left(\frac{\beta_2 u^c}{\theta}\right)_{;c}\tau_{ab} + 2\frac{\tau_{ab;c}u^c \beta_2}{\theta}  \right].
	\end{align}
	\end{subequations}
Whilst  the second bracket in \eqref{sflowii} is given by
	\begin{subequations}
	\begin{align}	
	\label{apd}
	\left(\frac{\alpha_0 \tau q^a}{\theta} \right)_{;a} & = q^a_{\ ;a}\frac{\tau \alpha_0}{\theta} + \tau_{;a}\frac{q^a \alpha_0}{\theta}+ \left(\frac{\alpha_0}{\theta}\right)_{;a}q^a \tau,\\
	\label{ape}
	\left(\frac{\alpha_1 q_b \tau^{ab}}{\theta} \right)_{;a} & = \tau^{ab}_{\ \ ;a}\frac{q_b \alpha_1}{\theta} + \left(\frac{\alpha_1}{\theta}\right)_{;a}q_b\tau^{ab} + q_{b;a}\frac{\tau^{ab}\alpha_1}{\theta}.
	\end{align}
	\end{subequations}
In order to factorise the last two equations, Israel (see \cite{israel}) introduced two extra thermodynamic coefficients $\gamma_0$ and $\gamma_1$ such that
	\beq
	\gamma_0 + \gamma_1 = 1.
	\eeq
Thus, it follows that \eqref{apd} and \eqref{ape} become
	\begin{subequations}
	\begin{align}
	\left(\frac{\alpha_0 \tau q^a}{\theta} \right)_{;a} & = q^a\left[ \tau_{;a} \frac{\alpha_0}{\theta} + \left(\frac{\alpha_0}{\theta}\right)_{;a} \tau \gamma_1\right] + \tau \left[q^a_{\ ;a}\frac{\alpha_0}{\theta} + \left(\frac{\alpha_0}{\theta} \right)_{;a}q^a \gamma_0 \right],\\
	\left(\frac{\alpha_1 q_b \tau^{ab}}{\theta} \right)_{;a} & = \tau^{ab} \left[ q_{b;a}\frac{\alpha_1}{\theta}+  \left(\frac{\alpha_1}{\theta} \right)_{;a} q_b \gamma_1\right] + q^a \left[\tau^b_{\ a;b}\frac{\alpha_1}{\theta} + \left(\frac{\alpha_1}{\theta} \right)_{b}\tau^b_{\ a}\gamma_0\right].
	\end{align}
	\end{subequations}
Finally, collecting all the terms, we obtain the expression for the entropy production;
	\begin{align}
	s^a_{\ ;a}\theta =& -\tau \left[u^a_{\ ;a} + \tau_{;a}u^a \beta_0 + \frac{1}{2}\left(\frac{\beta_0 u^a}{\theta}\right)_{;a}\tau \theta -q^a_{\ ;a}\alpha_0 - \left(\frac{\alpha_0}{\theta}\right)_{;a} q^a\theta  \gamma_0\right]\nn\\
			  & - q^a {\Bigg [}\theta_{;a}\frac{1}{\theta} + \dot u_a +\frac{1}{2}\left(\frac{\beta_1 u^b}{\theta}\right)_{;b} q_a \theta - \tau_{;a}\alpha_0 - \left(\frac{\alpha_0}{\theta}\right)_{;a}\tau\theta\gamma_1\nn\\
			  &\quad +q_{a;b}u^b\beta_1 - \tau^b_{\ a;b}\alpha_1 - \left(\frac{\alpha_1}{\theta}\right)_{;b}\tau^b_{\ a}\theta\gamma_0 {\Bigg ]}\nn\\
			  & -\tau^{ab}{\Bigg [}{\Bigg\langle}u_{a;b} +\frac{1}{2}\left( \frac{\beta_2 u^c}{\theta}\right)_{;c}\tau_{ab}\theta +\tau_{ab;c}u^c \beta_2 - q_{b;a}\alpha_1 -\left( \frac{\alpha_1}{\theta}\right)_{;a}q_b \theta \gamma_1  {\Bigg \rangle}{\Bigg ]}.
	\end{align}

In order to satisfy the second law constraint \eqref{seclaw}, the simplest - though not the most general - choice is to make each term positive. Thus, the Israel and Stewart theory makes the following identifications
	\begin{subequations}
	\begin{align}
		\tau   =& -\xi \left[u^a_{\ ;a} + \tau_{;a}u^a \beta_0 + \frac{1}{2}\left(\frac{\beta_0 u^a}{\theta}\right)_{;a}\tau \theta -q^a_{\ ;a}\alpha_0 - \left(\frac{\alpha_0}{\theta}\right)_{;a} q^a\theta  \gamma_0\right],\\
	\label{iscat}
	q^a=& -k\theta h^{ab} {\Bigg [}\theta_{;b}\frac{1}{\theta} + \dot u_b +\frac{1}{2}\left(\frac{\beta_1 u^c}{\theta}\right)_{;c} q_b \theta - \tau_{;b}\alpha_0 - \left(\frac{\alpha_0}{\theta}\right)_{;b}\tau\theta\gamma_1\nn\\
			  &\quad +q_{b;c}u^c\beta_1 - \tau^c_{\ b;c}\alpha_1 - \left(\frac{\alpha_1}{\theta}\right)_{;c}\tau^c_{\ b}\theta\gamma_0 {\Bigg ]},\\
	\tau^{ab} =& -2\eta h^{ac} h^{bd}{\Bigg\langle}u_{c;d} +\frac{1}{2}\left( \frac{\beta_2 u^e}{\theta}\right)_{;e}\tau_{cd}\theta +\tau_{cd;e}u^e \beta_2 - q_{d;c}\alpha_1 -\left( \frac{\alpha_1}{\theta}\right)_{;c}q_d \theta \gamma_1  {\Bigg \rangle}.
	\end{align}
	\end{subequations}


\subsubsection{Heat conduction}

For the purely heat conducting case we artificially turn off the viscous terms, {\it i.e.} we set $\tau=\tau^{ab}=0$. In this case, the energy-momentum tensor reduces to
	\beq
	T^{ab}=(\rho + p)u^a u^b + pg^{ab} + 2u^{(a}q^{b)},
	\eeq
and the entropy current is simply given by
	\beq
	s^a= sn^a + \frac{1}{\theta}q^a - \frac{u^a}{2\theta}\beta_1 q_b q^b.
	\eeq

The simplified version of the divergence of the entropy density flux is
	\beq
	\label{isheat}
	s^a_{\ ;a}\theta = -q^a \left[\theta_{;a}\frac{1}{\theta}+ \dot u_a +\frac{1}{2}\left(\frac{\beta_1 u^b}{\theta}\right)_{;b} q_a \theta +q_{a;b}u^b \beta_1 \right],
	\eeq
and we see that the quadratic ansatz \eqref{iscat} becomes 
	\beq
	\label{dis.cat}
	q^a= -k\theta h^{ab}\left[\theta_{;b} \frac{1}{\theta} + \dot u_b +\frac{1}{2}\left(\frac{\beta_1 u^c}{\theta} \right)_{;c} q_b\theta + q_{a;c}u^c \beta_1 \right].
	\eeq
This is the relativistic generalization of the Cattaneo equation given in section 2.2.2. It represents the central result of the purely heat conducting Israel \& Stewart model and it will serve as a point of comparison for the results in the next chapter. Note that  the last term in \eqref{dis.cat} is essentially a time derivative of the heat transport four-vector. Therefore,  its coefficient $\beta_1$ can be interpreted as  a `relaxation time' for the propagation of thermal disturbances. The subtle point we mentioned earlier lies in the fact that, although we assumed that the local equilibrium hypothesis holds\footnote{The energy density is a function of the particle number and entropy densities alone, $\rho = \rho(n,s)$.}, we have obtained a critical time-scale on which dissipation operates, a result which follows from an EIT point-of-view. Thus, the reader may wonder where did such a time scale come from? The answer is simple, we have removed the thermodynamic information of heat as a state variable at the expense of a complete set of coupling coefficients for the second order terms in the entropy flux. The remarkable feature of this approach is that it correctly reproduces an EIT result from the thermodynamic principles of LIT.



\subsection{Carter's  theory of dissipation}

The final section of this  chapter is devoted to the original work by Carter on a relativistic heat conduction.  Although it was not made explicit at the time, soon we will notice that Carter's ideas are rooted in those which gave rise to the the EIT programme. Furthermore, he presented an argument in which the Cattaneo equation \eqref{dis.cattaneo} follows directly from a Gibbs relation where the heat flux is included as a state variable. As before, we have tried to make every calculation explicit and consistent with the method discussed in the previous sections.


Let us now show that Carter's thermodynamic programme belongs to the class of EIT theories\footnote{We can readily see this from his original arguments on these matters. For completeness, we have included a transcript of a highly unknown reference where this fact was made explicit first. The reader will find a recollection of Carter's original ideas in  appendix A.}. Following the same methodology as in the previous sections, let us consider a single species fluid. Recalling the decomposition of a general energy momentum tensor with respect to the four-velocity of an observers moving with the particles of the fluid, we write [see equation \eqref{dec01}]
	\beq
	\label{cstress}
	T^{ab} = \rho u^a u^b + 2 q^{(a}u^{b)} + \pi^{ab}.
	\eeq 

In addition to the particle number density flow $n^a$,  let us introduce  an entropy density current $s^a$ - in general not parallel with $n^a$. We can decompose such a current by ({\it c.f.} section 3.4)
	\beq
	\label{centropy}
	s^a = s^*u^a + {j}^a,
	\eeq
where $s^*$\glo{$s^*$}{Entropy density as measured on the matter frame} is the entropy density measured in the Eckart (matter) frame,  and $j^a$ is the component of the entropy current  transverse to the matter flow.

As before, the equations of motion are given by the conservation laws \eqref{cons0} and \eqref{cons01} and the second law of thermodynamics is expressed by the positivity of the entropy production. Thus, we can express the equations of motion together with the second law in terms of the projections
	\begin{subequations}
	\begin{align}
	\label{cncons}
	n^a_{\ ;a} & = \dot n + u^a_{\ ;a} n = 0,\\
	\label{cscons}
	s^a_{\ ;a} & = \dot s + u^a_{\ ;a} s + {j}^a_{ ;a} \geq 0 .
	\end{align}
	\end{subequations}
 
Using the decomposition of $u_{a;b}$, equation \eqref{cdecom}, we can define the symmetric part of $u_{a;b}$ as
	\beq
	\tilde \sigma_{ab} = \frac{1}{3} u^c_{\ ;c}h_{ab} + \sigma_{ab}.
	\eeq   
Thus, the equations of motion imply a  local energy balance  equivalent to Eckart's expression \eqref{FL}
	\beq
	\label{ctu}
	T^b_{\ a;b}u^a   =  \dot\rho + u^b_{\ ;b}\rho + q^b_{\ ;b} + q^b \dot u_b  + \pi^{ab}\tilde\sigma_{ab} = 0.
	\eeq

Up to this point, the construction is identical to Eckart's model. It is here, where we need to introduce a thermodynamic assumption to extract the second law from the energy balance that Carter's model differs from Eckart's and Israel \& Stewart.

Let us  consider the possible functional dependence of the energy on the state variables.  In the special situation, when the medium is in strict thermal equilibrium, the energy density is a function of the particle number and entropy densities alone. However, in a general situation, we could also expect $\rho$ to be a function of the non-equilibrium  magnitude of the entropy transfer vector, {\it i.e.} 
	\beq
	\rho = \rho(n,s,j).
	\eeq

In an ordinary situation, we would consider states which are close to equilibrium. In such a case, $j^\q$ is very small quantity. Moreover, local isotropy requires that the dependence on $j^\q$ of the energy density should be of second order in the neighbourhood of equilibrium, that is
	\beq
	\rho= \rho(n,s,0) + \mathscr{O}({j}^2).
	\eeq 
Therefore, a general variation of the energy density takes the form
	\beq
	\label{dis.extendedgibbs}
	\d \rho = \mu \d n + \theta \d s + \mu_q \d j,
	\eeq
which we recognize as the extended Gibbs relation of section 2.2.3. Thus, Carter's theory of dissipation adopts an EIT point of view before imposing the second law of thermodynamics. The specific details of how the inclusion of the heat flux as a state variable leads to a relativistic version of Cattaneo equation are presented in the next chapter. Carter's regular model is a particular case of the general multifluid formalism for relativistic dissipation. For completeness, we finish our  discussion as in the original version of Carter's  manuscript. However, this \emph{incomplete} view does not present us with the final outcome of the theory. The reader has been warned.

 In order to take this last step, let us consider the Legendre-transformed energy density
	\beq
	\label{clt01}
	\hat\rho = \rho - \mu_\q j.
	\eeq
Differentiating with respect to the affine parameter of the observer's worldline gives 
	\beq
	\label{cenp}
	\dot{\hat\rho} = \mu \dot n + \theta \dot s + j^b\mathcal{L}_u[\mu_\q]_b - \sigma_{ab}\mu_\q^a {j}^b.
	\eeq
where we have made the substitution
	\beq
	\dot {\mu^\q}_b {j}^b = \mathcal{L}_{\rm u} \left[{\mu^q}_b {j}^b\right] - u_{a;b}\mu_\q^a j^b.
	\eeq

Let us further assume that	the heat flow is simply proportional to the transverse entropy current $j^a$, so we can write
	\beq
	\label{cheatf}
	q^a = \tilde \theta j^a_\s,
	\eeq
where we call $\tilde \theta$  the \emph{thermodynamic} temperature. Notice that we cannot tell if this is the temperature that appears in the extended Gibbs relation \eqref{dis.extendedgibbs}.

In analogy with Eckart's model, collecting the previous result and substituting into the local energy balance, one can show that the divergence of the entropy density current can be written as the sum of three terms
	\begin{align}
	s^a_{\ ;a}\theta =&-j^b\left[\tilde\theta_{;b} + \mathcal{L}_{\rm u}\left(\tilde\theta u_b + \mu^\q_b\right)\right]\nn\\ 
			  &+\left[h^{ab}\left(\mu n + \theta s - \hat\rho\right) + \mu_\q^a j^b - \pi^{ab}\right]\tilde\sigma_{ab}\nn\\
			  &+j_{\ ;b}^b\left(\theta - \tilde \theta\right).
	\end{align}

Once again, the simplest way to impose the second law of thermodynamics is to require  each term to be positive. Thus, we obtain a heat conduction equation of the form
	\beq
	\tilde\theta_{;b} + \mathcal{L}_u\pi^\q_b = - Y_{ab} q^a.
	\eeq
Here, the quantity $\pi^\q_b$ can be interpreted as momentum conjugate to the entropy density flux given by
	\beq
	\pi^\q_a = \tilde\theta u_a + \mu^\q_a,
	\eeq 
and the transverse tensor $Y_{ab}$ is the thermal resistivity.

Since the other two terms cannot be made strictly positive, it is required that they both vanish. This imposes the further restriction for the energy momentum tensor
	\beq
	\label{cstress00}
	\pi^{ab} = h^{ab}(\mu n +\theta s - \hat\rho) + \mu^a_\q j^b,
	\eeq
and allows us to conclude that the proportionality coefficient $\tilde\theta$ in our assumption for the heat flow is indeed the \emph{thermostatic} temperature of the fluid
	\beq
	\tilde\theta = \theta.
	\eeq
This last equality is also a consequence of the regularity ansatz. It is to severe a constraint for the relaxation time of thermal phenomena described by this early model. Indeed, this was a crucial part of the argument in the criticism of Hiscock and Olson of the regular model \cite{hiscolson}. In the next chapter, we construct from scratch a multifluid approach to heat conduction based on the ideas presented in this subsection. We will show that the regularity ansatz removes the information contained in the equation of state from the relaxation time. Therefore, finding a model which would violate local causality was not very difficult. Thus, as expected by Hiscock and Olson,  if the regularity ansatz is relaxed, we stand a better chance to obtain a consistent relativistic theory of heat.

			\chapter{Variational thermal dynamics}
\label{ch.canonical}

In this chapter we construct a  theory of two interacting fluids in General Relativity by means of an action principle. Such approach was pioneered by Taub \cite{taub} and later by Friedman \& Schutz \cite{fshutz} and championed by Carter \cite{carterquin} in the so called convective variational approach. Here the central role is played by a single  scalar function, the \emph{master function}, which we can understand as a generalisation of the Lagrangian for a purely conservative model. 

Variational principles for perfect fluids in General Relativity have been widely studied in the literature, see for instance \cite{annJDBrown} and references therein. We start by constructing the action from a Lagrangian density as a function  of all  the possible covariant combinations of  the kinematical variables, namely the flows $n^a$, $s^a$ and the metric $g_{ab}$. By considering an unconstrained variation of the action we obtain a general expression for the conjugate momenta. From this we realise that in the generic case the momenta are not aligned with the flows - we call this effect \emph{entrainment}. Later we will show that this is an essential ingredient to extend the analysis to dissipative systems.


To obtain the purely conservative equations of motion, we introduce the idea of a 3-dimensional \emph{material space} \cite{carterquin,carter30,kunzle,annJDBrown}, whose points represent idealised particles of the medium. We must associate a different material space for each particle species at each point of space-time. This  allows us to impose the current conservation constraints by ``pulling'' the closed volume 3-forms ``back'' into space-time, whose duals will be the conserved particle number density flow and the ``conserved piece'' of the entropy flow. 
By imposing these constraints we obtain the equations of motion in the form of a force balance equation corresponding to the local version of the energy conservation law.
 
We proceed with a thermodynamical analysis. There is an obvious choice of reference frame to perform such analysis. However,  by defining the relevant quantities as space-time scalars, the choice of  observers is irrelevant for the theory - but is of central importance for the correct interpretation.

 Finally by imposing the second law constraint on the equations of motion we obtain an entropy production rate which is in perfect agreement with the phenomenological one produced in the previous section.


\section[Convective construction]{Oldroyd-Carter rheometric (convective) construction}
\label{mscontr}

We start our discussion on canonical heat conduction by constructing the appropriate physical quantities. The main idea throughout this presentation is to follow the variational formalism as discussed in Chapter 2 while keeping track of the symmetry underlying the equations of motion \eqref{gf.Tcons2}. 

The variational approach to relativistic thermodynamics, as presented by Carter \cite{carternotes}, builds on the notion of a lower dimensional \emph{matter space} which represents the local configuration of an extended \emph{body} in spacetime \cite{ehlers}. This  provides an instructive way to identify physical quantities which are invariant under
the group of local transformations acting on spacetime which characterise the evolution of such a `body'.  

The set of invariant quantities of the motion represent the state of the medium - its intrinsic properties. Thus, the matter space description can suitably be applied to the study of equations of state, as discussed in \cite{carterrheometric}. We use the matter space to construct the notion of \emph{Lagrangian} variation in order to obtain the equations of motion of a system whose state is represented by \emph{conserved} particle currents.  

Consider a \emph{timelike} vector field $U$\glo{$U$}{Timelike vector field associated with a family of observers} whose integral curves  generate locally the one-parameter family of diffeomorphisms $\varphi_s\in {\rm Diff}{(\mathcal{M}})$. For a small enough region $\Omega$ of spacetime $\mathcal{M}$, these curves can be thought of as the particles of a medium, rigid or elastic, or even a fluid. Letting $\mathcal{G}_{U} = {\rm Diff}(\mathcal{M})|_{U}$ denote the one-dimensional sub-group\glo{$\mathcal{G}_{U}$}{The one-dimensional sub-group  of ${\rm Diff}(\mathcal{M})$ generated by $U$} of ${\rm Diff}(\mathcal{M})$ generated by $U$, we define the matter space as the quotient manifold\glo{$\mathscr{B}$}{Matter space}
	\beq 
	\mathscr{B}(\Omega,U) = \Omega/\mathcal{G}_{U},
	\eeq
whose points are the equivalence classes defined by the orbits of $\mathcal{G}_{U}$. Since the region $\Omega$ is four-dimensional and the orbits of $\mathcal{G}_{U}$ are the one-dimensional integral curves of $U$, it follows that the matter space $\mathscr{B}$ is a three dimensional manifold \cite{olver}. Following Carter \cite{carterrheometric}, we call $\mathcal{G}$ the \emph{rheometric} group.

Notice that we can construct as many matter spaces as we have timelike vector fields. Thus, anticipating the forthcoming discussion,  a two-fluid system generates rheometrical groups, $\mathcal{G}_{\n}$ and $\mathcal{G}_{\s}$ say, acting over the same region $\Omega$, and therefore two independent matter spaces $\mathscr{B}_\n$ and $\mathscr{B}_\s$ \cite{lrrnils}.

Although the matter space is a manifold of its own, its definition implies a projection map
	\beq
	\label{ms.proy}
	f: x^a \mapsto X^A(x^a), \qquad A={1,2,3},
	\eeq 
which takes an event  $x^a \in \Omega$ into its equivalence class $X^A\in \Omega/\mathcal{G}_{U}$ defined by the unique worldline passing though it. We can think of $X^A$ as local coordinates for the points in the matter space.

Quotient manifolds may have a very  limited range of  geometric structures defined on them. For instance, we cannot expect to have an unambiguously defined metric over $\mathscr{B}$, in fact, only in the case of a rigid congruence\footnote{We call a congruence rigid if its expansion and shear are identically zero.} is it possible to construct such a metric \cite{ehlers}. However, our main \emph{physical} ansatz is to assume that a canonical measure is well defined. Let us denote by $\n^\flat$ such a measure which, in the above coordinate representation, can be written as\glo{$n^\flat$}{Matter space volume form}
	\beq
	\label{ms.nflat}
	\n^\flat =  N_{ABC} \d X^A \wedge \d X^B \wedge \d X^C.
	\eeq
The interpretation of this volume form is obtained by means of the projection map \eqref{ms.proy}.

 The map $f$ naturally induces the pull-back $f^*$ from each cotangent space $T_{f(x)}\mathscr{B}$ to $T_x\mathscr{\Omega}$ and the push-forward $f_*$ from the tangent space at $T_x\Omega$ onto $T_{f(x)}\mathscr{B}$. Two remarks are in place \cite{ehlers}:
 	\begin{enumerate}
 		\item At each point $x\in\Omega$, the restriction of $f_*$ to the subspace of $T_x\Omega$ orthogonal to $U$ is a bijection.
 		\item The range of $f^*$ consists of the 1-forms which are annihilated by $U$. 
 	\end{enumerate}

The above remarks suggest that there is natural correspondence between $U$-orthogonal objects - tensors whose contraction with $u^a$ vanish - and tensors in the matter space. This is indeed the case, as shown by Oldroyd \cite{oldroyd} and Carter \cite{carterquin,carter30,carterrheometric}. However, the lack of a \emph{unique} metric tensor field in $\mathscr{B}$ allows the map to distinguish between raised and lowered indices\footnote{For a general non-rigid congruence, we can use the projection map to push forward the spacetime metric $g^{ab}$ at each spacelike hypersurface orthogonal to $U$ to obtain a \emph{family} of tensor fields on $\mathcal{B}$. For an elastic body, each individual tensor of the family defines a `state of strain'. We can extend this notion to the case of a fluid. However, we do not have \emph{a priori} a unique manner to identify one-forms and vector fields on $\mathcal{B}$. In the case of an elastic material, the identification is made, conventionally, through the pushed-forward metric at the `instant' the material was formed. For a detailed discussion about the state of strain of an elastic body see \cite{ehlers}.}.

  The pull-back of the canonical volume form \eqref{ms.nflat} is the spacetime 3-form\glo{$\n$}{Pulled back three-volume form the matter space $\mathscr{B}_\n$}
	\beq
	\label{ms.pb}
	f^* \n^\flat =  N_{ABC} X^A_{\ ;a} X^B_{\ ;b} X^C_{\ ;c} \d x^a \wedge \d x^b \wedge \d x^C =  n_{abc}\d x^a\wedge\d x^b\wedge\d x^c =\n.
	\eeq
By virtue of the definition of $\mathscr{B}$ and the compatibility between pull-backs and exterior derivatives, $\d f^* \n^\flat = f^* \d \n^\flat$, we see that \eqref{ms.pb} defines a closed form
	\beq
	\label{ms.closed}
	\d \n = f^* \d \n^\flat = \frac{\partial N_{ABC}}{\partial X^D} X^D_{\ ,d} X^A_{\ ,a} X^B_{\ ,b} X^C_{\ ,c}\ \d x^d \wedge \d x^a \wedge \d x^b \wedge \d x^c = 0.
	\eeq

It is easy to see that this is equivalent to some conservation law obtained by applying the dual operator given by the spacetime metric to \eqref{ms.closed},\glo{$\star$}{Hodge dual associated with the metric $\bf{g}$}
	\beq
	\label{ms.dual}
	\star \d \n = \frac{\sqrt{-g}}{4!} n^{abc;d}\epsilon_{abcd}=0 \quad \leftrightarrow \quad n_{[abc;d]} = 0,
	\eeq
which follows from the same symmetry reasoning. Thus, equation \eqref{ms.dual} defines the conserved current
	\beq
	\label{ms.flow}
		n^a_{\ ;a}=0 \quad {\rm where} \quad n^a = \frac{1}{3!} n_{bcd}\epsilon^{abcd} = n u^a, \quad \text{and} \quad u^a u_a = -1.
	\eeq
 
 We can interpret the canonical measure $\n^\flat$ on the matter space $\mathscr{B}$ as the number density $n$ of \emph{conserved} particles whose world-lines are a subset of the flow generated by the normalised vector field $U$ in $\Omega$. We can say, in a less formal manner, that $U$ generates the conserved particle flow.

It may seem that this construction obscures a fact that we could have taken for granted as a starting point. However, the true value of using a matter space to describe the motion of an elastic or fluid medium will shortly become evident when we use the \emph{relevant}  field variations in the action principle.

Notice that functions (0-forms) defined on $\mathscr{B}$, by construction, are invariants of the flow generated by $U$. Conversely, any function on spacetime which is invariant under the action of $\mathcal{G}_{U}$ has a natural counterpart in $\mathscr{B}$. Thus, there is a one-to-one correspondence between $\mathcal{G}_{U}$-invariant functions in $\Omega$ (functions on spacetime which are preserved by the flow) and arbitrary functions in $\mathscr{B}$ \cite{olver}. In particular, the coordinate functions $X^A$ are dragged by the flow  generated by the current \eqref{ms.flow}. We can easily see this, by the explicit calculation
	\begin{align}
	\label{ms.xdrag}
	\mathcal{L}_n X^A  & = X^A_{\ ,a}n^a  = X^A_{\ ,a} \frac{1}{3!} n_{bcd}\epsilon^{abcd}  \nn\\
			     & = \frac{1}{3!} N_{BCD} \epsilon^{abcd} X^A_{\ ,a} X^B_{\ ,b} X^C_{\ ,c} X^D_{\ ,d} = 0,
	\end{align}
where the last equality follows from the commutative property of partial derivatives and the anti-symmetry of ${\bf \epsilon}$. The study of more general rheometric invariant objects is described in detail in \cite{carterrheometric}. Our interest here is to use the matter space to assist in the construction of an action principle for thermodynamics following the reasoning described in chapter 2.

\subsection{Lagrangian and Eulerian variations}

In the construction of the variational principle for general relativity, we used a one-parameter family of diffeomorphisms $\varphi_s$ locally generated by an arbitrary vector field $\xi$, acting on spacetime fields - including the metric - at \emph{fixed points} in spacetime whose coordinate representation plays no role. In the forthcoming exposition of a variational principle for relativistic thermodynamics, however,  we want to follow Carter's work \cite{carter30}  and introduce two different types of variation. 

The first type of variation we wish to consider corresponds to those whose effect on the fields in $\Omega$ is described in terms of a \emph{coordinate system} which is itself dragged by the displacement vector $\xi^a=\delta x^a$. We call such variations \emph{Lagrangian} and denote them by $\delta_\ell$. In particular, as we will explain, Lagrangian variations leave the projection map $f$ fixed \cite{carterrheometric,carternotes}.

The second kind arise whenever we have a definite prescription for identifying  the points of spacetime and we consider a variation of the fields at a \emph{fixed} spacetime point. Such a prescription is, necessarily, not generally covariant since, as discussed in Section \ref{coincidences}, it gives a physical meaning to the spacetime point at which the variation is  taken. In this case, a general perturbation cannot be required to preserve the projection map and, therefore, we need to make a distinction between this \emph{Eulerian} type of variations from their Lagrangian counterpart. Let us denote this type of variations by $\delta_\mathsf{e}$.

For any field quantity $\Psi$ and Lagrangian displacement $\xi$ generating the one-parameter family of diffeomorphisms $\varphi_s$, as in \eqref{gf.onepf}, we have that the difference between Lagrangian and Eulerian variations is the Lie-derivative \cite{carter30,carterrheometric,carternotes,lrrnils}\glo{$\delta_{\ell}$}{Lagrangian variation}\glo{$\delta_{\mathsf{e}}$}{Eulerian variation}
	\beq
	\label{ms.lag-eu}
	(\delta_\ell - \delta_{\mathsf{e}})\Psi = \mathcal{L}_\xi \Psi. 
	\eeq      
The explanation for this expression - found in many treatments of relativistic elasticity - is sketched below by making a distinction between active and passive diffeomorphisms. The spirit of the argument is to facilitate the interpretation of \eqref{ms.lag-eu} and not to provide a rigorous proof for it. 

\subsubsection{Active and passive diffeomorphisms}

 Although in the course of this work we have been using extensively the idea of diffeomorphisms of spacetime, we wish to address a subtle but relevant point; the difference between those which we refer as \emph{active} from the ones we call \emph{passive} \cite{rovelliQG}.   

An \emph{active} diffeomorphism $\varphi$ is a differentiable map with differentiable inverse from  $\mathcal{M}$ into itself. This is to be distinguished from a \emph{passive} diffeomorphism $\phi$, which is a similar map from a \emph{coordinate system} for an open region of spacetime into another coordinate system. A coordinate system for an open subset  $\Omega\subset \mathcal{M}$ is also an invertible differentiable map assigning to each point in the interior of $\Omega$ its coordinate representation in $\mathbb{R}^d$. Thus, we have the following mappings: 
	\begin{enumerate}
		\item Active diffeomorphism: $\varphi: q \mapsto p=\varphi(q)$, where  $p,q \in \Omega$.
		\item Coordinate system: $\vartheta: p \mapsto x = \vartheta(p)$, where $p\in\Omega$ and $x \in \mathbb{R}^d$.
		\item Passive diffeomorphism: $\phi: x' \mapsto x=\phi(x')$, where $x',x \in \mathbb{R}^d$.
	\end{enumerate}

For simplicity reasons, we consider the case of a scalar field only. A scalar field $\Psi$ is a function on spacetime. An active diffeomorphism defines a \emph{new} scalar field at the \emph{same} spacetime point
	\beq
	\tilde\Psi(q)=\Psi\left(\varphi(q)\right).
	\eeq 
Also, a scalar field $\Psi$ on spacetime determines a function $\psi$ on $\mathbb{R}^d$ defined by
	\beq
	\psi(x)=\Psi\left(\vartheta^{-1} (x) \right),
	\eeq
which through a passive diffeomorphism $\phi$ defines  \emph{another} scalar field in a {new} coordinate representation  given by
	\beq
	\psi'(x') = \psi\left(\phi(x')\right).
	\eeq
This can be summarised by the commutative diagram

	\begin{equation}
	\label{ms.diagram}
	\begin{diagram}
		\node[7]{\text{\tiny\it q}}  \node[12]{\text{\tiny\it p}}\\
		\node[7]{\Omega} \arrow[12]{e,t}{\varphi} \arrow[12]{s,l}{\vartheta} \arrow[5]{se,t}{\tilde{\Psi}}
			\node[12]{\Omega} \arrow[12]{s,r}{\vartheta} \arrow[5]{sw,t}{\Psi} \node[8]{\text{\small Active}}\\[6]
		\node{\text{\small\it Lagrangian}} \node[12]{\mathbb{R}} \\[6]
		\node[7]{\mathbb{R}^d} \arrow[12]{e,b}{\phi} \arrow[5]{ne,b}{\psi'} \node[12]{\mathbb{R}^d} \arrow[5]{nw,b}{\psi} \node[8]{\text{\small Passive}}\\
		\node[7]{\text{\tiny\it x'}}  \node[12]{\text{\tiny\it x}}\\[2]
		\node[13]{\text{\small\it Eulerian}}
	\end{diagram}
	\end{equation}

Let us now consider the one-parameter families of  active and passive diffeomorphisms locally generated by $\xi$ and $\zeta$ respectively. As in section \ref{coincidences}, let $\varphi_0={\rm id}\mathcal{M}$ and $\phi_0={\rm id}\mathbb{R}^d$. There are two ways we can compare the different numerical values of the fields $\tilde\Psi$ and $\Psi$. 

First, we can use the upper part of the diagram to visualise the setting we discussed in section \ref{coincidences}. There,  we defined the variation of the field $\Psi$ [see equation \eqref{gf.varia}] in terms of the one-parameter family $\varphi_s$.  Following the top cycle of \eqref{ms.diagram} we construct the difference of two \emph{distinct} fields, $\Psi$ and $\tilde\Psi$, at the \emph{same} spacetime point,
	\beq
	\label{ms.activelie}
	\Delta \Psi = \tilde\Psi(q) - \Psi(q) = \Psi(\varphi_s(q)) - \Psi(q).
	\eeq
We can divide this variation by the change in the parameter which, when evaluated at $s=0$,  is indeed  the Lie-derivative of the field $\Psi$ with respect to the vector field $\xi$ at the point $q$. Therefore, \eqref{ms.activelie}  reproduces the RHS of \eqref{ms.lag-eu}.

The other way of comparing the two fields is by taking the lower route of \eqref{ms.diagram}. Starting at some point $q\in\Omega$, we can look at the difference between the value of the field $\tilde\Psi(q)$ and its associated value in terms of a coordinate representation $\psi'(x')$ after being affected by some passive diffeomorphism $\phi_{t_0}$  
	\beq
	\label{ms.pass1}
	\tilde{\Psi}(q) - \psi(x) = \psi\left((\phi_t \circ \vartheta)(q)\right) - \psi(\vartheta(q)).
	\eeq
Then we can take another point $p\in\Omega$ and proceed in the opposite direction
	\beq
	\label{ms.pass2}
	\Psi(p) - \psi'(x') = \psi'((\phi_{-t}\circ \vartheta)(p)) - \psi'(\vartheta(p)).
	\eeq
The comparison of the two different paths gives the result
	\begin{align}
	\label{ms.pass3}
		\check\Delta \Psi  & = \left[\tilde \Psi(q) - \psi(x)\right] - \left[\Psi(p) - \psi'(x')\right]\nn\\
					&= \left[\tilde\Psi(q) - \Psi(p)\right] - \left[\psi'(x')-\psi(x)\right].
	\end{align}
Notice the similarity between the first term in \eqref{ms.pass3} and the intermediate expression in \eqref{ms.activelie}. Despite the suggestive look, they are not the same; they are evaluated at \emph{different} spacetime points. In the limit when the parameter $t$  tends to zero, this defines the Lagrangian variation of $\Psi$
	\beq
	\label{ms.lag01}
	\delta_\ell \Psi \overset{!}{=} \Psi\left((\vartheta^{-1}\circ \phi_t \circ \vartheta)(q)\right) - \Psi(p).
	\eeq
Now we can explicitly see how the ``coordinates are dragged'' by the \emph{Lagrangian displacement} $\zeta$. Looking at the first term of \eqref{ms.lag01}, we realise that each diffeomorphism in the family acts directly on the coordinate system $\vartheta$ associated with the point $q$.

The second term in \eqref{ms.pass3} is easier to interpret. It is simply the difference between two  representations of the field at the same point in $\mathbb{R}^d$, and, therefore, at the same spacetime point.  Thus, this term corresponds to the Eulerian variation
	\beq
	\label{ms.eul01}
	\delta_{\mathsf{e}}\Psi \overset{!}{=}  \psi'(x') - \psi(x)= \psi(\phi_t(x')) - \psi(x') = \Psi\left((\vartheta^{-1}\circ \phi_t)(x')\right) - \Psi(\vartheta^{-1}(x')).
	\eeq
Note that in this case the explicit choice of a coordinate system prevents \eqref{ms.eul01} from being equal to the general covariant expression \eqref{ms.activelie}.  

The final point to notice is that both expressions, \eqref{ms.lag01} and \eqref{ms.eul01}, compare values of the field $\Psi$ \emph{at} spacetime locations, independent of the coordinate choice, thus \eqref{ms.pass3} is a generally covariant expression. However, in order to evaluate \eqref{ms.lag01} in the limit when the parameter $t$ is zero, we need the condition $\vartheta^{-1} \circ \phi_0 \circ \vartheta = {\rm id}\mathcal{M}$ independently of the coordinate system we choose. This bounds the displacement $\zeta$ to the vector field $\xi$ generating the family $\varphi_s$. Therefore, in the infinitesimal limit, \eqref{ms.activelie} coincides with \eqref{ms.pass3} and, using the definitions for the Lagrangian and Eulerian variations, \eqref{ms.lag01} and \eqref{ms.eul01},  we obtain the expression for the Lie derivative \eqref{ms.lag-eu}.

\subsection{Eulerian variation of conserved currents}
\label{eulcurr}

From the previous discussion on the rheometric invariance of the coordinate functions of $\mathscr{B}$, an arbitrary variation of $X^A$ is of purely Eulerian type and  is simply given by 
	\beq
	\label{ms.pert0}
	\delta_\mathsf{e} X^A  = \left[\delta_\ell -\mathcal{L}_\xi\right]X^A= -\mathcal{L}_\xi X^A = -\mathcal{L}_{\xi_\perp}X^A = -X^A_{\ ;a}\xi^a_\perp.
	\eeq
Thus, in general, changes of rheometric invariant objects due to an arbitrary coordinate displacement $\xi$ are solely given in terms of displacements orthogonal to the flow. In this case, we can freely remove the $\perp$ sub-script from \eqref{ms.pert0}, knowing that the coordinate variations parallel to the flow will not contribute to the Eulerian perturbation.

 Let us consider the Lagrangian and Eulerian variations of the objects we have defined so far. A Lagrangian variation of the closed 3-form \eqref{ms.pb} is given by a displacement tangent to the flow of $U$
	\beq
	\label{ms.eun1}
	\delta_\ell \n = \mathcal{L}_{\lambda U}\n = \lambda \mathcal{L}_U \n,
	\eeq
where $\lambda$ is some parameter of the flow. From the Cartan identity for Lie-differentiation of $r$-forms [see equation \eqref{kin.cartan}] we have
	\beq
	\label{ms.eun1.5}
	\mathcal{L}_U \n = \d (U \rfloor \n) = \d (\xi^d \frac{\partial}{\partial x^d} \rfloor n_{abc} \d x^a \wedge \d x^b \wedge \d x^c).
	\eeq
The $a$-contraction on the RHS of \eqref{ms.eun1.5} is
	\beq
	\label{ms.eun2}
	[\delta_\ell \n]_a  = \d (u^a n_{abc} \d x^b \wedge \d x^c ) = (u^a n^d \epsilon_{abcd})_{;e}\d x^e \wedge \d x^b \wedge \d x^c= 0.
	\eeq
This follows from our ansatz \eqref{ms.nflat}, which implies the conserved current in the form \eqref{ms.flow} and a symmetry argument. In the same manner, the contractions $[\delta_\ell \n]_b$ and $[\delta_\ell \n]_c$ are identically zero. Thus, the Lagrangian perturbation of the pulled-back volume  form \eqref{ms.eun1} vanishes, as expected from the definition of Lagrangian variations and rheometric invariance.  

The Eulerian variation of $\n$ follows from the above discussion and the definition \eqref{ms.lag-eu}
	\beq
	\label{ms.pert}
	\delta_{\mathsf{e}} \n = -\mathcal{L}_{\xi} \n = - \d (\xi \rfloor \n).
	\eeq
Using the same contraction decomposition as in \eqref{ms.eun2}, the $a$ contraction is
	\beq
	[\delta_\mathsf{e}\n]_a = - (\xi^a n_{abc})_{;d}\d x^d \wedge\d x^b \wedge \d x^c = -(\xi^a_{\ ;d} n_{abc} + n_{abc;d}\xi^a)\d x^d \wedge \d x^b \wedge \d x^c.
	\eeq
It is not difficult to see that by adding up all the contractions, and shuffling the indices in a suitable order, we obtain the well known  coordinate expression \cite{langlois,lrrnils} 
	\beq
	\delta_\mathsf{e} n_{abc} = -\left(3\xi^d_{\ ; [ a}n_{|d|bc]} + n_{abc;d}\xi^d\right),  
	\eeq
where the index between the bars is excluded from the antisymmetrisation bracket. Thus, using the definition  \eqref{ms.flow} and contracting with the volume form $\epsilon_{abcd}$, whose Eulerian variation  is \cite{lrrnils}
	\beq
	\delta_\mathsf{e} \epsilon_{abcd} = \frac{1}{2}\epsilon_{abcd} g^{rs}\delta_\mathsf{e} g_{rs},
	\eeq
we obtain a key result for our forthcoming variational principle; the Eulerian variations of the conserved current
     \beq
     \label{ms.eulfin}
     \delta_\mathsf{e} n^a = - \mathcal{L}_{\xi}n^a - n^a\left(\xi^b_{\ ;b} + \frac{1}{2}g^{bc}\delta_\mathsf{e} g_{bc}\right),
     \eeq
cannot be varied independently of the metric. This is, however, not surprising. An arbitrary variation of the coordinates will not preserve the projection map in general.


\section{The two-fluids action principle}
\label{twofluidsa}

We start this section by considering an unconstrained variation of the action for a system of consisting of two fluids.  This  will exhibit the general formalism contained in  a multi-fluid variational principle. It will show that,  in order to obtain the correct dynamics, we will need to impose a set of constraints on the \emph{world-line} variation of the currents. It will also serve as further motivation to use the mathematically more elaborate, but physically deeper,  assumption of a well defined volume form for two independent matter spaces in the case of the study of conserved currents.

\subsection{The unconstrained variation}

  Let us define the relativistic invariant Lagrangian-type density $\Lambda$ as the integrand for  the matter sector of the Einstein-Hilbert  action \eqref{gf.matterL}. We refer to this density as the \emph{master function}. The symmetry group of general relativity requires the action to be generally covariant, therefore, assuming that the system is isotropic, the master function can only depend on the scalars we can form from the two fluxes and the metric. As customary, following the notation previously introduced, we use $n^a$ and $s^a$ to denote the particle number and entropy density currents respectively. Thus, we have the three scalars
	\begin{subequations}
	\label{vf1.dens}
	\begin{align}
		\label{dens1}
				n^2 =& -g_{ab}n^a n^b,\\
		\label{dens2}
				s^2 =& -g_{ab}s^a s^b,\\
		\label{dens3}
				 j^2 =& -g_{ab}n^a s^b,
	\end{align}
	\end{subequations}
and the matter action [see \eqref{gf.matterL}]
	\beq
	\label{var1.action}
	S_m(\Omega) = \int_{\Omega} \Lambda(n,s,j) \sqrt{-g}\d\Omega.
	\eeq

An unconstrained variation leads to
	\beq
	\label{var1}
	\delta \Lambda = \frac{\partial \Lambda}{\partial n}\delta n + \frac{\partial \Lambda}{\partial s} \delta s + \frac{\partial \Lambda}{\partial j} \delta j.
	\eeq
	
Using equations \eqref{vf1.dens}, we can change the passive density variations for dynamical variations of the world-lines generated by the fluxes and the metric. That is, 
	\begin{subequations}
	\begin{align}
	\delta n =&-\frac{1}{2n}[2g_{ab}n^a\delta n^b + n^a n^b \delta g_{ab}],\\
	\delta s =&-\frac{1}{2s}[2g_{ab}s^a\delta s^b + s^a s^b \delta g_{ab}],\\
	\delta j =& -\frac{1}{2j}[g_{ab}(n^a \delta s^b + s^b \delta n^a) + n^a s^b \delta g_{ab}],
	\end{align}
	\end{subequations}
together with the fact 
	\beq
	\frac{1}{2n}\frac{\partial \Lambda}{\partial n} = \frac{\partial \Lambda}{\partial n^2},\quad%
	\frac{1}{2s}\frac{\partial \Lambda}{\partial s} = \frac{\partial \Lambda}{\partial s^2},\quad%
	\frac{1}{2j}\frac{\partial \Lambda}{\partial j} = \frac{\partial \Lambda}{\partial j^2},
	\eeq	
we find that the variation \eqref{var1} becomes
	\begin{align}
	\label{var2}
	\delta \Lambda = \left[-2\frac{\partial \Lambda}{\partial n^2}n_a -\frac{\partial \Lambda}{\partial j^2}s_a \right]\delta n^a & + \left[ -2\frac{\partial \Lambda}{\partial s^2}s_a -\frac{\partial \Lambda}{\partial j^2}n_a \right]\delta s^a\nn\\
		    & + \left[-\frac{\partial \Lambda}{\partial n^2}n^an^b - \frac{\partial \Lambda}{\partial s^2}s^a s^b - \frac{\partial \Lambda}{\partial j^2}n^a s^b\right]\delta g_{ab}.
	\end{align}

From \eqref{var2}, we can read off the conjugate momenta associated with each of the fluxes\glo{$\mu_a$}{Conjugate momentum to the particle number density current}\glo{$\theta_a$}{Conjugate momentum to the entropy density current}
	\begin{subequations}
	\label{var1.momenta}
	\begin{align}
	\label{pin}
	\mu_a=\frac{\partial \Lambda}{\partial n^a} = & g_{ab}(\Bn n^b + \Ans s^b)\\
	\label{pis}	
	\theta_a=\frac{\partial \Lambda}{\partial s^a} = & g_{ab}(\Bs s^b + \Ans n^b)
	\end{align}
	\end{subequations}
where we have introduced the coefficients
	\beq
	\label{var.coefs}
	\mathcal{B}^\n\equiv -2 \frac{\partial \Lambda}{\partial n^2}, \quad \mathcal{B}^\s\equiv -2 \frac{\partial \Lambda}{\partial s^2}, \quad \mathcal{A}^{\n\s}\equiv-\frac{\partial \Lambda}{\partial j^2}.
	\eeq
	
The conjugate variables \eqref{pin} and \eqref{pis} demonstrate the fundamental role of the master function \cite{carternotes}. The distinct role of the fluxes and their conjugate momenta are mostly omitted in the fluids literature. A key advantage of the variational approach is that these quantities are immediately determined by the form of the master function. Furthermore, it is clear that the momenta are not aligned with the respective currents, owing to the fact that the master function depends on the relative flow \eqref{dens3}. This is a physically relevant effect. Fundamentally, there is no argument to rule out the dependence of the master function on the relative flow. In fact, this coupling is associated with the \emph{entrainment} effect that is known to be of central importance in multi-fluid systems \cite{lrrnils}.

In the case of a super-fluid neutron star core, the entrainment arises owing to the strong interaction and couples the neutron and proton fluxes. In the heat problem we want to address, we will show that the entrainment between matter and entropy is associated with the thermal relaxation of the medium.

It remains to identify the coefficient for the metric variation. From the discussion in chapter 2, it is related to the energy momentum tensor through \eqref{gf.emtensor}. From the symmetry of $\delta g_{ab}$ and the definitions \eqref{pin} and \eqref{pis}, we can re-write the variation \eqref{var2} in terms of the fluxes and their momenta to obtain
	\beq
	\label{var01}
	\delta \Lambda = \mu_c \delta n^c + \theta_c \delta s^c +\frac{1}{2}g^{bc}(\mu_c n^a + \theta_c s^a)\delta g_{ab}.
	\eeq
It is not difficult to see that the metric contraction with each  individual quantity inside the brackets in \eqref{var01}, $\mu_c n^a$ and $\theta_c s^a$, is symmetric only when the entrainment coefficient $\Ans$ is identically zero 
	\begin{subequations}
	\begin{align}
	\label{var.asym}
	\mu^{[a}n^{b]} &= \frac{1}{2}[\Ans s^a n^b -\Ans s^b n^a ],\\
	\theta^{[a}s^{b]} &= \frac{1}{2}[\Ans n^a s^b -\Ans n^b s^a ].
	\end{align}
	\end{subequations}
However, it is evident that these terms satisfy the Noether identity   
	\beq
	\mu^{[a}n^{b]} + \theta^{[a}s^{b]} = 0,
	\eeq
regardless of entrainment. This should indeed be the case if we expect to produce a stress-energy tensor from the variation of the master function with respect to the metric.

From \eqref{var01} and the variational definition \eqref{gf.emtensor}, we can read off the stress-energy tensor
	\beq
	\label{var.stensor}
	T^{ab} = \mu^a n^b + \theta^a s^b - \Lambda(n,s,{\bf g}) g^{ab}.
	\eeq 
The equations of motion are given by a \emph{variational identity} [see \eqref{gf.Tcons2}] as the local conservation law of the stress-energy tensor \eqref{var.stensor}.  If the variation of the fluxes, $n^a$ and $s^a$, were left un-constrained, such variational identity would require  the momenta \eqref{pin} and \eqref{pis} to vanish in all cases. To show this, we explicitly evaluate the local conservation law, and then note that this implies the orthogonality between the currents and the `forces' given by 
	\beq
	\label{var.div1}
	T^{\ b}_{a\ ;b} n^a= \mu_{a;b} n^b n^a + \theta_{a;b}s^b n^a + 2 n^b_{\ ;(b} \mu_{a)} n^a  + 2 s^b_{\ ;(b}\theta_{a)} n^a = 0,
	\eeq
with an analogous expression for $T_{a\ ;b}^{\ b}s^a$. Even in the simplest case, when both currents are aligned, the only possibility for these equations to hold is that both momenta are identically zero. In particular, this is an indication that the particle number density flux has only three independent components and, therefore, the only compatible dynamics with an unconstrained variation is trivial. Thus, the only hope to obtain non-trivial dynamics from this type of Lagrangian master function, where  equation \eqref{var.div1} follows as a variational identity, is to impose some kind of constraints on the variations of the currents. In particular, if we assume from the onset that at least one of the fluxes coming into the action principle is conserved, we can fully exploit the matter space formalism introduced in the previous section. Furthermore, we can use it to explore the rheometric symmetry of the system.

\subsection{The constrained variation}

 Let us assume that the particle number density and entropy currents are conserved. This corresponds to the existence of two different matter spaces  together with their respective volume forms. This imposes the constrained Eulerian variation of each current in the form \eqref{ms.eulfin}. Thus, assuming two different Lagrangian displacements, $\xi_\n$ and $\xi_\s$, we have
	\begin{subequations}
	\label{cv.n}
	\begin{align}
	\delta_{\mathsf{e}} n^c &= -\mathcal{L}_{\xi_\n}n^c - n^c \left(\xi^a_{\n;a} + \frac{1}{2} g^{ab}\delta_{\mathsf{e}} g_{ab} \right),\\
	\delta_{\mathsf{e}} s^c &= -\mathcal{L}_{\xi_\s}s^c - s^c \left(\xi^a_{\s;a} + \frac{1}{2} g^{ab}\delta_{\mathsf{e}} g_{ab} \right).
	\end{align}
	\end{subequations}

Some feeling of discomfort may arise from the use of equations \eqref{cv.n} directly into the variation of the master function. The current displacements in \eqref{var01} are of the general form \eqref{gf.varia}, whereas equations \eqref{cv.n} correspond to the Eulerian part of the general variation \eqref{ms.lag-eu}. However,  as far as the variation principle is concerned, there is no difference in choosing one kind of variation from the other.  The reason for this is given by Carter \cite{carter30} in the following argument. Consider the Lagrangian and Eulerian variations of the action  \eqref{var1.action} produced by a general displacement $\xi$. Then, using formula \eqref{ms.lag-eu}, their difference  is given by
	\beq
	\left(\delta_\ell - \delta_\mathsf{e}\right) S(\Omega) = \int_\Omega \mathcal{L}_\xi \left[\Lambda \sqrt{-g}\right]\d\Omega.
	\eeq
The integrand is straightforward to evaluate 
	\beq
	\mathcal{L}_\xi [\Lambda \sqrt{-g}] = \Lambda_{;a}\xi^a \sqrt{-g} + \xi^a_{\ ;a}\Lambda\sqrt{-g} = (\Lambda \xi^a)_{;a}\sqrt{-g},
	\eeq
where we have made use of the formula $\mathcal{L}_\xi \sqrt{-g} = \sqrt{-g}\xi^a_{\ ;a}$. Therefore, using Stokes theorem, we have
	\beq
	\left(\delta_\ell - \delta_\mathsf{e}\right) S(\Omega) = \int_{\partial \Omega} \Lambda \xi^a \d \Sigma_a,
	\eeq
where $\d\Sigma_a$ denotes the surface volume element of the boundary $\partial \Omega$. Thus, from our regular assumption, the displacement $\xi$ vanishes on the boundary of the region $\Omega$ and, therefore, the action principle does not distinguish between Eulerian and Lagrangian variations.

Before we continue with our variation analysis, let us make one further assumption; we consider the metric as a \emph{passive} field. By this we mean that, to linear order, a Lagrangian variation of the metric  is zero. Thus, the metric acts as a \emph{predetermined} background. This does not contravene general covariance in the sense that the equations of motion  will still follow as a Noether identity, but their symmetry  may  not be as broad as ${\rm Diff}\mathcal({M})$ but a sub-group of it.

Substituting  the Eulerian variations \eqref{cv.n} into equation \eqref{var01}, the  variation of the master function  becomes
	\begin{align}	
	\label{cv.var01}
	\delta_{\mathsf{e}} \Lambda = \left(-n^c_{\ ;b}\xi^b_\n + 2 \xi^{[c}_{\n;b}n^{b]} \right) \mu_c &+ \left(-s^c_{\ ;b}\xi^b_\s + 2 \xi^{[c}_{\s;b}s^{b]} \right)\theta_c \nn\\
	 & + \frac{1}{2}\left[\mu^a n^b + \theta^a s^b - \right(\mu_c n^c + \theta_c s^c \left)g^{ab} \right]\delta_\mathsf{e} g_{ab}.  
	\end{align}
It is straightforward to show that the first term in each bracket of \eqref{cv.var01}, contracted with the corresponding momentum,  can be written as
	\begin{subequations}
	\begin{align}
	-n^a_{\ ;b}\xi^b_\n \mu_a &= -\left(2\mu_{[a;b]}n^b + n^b_{\ ;b}\mu_a \right)\xi^a_\n - 2\xi^{[a}_{\n;b}n^{b]}\mu_a + \left( 2 \xi^{[a}_\n n^{b]} \mu_a\right)_{;b},\\
	-s^a_{\ ;b}\xi^b_\s \theta_a &= -\left(2\theta_{[a;b]}s^b + s^b_{\ ;b}\theta_a \right)\xi^a_\s - 2\xi^{[a}_{\s;b}s^{b]}\theta_a + \left( 2 \xi^{[a}_\s s^{b]} \theta_a\right)_{;b}.
	\end{align}
	\end{subequations}
Thus, the constrained variation \eqref{cv.var01} can be written in terms of the two different displacement vectors for the fluxes, $\xi_\n$ and $\xi_\s$, together with a general displacement for the Eulerian variation of the metric and a divergence term which does not contribute to the variational principle. 

Using equation \eqref{ms.lag-eu} and the assumption of a passive metric we have 
 	\beq
 	\delta_\mathsf{e} g_{ab} = - 2 \xi_{(a;b)},
 	\eeq 
and after the usual integration by parts, ignoring all the boundary terms, the Eulerian variation of the action \eqref{var1.action} is [compare with \eqref{gf.varsm}]
	\beq
	\delta_\mathsf{e} S(\Omega) = \int_\Omega \left[-f_a^\n \xi^a_\n - f_a^\s \xi^a_\s - T^{\ b}_{a\ ;b} \xi^a \right] \sqrt{-g}\d \Omega,
	\eeq
where we introduce the notation for the `forces'
	\begin{subequations}
	\label{cv.forces}
	\begin{align}
	\label{cfn}
	f^\n_a &= 2\mu_{[a;b]}n^b + n^b_{\ ;b}\mu_a,\\
	\label{cfs}
	f^\s_a &= 2\theta_{[a;b]}s^b + s^b_{\ ;b}\theta_a,
	\end{align}
	\end{subequations}
and the energy-momentum tensor is found  directly from \eqref{cv.var01} to be 
	\beq
	\label{cemtensor}
	T_a^{\ b} = \mu_a n^b + \theta_a s^b + \Psi \delta_a^{\ b},
	\eeq
where we define the generalised pressure, $\Psi$, as a function of the conjugate momenta by
	\beq
	\label{cpress}
	\Psi = \Lambda - \mu_a n^a - \theta_a s^a.
	\eeq

Using \eqref{cemtensor} and \eqref{cpress} only, it is an easy  task to show that the divergence of the  energy-momentum tensor can be written as a ``sum of forces''
	\beq
	\label{cv.div}
	T^{\ b}_{a\ ;b} = f^\n_a + f^\s_a.
	\eeq 
Furthermore, as stated in chapter 2, if the matter fields satisfy their own equations of motion, that is, if each of the forces \eqref{cv.forces} are identically zero, the local conservation law for the energy-momentum tensor follows as an identity [see \eqref{gf.Tcons2}]. In particular, this implies the \emph{force balance}
	\beq
	\label{cv.fbal}
	f^\n_a = - f^\s_a.
	\eeq 
Thus, the equations of motion are given by the system \eqref{cv.div} together with with \eqref{cv.fbal}. It is worth mentioning that the reasoning followed to obtain those equations also applies to the unconstrained variation \eqref{var01}. However, the only solution to those is trivial. That is because we needed to impose the orthogonality condition with respect to the particle number density flux. In the case of the constrained equations of motion \eqref{cv.div}, the orthogonality follows as an identity consequence of the symmetry of \eqref{cfn}
	\beq
	\label{cv.nfcont}
	f^\n_an^a = 2\mu_{[a;b]}n^b n^a = 0,
	\eeq
where we use the conservation law \eqref{ms.flow} to eliminate the second term of \eqref{cfn}. Therefore, by virtue of the force balance, equation \eqref{cv.fbal}, the particle flux is also orthogonal to $f^\s_a$ and hence 
	\beq
	\label{cv.nTcont}
	T_{a\ ;b}^{\ b}n^a = 0
	\eeq
holds in all cases. 

In the strict variational sense, we should have an analogous identity for the entropy flux $s^a$. This would imply that the particle number density current and the entropy flux are parallel, as can be seen by the symmetry of both equations \eqref{cv.forces} when contracted with each of the fluxes provided both currents satisfy a conservation law of the type \eqref{ms.flow}. Furthermore, it means that both currents are dual to two, in general distinct, spacetime volume 3-forms pulled back from their respective matter spaces. In this case only, both matter spaces coincide and the canonical volume form becomes degenerate. However, as far as the variational principle is concerned, the dynamics of the system are implied by the vanishing of \eqref{cv.forces} regardless of any production term. It is the identity \eqref{cv.nfcont}  that implies the non-trivial behaviour of the system, which follows from the conservation law of one of the currents only. This is the key point we analyse in detail in the coming sections. 

Let us make a final remark about the symmetry  group of the equations of motion implied by the identity \eqref{cv.nTcont}. The orthogonality between the particle number density current and the divergence of the stress-energy tensor shows that both forces, $f^\n$ and $f^\s$, are $U$-orthogonal tensors and, therefore, their evolution has a suitable description in the matter space corresponding to the particles. 

\section{Thermodynamic interpretation}
\label{ThI}

The action principle and the equations of motion we have derived are those of an arbitrary two-fluid system. So far the labels $\n$ and $\s$,  denoting particle number density current and entropy flux, have just been ad-hoc tags to distinguish between two generic fluid species. This section is devoted to show that, indeed, these tags can be given the thermodynamic meaning they express.

As noted in the last remark in the previous section, when the particle number density current satisfies the conservation law \eqref{ms.flow}, the forces \eqref{cv.forces} are $U$-orthogonal\footnote{Here $U$ corresponds to the normalised four-velocity associated with the particle number density flux. See discussion in Section 5.1.}.  We will assume from the outset that this is the only conserved current.  This suggests that it is natural to \emph{choose} observers moving with the \emph{matter flow} [see \eqref{ms.flow}]. This coincides with the choice made by Eckart, as we have seen in the previous chapter. In the following discussion we use the terms \emph{matter} and \emph{Eckart} frame to express the same observer choice. More complex settings, e.g. when dealing with more than one conserved current (excluding entropy) or when reactions are present, make the choice of frame less obvious. We will extend this discussion further at the end of the section.

From the previous discussion, it is clear that when the entropy flux is not conserved, it cannot be parallel to the matter flow. Thus, it can be decomposed into  \emph{co-moving}  and  $U$-orthogonal components. The entropy flux is then expressed as 
	\beq
	\label{thermal01}
	s^a = s^* (u^a + w^a), \quad s^*= -s^au_a,
	\eeq
where $s^*$ is the projection of the entropy current  into the matter frame  and $w^a$\glo{$w^a$}{Relative velocity between the matter and entropy frames} is the relative velocity between the two frames which, by definition, satisfies the orthogonality condition $w^a u_a=0$. 

Let us denote the normalised entropy four velocity by $u^a_\s$. In this frame the entropy flux is written simply as   $s^a=s u^a_\s$, where $s$ is given by \eqref{dens3}. From this and \eqref{thermal01}, we see that 
	\beq
	s^*= s\left(u^a_\s u_a \right) = s\gamma, \quad {\rm where} \quad \gamma = |u^a + w^a| = (1-w^2)^{-1/2},
	\eeq
is the redshift associated with the relative motion of the two frames.

Analogously, we can decompose the momenta \eqref{var1.momenta} into `co-moving' and `orthogonal' pieces. Substituting \eqref{thermal01} into the momenta \eqref{var1.momenta} one obtains
	\begin{subequations}
	\label{tp.momdecomp}
	\begin{align}
	\label{tp.mudecomp}
	\mu_a 		& = \left(\Bn n + \Ans s^* \right)u_a + \Ans s^* w_a = \mu^* u_a + \mu^\natural w_a,\\
	\label{tp.thetadecomp}
	\theta_a	& = \left(\Bs s^* + \Ans n  \right)u_a + \Bs s^* w_a = \theta^* u_a + \theta^\natural w_a.
	\end{align}
	\end{subequations}
Here we  denote  with a `star' the components of the momenta \eqref{pin} and \eqref{pis} aligned with  $U$ and  with a `natural' symbol the pieces  which are orthogonal.  Thus, we have the definitions\glo{$\mu^*$}{Chemical potential measured in the matter frame}\glo{$\theta^*$}{Thermal momentum measured in the matter frame (temperature)}  
	\begin{subequations}
	\label{ti.parmomenta}
	\begin{align}
	\label{mustar}
	\mu^*		& \equiv -u^a \mu_a = \Bn n + \Ans s^*  	\quad \text{and} \quad \mu^\natural \equiv \Ans s^*,\\
	\label{thetastar}
	\theta^*	& \equiv -u^a \theta_a = \Ans n + \Bs s^*	\quad \text{and} \quad \theta^\natural \equiv \Bs s^*.
	\end{align}
	\end{subequations}

It is the aim of this subsection to show that the `starred' quantities in \eqref{ti.parmomenta} are the ones susceptible to a thermodynamic interpretation. 

When dealing with a generic multi-fluid system, the components of each  momentum measured in its associated frame - the one defined by the current it is conjugate to - correspond to the \emph{chemical potential} of that species. In the case of the entropy flux, we use the tag \emph{thermal} for its conjugate momentum, equation \eqref{pis}. It is reasonable that the  projection of the thermal momentum \eqref{thetastar} into the matter frame should correspond to the temperature an observer moving with the flow would measure. In the following subsection, we argue that the quantity $\theta^*$ indeed satisfies the definition of temperature in the Gibbs sense; as the rate of change of the energy with respect to the entropy while keeping \emph{all} other thermodynamic variables fixed.

\subsection{The temperature problem}    

Thermodynamic properties such as temperature or pressure are uniquely defined  only in the case when the system under consideration is in equilibrium. Intuitively this makes sense since, in order to carry out a measurement, the measuring device must have time to reach equilibrium with the system. Such measurement is only meaningful as long as the time scale required to obtain a result is shorter than the evolution time of the system. This, however, does not prevent a generalization of the various thermodynamic concepts. The procedure may not be unique, but one should at least require the generalized concepts to be internally consistent within the chosen \emph{extended} thermodynamics model.

In our present discussion we are facing a larger problem. We must find an internally consistent set of thermodynamic variables to describe the state of the system when it is out of equilibrium. This approach should allow us to infer the \emph{thermodynamic} temperature  from an extended Gibbs relation in terms of this set of variables. Furthermore, since there are no well established transformation laws for the temperature (see \cite{landsnature}), we need to show that when the analysis is carried out in the matter frame, the thermodynamic temperature corresponds exactly to the \emph{dynamical} temperature \eqref{thetastar}.  Therefore, in the spirit of physical interpretation, unless stated otherwise, we shall  consider all physical measurements to be taken in - and with respect to - the matter frame defined by the vector field $U$. We attach a star symbol to distinguish any quantity defined in this manner.

Let us start by re-writing the variationally defined two-fluid stress-energy tensor \eqref{cemtensor} into its more common dissipative form  [see definitions \eqref{dec01} and \eqref{stress}].  Following the general decomposition \eqref{dis.dec}, the energy-momentum tensor \eqref{cemtensor} takes the form
	\beq
	\label{tp.decomp}
	T_{a}^{\ b} = \rho^* u_a u^b + q_a u^b + u_a q^b + P_{a}^{\ b}.
	\eeq	
Here, the energy density $\rho^*$, the heat flux $q^a$ and the purely spatial stresses $P^{ab}$ can be obtained by direct computation [see \eqref{eck01}-\eqref{eck03}]
	\begin{subequations}
	\begin{align}
	\label{cend01}
	\rho^*  &= \mu^* n + \theta^* s^* - \Psi,\\
	\label{tp.heat}
	q^a & = \theta^* \sigma^a,\\
	\label{tp.stress}
	P_a^{\ b} & = p_a \sigma^b + \Psi h_a^{\ b},
	\end{align}
	\end{subequations}
where $h_a^{\ b}$ is the projection into the space orthogonal to $U$ [see definition \eqref{dis.proj}] and where we have defined the quantities
	\begin{subequations}
	\label{ti.transverse}
	\begin{align}
	\sigma^a 	&= s^* w^a,  \\
	\label{tp.transp}
	p_a		&= \theta^\natural w_a,
	\end{align} 
	\end{subequations}
which are simply the \emph{spacelike} components of the entropy flux and the thermal momentum respectively.

It is, however,  physically more interesting to note that the energy density \eqref{cend01} could equally have been  obtained as a Legendre-type transform on the master function in terms of the quantities \eqref{ti.transverse}. To see this, we also note that the definition of the generalised pressure, equation \eqref{cpress}, is also a Legendre-type transform on the master function\glo{$\Psi$}{Generalised pressure} 
	\beq
	\label{tp.pres1}
	\Psi(\mu,\theta^*,\sigma) = \Lambda + \mu n + \theta^* s^* - p \sigma  
	\eeq
where $\sigma = s^*\left(w_a w^a\right)^{1/2}$ and $p = \Bs \sigma$. Thus, substituting \eqref{tp.pres1} into \eqref{cend01}, the energy density becomes 
	\beq
	\label{tp.rhostar}
	\rho^*(n, s^*, p) = -\Lambda + p_a \sigma^a.
	\eeq
This result is in complete analogy with \eqref{clt01}. In fact, this relation informs us that the definitions \eqref{ti.transverse} are indeed key variables in our thermodynamic model. Therefore, by virtue of \eqref{tp.rhostar} and \eqref{tp.pres1},  the variation of the energy density
	\beq
	\label{can.gibb}
	\d \rho^* = \mu^* \d n + \theta^* \d s^* + \sigma^a \d p_a,
	\eeq
shows that indeed, the dynamical temperature \eqref{thetastar} \emph{is} the thermodynamic temperature as measured in the matter frame $U$. 

Another point worth noting at this stage is that the variational analysis leads naturally to the presence of shear terms in the energy momentum tensor, \eqref{tp.stress}. Such terms are usually associated with viscous stresses and it is interesting to note that they arise even when we consider the \emph{pure} heat conduction problem. 

To close our discussion on the thermodynamic interpretation of the model, let us note that in the limit when the two currents are parallel, $\sigma^a = p_a = 0$, the energy density \eqref{tp.rhostar} becomes $\rho=-\Lambda$ and the starred notation becomes redundant. In this case, the quantities \eqref{mustar} and \eqref{thetastar} become the usual chemical potential and thermostatic temperature respectively 
        \begin{align}
        \mu & =\frac{\partial \rho}{\partial n} = -\frac{\partial \Lambda}{\partial n}=-2n\frac{\partial \Lambda}{\partial n^2}=n\Bn,\\
        \theta& =\frac{\partial \rho}{\partial s} = -\frac{\partial \Lambda}{\partial s}=-2s\frac{\partial \Lambda}{\partial s^2}=s\Bs.
        \end{align}
Thus, equation \eqref{cend01} becomes the familiar \emph{equilibrium} Euler relation
        \beq
        \label{can.gp}
        \rho + \Psi = \mu n + \theta s,
        \eeq
which gives a consistent and unambiguous thermodynamic interpretation of the two-fluid model, at least in the case when the relative flow vanishes. Furthermore, this is an indication that the non-equilibrium case is effectively characterised by the misalignment of the two fluxes. However, a correct understanding of the nature of equilibrium can only be achieved by means of the second law of thermodynamics; through the entropy production.

\section{Rheometric covariant thermodynamics}

Now we are about to present the central result of this chapter. Its simplicity  will be a direct consequence of the dynamical multi-fluid programme introduced in  section \ref{twofluidsa}, combined with the ability to draw thermodynamic conclusions from such approach. 

We have seen that - independently of any observer - the conservation of the particle number density current \eqref{ms.flow} implies that each of the forces \eqref{cv.forces} are $U$-orthogonal tensors. Also, the result from the previous section shows that the multi-fluid formalism has a natural thermodynamic interpretation when the physical measurements are taken in the matter frame $U$.

Let us make a pertinent remark about the role of general covariance in a generic theory of dissipation.  It is  clear that \eqref{tp.decomp} is a mere decomposition of the manifestly general covariant tensor \eqref{cemtensor} relative to a particular observer; the matter flow $U$ in our case\footnote{This remark trivialises a point often made by Garcia-Colin and collaborators  \cite{colin05} in their efforts to merge the principles of linear irreversible thermodynamics with those of relativistic fluids. They argue that 
	\begin{quote}
	...this [the inclusion of the heat flow in the stress-energy tensor] is against the tenets of the general theory of relativity, the stress energy tensor includes only all forms of mechanical energy...Heat cannot be incorporated into its structure...\cite{colin05}
	\end{quote}
From our discussion, it is clear that the presence of the transverse energy flow \eqref{tp.heat}  in a particular expression for the energy-momentum tensor is a consequence of the choice of  observer.   We could have equally chosen  another observer whose four-velocity is aligned  with a time-like eigenvector of the stress energy tensor at each spacetime point. In such case, the decomposition \eqref{tp.decomp} would be free from terms containing $q^a$. However, such an observer will see (in a general case) a net particle flux in its rest frame. Historically, this is the difference between the Eckart (matter) frame and Landau  descriptions.}.  For a manifestly general covariant theory of matter  - defined as in section \ref{coincidences} - the \emph{internal} variables, e.g. heat flow or shear stresses, play a passive role in the dynamics; they are implicit in the equations of motion. It is the specific choice of  observer that gives them a definite physical interpretation; their meaning is bound to such a choice.

It may seem that the above remark represents a backwards step in our pursuit a covariant theory of thermodynamics. Quite the opposite, the symmetry  of the action \eqref{var1.action} is the complete group of diffeomorphisms of spacetime. However, we have shown that in order to obtain non-trivial dynamics in the \emph{physical} case where one has a conservation law of the form \eqref{ms.flow}, the relevant symmetry is that of the rheometric group $\mathcal{G}_U$. Thus, although the dynamics is clearly independent of any possible choice of  observer [see \eqref{cv.forces}, \eqref{cv.div} and \eqref{cv.fbal}], by construction, the most suitable family of observers from a physical and geometric point of view are those moving with the matter flow. In this sense, \emph{local} deviations away from  thermodynamic equilibrium have a natural description in the matter space $\mathscr{B}(\Omega,U)$. 

The study of departures from equilibrium is objectively characterised by a non-vanishing entropy production. As discussed earlier in this chapter, we will use the explicit freedom present in the `force' equation \eqref{cfs} to explore the dynamical consequences of a positive entropy production term. In the rest of this section, we will divide our thermodynamic analysis in two parts. First we deal with the internal `rheometric' dynamics, taking place in the matter space $\mathscr{B}$. Then, in the second part, we shall discuss the implications of thermodynamic departures from equilibrium in the `objective' covariant dynamics of the system.

\subsection{Internal dynamics: the relativistic Cattaneo equation}

The aim of this subsection is to show that the internal thermodynamic evolution of a material system - represented by the particle number density current - leads in a natural way\footnote{By now it should be clear that, when dealing with non-equilibrium thermodynamics, we understand `natural' to mean the simplicity motivated quadratic ansatz for the entropy production. It cannot be overemphasised that such ansatz is the simplest - not the most general - way to ensure that the second law holds.} to a relativistic analogue of the Cattaneo equation [see \eqref{dis.cattaneo}]. 

Let us first note that by making use of the definition of the transverse energy flow \eqref{tp.heat}, we can re-express the entropy flux \eqref{thermal01} as\glo{$q^a$}{Heat flow}
	\beq
	\label{rct.entropy}
	s^a = s^* u^a + \frac{1}{\theta^*}q^a.
	\eeq

The conjugate momentum \eqref{pis} can also be decomposed in terms of the matter four velocity $U$ and the heat flow $q$. Although the calculation is straightforward, we wish to alert that such decomposition highlights a feature which is  of prime importance to guarantee the stability and causality of the theory; entrainment. As a reminder, entrainment is a generic effect in multi-fluids systems which tilts the momenta relative to their respective currents, i.e. in a generally coupled multi-fluids system, currents and momenta are not `aligned'. One way of characterising such effect is given by the scalar quantity 
	\beq
	a \equiv (\theta^\natural - \theta^*) s^* = -\Ans n s^*.
	\eeq
Thus, using once again the expression for the heat flow \eqref{tp.heat}, it is easy to see that the decomposition of the thermal momenta \eqref{tp.thetadecomp} can be written as 
	\beq
	\label{rct.thermal}
	\theta_a = \theta^* u_a + \left(\frac{1}{s^*}  + \frac{\theta^\natural - \theta^*}{s^* \theta^*} \right) q^a.
	\eeq
Finally, in order to preserve the clarity of the analysis, let us introduce the shorthand notation for the quantity inside the brackets in the above equation
	\beq
	\label{can.beta}
	\beta \equiv \frac{1}{s^*}  + \frac{\theta^\natural - \theta^*}{s^* \theta^*}  = \frac{1}{s^*} - \frac{\Ans n}{s^* \theta^*}.
	\eeq

It is worth noting that entrainment makes the decomposition \eqref{rct.thermal} non-trivial. If we ignore the entrainment coupling between the currents in the Lagrangian density by setting $\Ans = 0$, it follows from the definitions \eqref{thetastar}  that  $\theta^\natural = \theta^*$.  This particular case was considered by Carter \cite{carter1988} and, from the analysis made by Hiscock and Olson \cite{hiscolson}, we know that it leads to a model that exhibits stability and causality problems. We will shortly return to this point.

In terms of the decompositions for the entropy flux and thermal momenta, equations \eqref{rct.entropy} and \eqref{rct.thermal}, and by virtue of the orthogonality of the force \eqref{cfs} with the normalised vector field $U$ tangent to the particle world-lines, the entropy creation rate is found to satisfy
	\begin{align}
	\label{cheat04}
	s^b_{\ ;b} \theta^* & = 2\theta_{[a;b]} u^a s^b  = 2\theta_{[a;b]}u^a q^b \frac{1}{\theta^*}.
	\end{align}
Expanding the RHS of \eqref{cheat04} immediately leads to
	\beq
	\label{can.entprod}
	s^a_{\ ;a} = -\frac{1}{\theta^*}q^a \left[\theta^*_{;a}\frac{1}{\theta^*} + \dot u_a - q_{c;a}\frac{u^c \beta}{\theta^*} + \dot q_a \frac{\beta}{\theta^*} + \beta_{;c} \frac{u^c q_a}{\theta^*} \right].
	\eeq

As we have been doing in the previous chapter, the simplest assumption one can make to secure the positivity of \eqref{cheat04} is to introduce the \emph{constitutive} equation for the heat flow
	\beq
	\label{can.cat}
	q^a= - \kappa \theta^* h^{ab} \left[\theta^*_{;b}\frac{1}{\theta^*} + \dot u_b - q_{c;a}\frac{u^c \beta}{\theta^*} + \dot q_b \frac{\beta}{\theta^*} + \beta_{;c} \frac{u^c q_a}{\theta^*} \right],
	\eeq
which we can immediately re-arrange to obtain a Cattaneo-like relation
	\beq
	\label{can.cattaneo}
	2\tau q^{[a;b]}u_b + q^a = -\tilde{\kappa} h^{ab}\left[\theta^*_b + \theta^* \dot u_b \right].
	\eeq
Here we have introduced the \emph{effective} thermal conductivity
	\beq
	\tilde{\kappa} = \frac{\kappa}{1 + \kappa \dot \beta},
	\eeq
and the thermal relaxation time
	\beq
	\label{can.reltime}
	\tau = \frac{\kappa \beta}{1+\kappa \dot\beta}.
	\eeq
It is worth noting that, if $\beta$ varies on a time scale $\tau_\beta$ which is long compared with the relaxation time, then
	\beq
	\kappa \dot \beta \sim \frac{\kappa \beta}{\tau_\beta} \sim \frac{\tau}{\tau_\beta} \ll 1,
	\eeq
and in this case one would simply have
	\beq
	\tilde\kappa \sim \kappa \quad \text{and} \quad \tau \sim \kappa\beta.
	\eeq


Thus we see that the dynamics obtained from the constrained variation of the two-fluid Lagrangian as described by an observer moving together with the matter flow, not only has a natural thermodynamic interpretation, but it provides us with the key constitutive equation to obtain a hyperbolic theory of heat conduction. This was already the case of Carter's regular model, however, it is now clear  that the omission of entrainment would make the relaxation time, equation \eqref{can.reltime} together with \eqref{can.beta}, independent of the matter description as given by the master function or through an equation of state. This disadvantage was the main tool used by Hiscock and Olson \cite{hiscolson} to show that, despite being hyperbolic, linear perturbations of realistic matter models  (a non relativistic Boltzmann gas and a strongly degenerate Fermi gas) violate causality when described in terms of the regular model. Hence, the variational approach can be used to construct a hyperbolic theory of dissipation, but  entrainment is the key feature to guarantee causality. The linear stability and causality analysis of a general two-fluid system will be developed in the next chapter.

\subsubsection{Thermal equilibrium and the Tolman-Ehrenfest effect}

So far we have been interested on the internal dynamics of our two-fluid construction. The central idea has been to relax the conservation law \eqref{ms.flow} for the entropy current and analyse the variational dynamics governed by the Noether identity $T^{ab}_{\ \ ;b} =0 $, given by the pair of equations \eqref{cv.div} and \eqref{cv.fbal}. We have seen in Chapter 2 that thermal equilibrium is defined to be the state for which the entropy production vanishes. Therefore,  as we have previously argued, the misalignment of the two currents $n^a$ and $s^a$ corresponds to a generic non-equilibrium situation. 

In the present case, the only source of entropy is heat [{\it c.f.} equations \eqref{can.entprod} and \eqref{can.cat}]. Therefore, thermal equilibrium is equivalent to the vanishing of the heat flow. In this case, the entropy is carried along with the matter and we obtain the usual Euler relation \eqref{can.gp}.

By adding the force equations \eqref{cfn} and \eqref{cfs}, we find that
	\beq
	\left(n\mu + s\theta \right) \dot u_a + h^b_{\ a}\left(\mu_{;b}n + \theta_{;b}s \right) = 0.
	\eeq 
It follows directly from the Euler relation \eqref{can.gp} and the Gibbs relation for the pressure (analogous to \eqref{can.gibb} setting the last term to zero) that
	\beq
	h^b_{\ a} \left[P_{;b} + (P + \rho)\dot u_b\right] = 0
	\eeq
from which we immediately obtain
	\beq
	\label{te.tol01}
	\left(P + \rho\right)\dot u_a = -h^b_{\ a}P_{;b}.
	\eeq
As expected, we have the usual relation between the acceleration and the pressure gradient. Finally, the entropy momentum equation can be cast in the form
	\beq
	h^b_{\ a}\left[\theta_{;b} + \theta \dot u_b \right] = 0.
	\eeq
or, equivalently
	\beq
	\label{te.tol02}
	\theta \dot u_a = -h^b_{\ a}\theta_{;b}.
	\eeq
Comparing equations \eqref{te.tol01} and \eqref{te.tol02}, we see that 
	\beq
	\frac{1}{P + \rho} h^b_{\ a}P_{;b} =\frac{1}{\theta}h^b_{\ a}\theta_{;b}.
	\eeq
Thus, the equilibrium dynamics are characterised by a balance law between temperature and pressure gradients. It is worth noting that, even in the case of thermal equilibrium, the acceleration term plays a central role in the dynamics. 

Let us recall a result from section 3.4.1. In our previous discussion about thermal equilibrium [{\it c.f.} equations \eqref{equilibrium}] we conclude that in order to attain such states, a necessary condition is the existence of a timelike Killing vector field. In this case, the spacetime is said to be \emph{stationary}. Therefore, in coordinates adapted to such a Killing field ($g_{ab,0}=0$), the four-acceleration can be expressed in terms of the connection coefficients as
	\beq
	\label{tol.ac1}
	\dot u^a = -\Gamma^a_{00}, \quad \text{with} \quad \Gamma^a_{00} = -\frac{1}{2}g_{00,b}g^{ab}.
	\eeq 
 Recalling that the acceleration is orthogonal to the matter flow,  the entropy momentum equation \eqref{te.tol02} can be written as
	\beq
	\label{tol.temp}
	\theta_{,k}g^{ik} = -\frac{1}{2}g_{00,k}g^{ik} \theta
	\eeq
In the Newtonian limit, $g_{00}$ is associated with the gravitational potential $\Phi_g$ by the expression
	\beq
	\label{tol.g00}
	g_{00} \overset{!}{=} - \left(1 + \frac{2\Phi_g}{c^2} \right)
	\eeq
where we have restored the value of the speed of light to $c$. Substituting \eqref{tol.g00} back in \eqref{tol.temp} one immediately obtains the Tolman-Ehrenfest effect
	\beq
	\frac{\nabla \theta}{\theta} = \frac{\vec{g}}{c^2},
	\eeq 	
where $\vec{g}$ is the Newtonian gravitational acceleration.

\subsection{Further remarks on the internal thermal dynamics}    

Before moving forward in our analysis of the second law, it will be useful to consider a slightly different point of view. 

As a first step, we recall that the entropy force $f_a^\s$ is $U$-orthogonal. Thus, contracting $f^\n_a$ from equation \eqref{cfn} with $u^a$ and using the result in the force $f^\s_a$, equation \eqref{cfs}, we arrive at the rather elegant expression
	\beq
	\label{can.nilsf}
	\theta^* f^\s_a = - 2 u^c s^b \left(\theta_{[c}\theta_{a];b} + \theta_{b;[c}\theta_{a]} \right).
	\eeq
This expression is manifestly $U$-orthogonal. Moreover, it emphasizes the relevance of the entropy conjugate momentum $\theta_a$. However, if we want to gain insight into the key factors that contribute to the force, then we need to expand 
\eqref{can.nilsf}. To do this, we contract the entropy force \eqref{cfs} with the entropy flux \eqref{rct.entropy} to obtain
	\beq
	\label{can.nilsep}
	\left(s^a\theta_a\right)\Gamma_\s = s^a f_a^\s = \frac{1}{\theta^*} q^a f_a^\s.
	\eeq
This shows that the entropy production $\Gamma_\s = s^a_{\ ;a}$ only depends on the component of the entropy force that is parallel to the heat flux. In general, given that the entropy force has no components in the direction of the particle flow, we can make the decomposition
	\beq
	\label{can.nilsdec}
	f^\s_a = f^\parallel q_a + f^\perp_a.
	\eeq
From \eqref{can.nilsep}, it is clear that $f^\perp_a$ cannot contribute to the entropy production. Thus, there are two degrees of freedom in the entropy dynamics which are not constrained by the second law. This is an important point since there is no obvious way to distinguish the viability of models with different forms of $f^\perp_a$.

It is straightforward to compute the values of $f^\parallel$ and $f^\perp_a$. The former can simply be read out from \eqref{can.nilsep} as
	\beq
	\label{can.fqpar}
	f^\parallel  = -\frac{1}{\theta^*}\left[\frac{\beta q^2 - s^* {\theta^*}^2}{q} \right] q^b \left[\theta^*_{\ ;b}\frac{1}{\theta^*} + \dot u_b + \dot \beta \frac{q_b}{\theta^*} + 2q_{[b;c]}\frac{u^c \beta}{\theta^*} \right].
	\eeq
This component should vanish when there is no heat transport. Again, there are a number of ways in which we could impose this condition. The simplest corresponds to our previous ansatz \eqref{can.cat}, leading to our generalization of Cattaneo equation \eqref{can.cattaneo}. This is not a surprising result. However, it should be noticed that in this case we were never concerned about the second law at all! We arrived at this result solely by requiring that \eqref{can.fqpar} vanishes in the absence of heat. As we have seen before, such an ansatz guarantees the positivity of the entropy production $\Gamma_\s$ in \eqref{can.nilsep}.
 
To compute the perpendicular component $f^\perp_a$, let us introduce the projector orthogonal to $q^a$
	\beq
	\perp^a_{\ b} = \delta^a_{\ b} - \frac{1}{q^2} q^a q_b.
	\eeq
Then, using ansatz \eqref{can.cat}, it follows that
	\beq
	\label{can.nilsperp}
	f^\perp_a = -2 s^*\theta^*h^c_{\ a}\perp^b_{\ c} \left[\beta_{;b}q^2 + \frac{1}{2}q^2_{\ ;b}\beta - q_{b;d}q^d\beta \right].
	\eeq
We see that the variational approach leads to the presence of terms that, even though they involve the heat flux, are not associated with entropy production. As far as we are aware, the dynamical role of these terms has not been discussed in detail in the literature even though similar terms are (as we will soon see) also included in the Israel \& Stewart formalism. The variational model leads to these terms taking a specific form. In particular, those in equation \eqref{can.nilsperp} are all quadratic in $q^a$, the deviation from equilibrium. At this order, the most general case would allow a force of the form
	\beq
	f^\perp_a = h^c_{\ a}\perp^b_{\ c} \left[A_b q^2 + (q^2)_{;b} B + q_{b;d}q^d C\right],
	\eeq
with $A_a$, $B$ and $C$ unspecified coefficients. There may also, in principle, be first order terms. Clearly, equation \eqref{can.nilsperp} represents a particular case where all the coefficients follow from $\beta$. Hence, the variational model is a particular example of the general class of permissible theories. The fact that all these models satisfy the second law of thermodynamics means that we cannot express a preference at this point. A very interesting question concerns whether there are situations where $f^\perp_a$ has a distinguishable effect on the dynamics of the system.

\subsubsection{Comments on the four-acceleration}

The key result of this section is the relativistic extension of Cattaneo equation. We obtained this by examining the dynamics of the two-fluid model in the Eckart frame, where the system has a natural thermodynamic interpretation and the state of the system is well described by the matter space.

To facilitate a comparison with earlier results, let us recall the form of the relativistic version of Fourier's law as derived by Eckart and the constitutive equation for heat transport obtained by Cattaneo [{\it c.f.} equation \eqref{heat01} and \eqref{dis.cattaneo}]
	\begin{align}
	q^a &= - \kappa h^{ab}\left[\theta_{;b} + \theta \dot u_b \right],\nn\\
	 q^i & = - \kappa \frac{\partial}{\partial x^i} T -\tau \frac{\partial}{\partial t} q^i \qquad i=1,2,3.\nn
	\end{align}
The structure of the variational constitutive relation, equation \eqref{can.cattaneo} combines the features present in both of the above equations, including the acceleration term which in the variational context arises as a consequence of the equations of motion.

The role of the four acceleration can naturally be interpreted as being due to an effective `mass' per unit entropy given precisely by the temperature. As we will see later, this term has no counterpart in the Newtonian problem. Formally, this term originates from the local energy balance, equation \eqref{FL}. As we have discussed in the previous chapter [{\it c.f.} the discussion after equation \eqref{sc0}], it results from the fact that the infinitesimal 3-spaces orthogonal to the matter world lines are not parallel, but relatively tipped over because of the curvature of the world line. This leads to the interpretation of the four-acceleration contribution in terms of the effective \emph{inertia of heat}.

\subsection{Equivalence with Israel \& Stewart}

Since the two approaches are based on different strategies, any comparison between the variational model and the Israel \& Stewart results must be done carefully. Notably, the Israel \& Stewart model is based on an expansion including terms up to second order in the deviation from equilibrium in the entropy flux.

Meanwhile, the variational analysis did not involve such an expansion. As a result, the final equation for the heat flux, equation \eqref{can.cattaneo}, contains higher order terms. The corresponding  Israel \& Stewart version 	[{\it c.f.} equation \eqref{iscat}] reduces to
	\beq
	\label{iscatq}
	q^a = \check\kappa T h^{ab}\left[T_{;b} + T \dot u_b + T \beta_1 \dot q_b + \left( \frac{\beta_1}{2T}u^c \right)_{;c}T^2 q_b \right]
	\eeq
when we neglect the viscous terms $\tau$ and $\tau^{ab}$. We observe that equation is manifestly linear in $q^a$.

In order to compare the results, we focus on the linear deviation from equilibrium of the temperature $\theta^*$. Thus, we have
	\beq
	\theta^* = T + \mathcal{O}(q^2),
	\eeq
and it follows that the generalised Cattaneo equation \eqref{can.cattaneo} can be approximated by
	\beq
	\tau \left(\dot q^a + u^{c;a}q_c \right) + q^a \simeq -\tilde\kappa h^{ab}\left(T_{;b} + T\dot u_b \right).
	\eeq
	
Here it is worth noting that in the second term in the first bracket, we could use the standard decomposition of $u_{a;b}$ in terms of the expansion, shear, vorticity and acceleration. This would lead to terms that were explicitly excluded from the Israel \& Stewart model at the point where we focussed on the case with $\tau^{ab} =\tau =0$. Basically, the variational analysis leads to the presence of terms that couple the heat flux to the shear and expansion of the flow. As these were artificially excluded from the analysis that led to equation \eqref{iscatq}, we cannot count this as a difference between the two models. In fact, the full comparison carried out by Priou show that these terms agree in the two descriptions \cite{priou1}.

Keeping terms up to second order (treating the shear and the divergence of $u^a$ as first-order quantities), equation \eqref{iscatq} can be written as
	\beq
	\tau_I\dot q^a + q^a \simeq -\check\kappa_Ih^{ab}\left(T_{;b} + T\dot u_b \right),
	\eeq
where the Israel \& Stewart thermal relaxation time  is given by
	\beq
	\tau_I = \beta_1\check\kappa T \left[1 + \left(\frac{\beta_1}{T} \right)_{;c}u^c\frac{\check\kappa T^2}{2} \right]^{-1},
	\eeq
and the thermal conductivity is
	\beq
	\check\kappa_I = \check\kappa \left[1 + \left(\frac{\beta_1}{T} \right)_{;c}u^c\frac{\check\kappa T^2}{2} \right]^{-1}.
	\eeq

Now it is clear that the two equations for the heat flux are formally identical, and we can ‘identify’ the parameters in the two models. The upshot of this is that the models will only produce different results at higher order deviations from equilibrium. Given that this regime is hardly tested at all, we cannot at this stage comment on which of the two descriptions (if either) may be the most appropriate. Having said that, it is clear that the variational approach is formally elegant and the fact that it also applies  far from equilibrium (at least in principle) may be relevant. An interesting question concerns whether there are situations where the, rather specific, set of higher order terms predicted by the variational analysis affect the non-linear dynamics.

\subsection{Covariance and entropy production: an electrodynamic analogue}

To close this section, let us discuss an analogy with electrodynamics. Consider the  motion of a congruence of test charged particles inside a region $\Omega$ of a fixed spacetime background under the influence of an electromagnetic field whose Faraday tensor is $F^{ab}$. The equations of motion for such a particle are given by
	\beq
	\label{canf.fin01}
	u^a_{\ ;b}u^b = q F^a_{\ b} u^b,
	\eeq
where $u^a$ represents the normalised timelike vector field tangent to the particles' worldlines and $q$ represents its charge density.

In  classical electrodynamics, a source of the electromagnetic field acts effectively as a singularity, {\it i.e.} as missing point in the interior of  $\Omega$. Thus, for a source-free electromagnetic field,  $\Omega$ must be a simply connected region and the homogeneous Maxwell's equations,
	\beq
	\d F = 0,
	\eeq
imply the existence of a a gauge dependent potential one-form $A$ such that
	\beq
	F = \d A.
	\eeq
In this case, we can write the equations of motion \eqref{canf.fin01} as the `force' equation
	\beq
	f^{\rm em}_a = 2A_{[a;b]} j^b,
	\eeq
where $f^{\rm em}_a$ stands for the electromagnetic force and  we have defined the charge density current by
	\beq
	j^a = q u^a.
	\eeq	
Thus, the electromagnetic force has the exact same form as that of the force equation of a multifluid system whose individual components are conserved [{\it c.f.} equations \eqref{cv.forces} with vanishing production terms]. In particular, for our two-fluid case, each of the conjugate momenta $\mu_a$ and $\theta_a$, play the role of the potential one-form for electromagnetism. Moreover, has we have argued at the end of section 5.2.2, when more than one fluid species is present the dynamics allows for one of the species to have non-vanishing production, as in the case of \eqref{cfs}
	\[
	f^\s_a = 2\theta_{[a;b]}s^b + s^{b}_{\ ;b} \theta_a.
	\]
This is the fact that we use to obtain an expression for the entropy production. There is a striking feature in this: What would be the analogous electromagnetic expression of the entropy force \eqref{cfs}? We can write down the answer straight away
	\beq
	f^{\rm em}_a = 2 A_{[a;b]}j^b + j^b_{\ ;b} A_a.
	\eeq
Note that, in this case, the gauge potential would have an observable effect in the motion of the test particles! Fortunately, we can invoke the variational principle from which \eqref{canf.fin01} was derived and verify that charge conservation in $\Omega$ implies that $j^a_{\ ;a} = 0$, removing every gauge dependence from the dynamics.

In electrodynamics, gauge invariance means that the potentials have no observable effects; only a potential difference is meaningful.  In the same manner, in special relativity Lorentz invariance tells us that velocities on their own are meaningless, the physically relevant information is contained in  relative velocities. Furthermore, in general relativity, general covariance removes the physical significance of individual spacetime points, only coincidences are important. Thus, let us close this chapter with the following question:  What is happening here, that allows the thermal momenta to have an effect on the dynamics away from equilibrium?\footnote{Perhaps a principle of restricted covariance  is forced upon us by the dynamics of thermal systems away from equilibrium.  The answer to this question may give a physical meaning to the covariance of our dynamical laws. For a discussion on the physical significance of general covariance, the reader is referred to \cite{matravers,tavakol}.}



			\chapter[Adiabatic perturbations]{Adiabatic perturbations of equilibrium}

In the preceding chapter we developed a theory of heat conduction from a generic two-fluid variational principle. It would be natural then to expect that some generic features of a two-fluid system should appear in the context of our approach to relativistic dissipation. In particular, in this chapter we study the onset and  development of instabilities arising from the  interaction between the fluids, either through their chemical coupling or  entrainment. 
 
We address the issues of stability and local causality of adiabatic perturbations, that is, we look for a criterion to be imposed on a given matter model (equation of state) in order to maintain the sound speeds bounded by  that of light.  Thus, this chapter presents an analysis of the local propagation of plane waves on a given background space-time. 

Compared to the standard single-fluid analysis, a multi-fluid system has more degrees of freedom; we need to account for relative flows between the different components. This is an essential requirement for the two-stream instability \cite{AnComPrix}. This is a generic phenomenon that does not require particular fine-tuning to be triggered, nor is limited to any specific physical system. The only requirements are that there must be a relative (background) flow and a coupling between the
fluids. Such instabilities are known to exist for a variety of configurations. For example, in shearing motion at
an interface, we have the well known Kelvin--Helmholtz instability.  This chapter presents a covariant generalization of the two-stream instability analysis to relativistic situations.

The stability and causality analysis of relativistic theories of dissipation as presented by Hiscock and Lindblom \cite{hiscock01} was tailored to suit the variables of the Israel \& Stewart expansion. One of their main results, as we have previously discussed ({\it c.f} chapter 4), was that of the generic instabilities of `first-order' theories. To close this chapter, we provide a different view regarding the stability of such theories. Building on the two-fluid construction of the previous chapter, we carry out a `consistent' first-order  analysis and show that one can have stable first-order models. However, when the relaxation time is varied over the dissipation time-scale, the system under consideration becomes unstable.

\section{Linear perturbations of thermal equilibrium}

In this section we deal with linear adiabatic perturbations of equilibrium states, {\it i.e.} those occurring without heat transfer. 

From the discussion in the previous chapter, the thermal dynamics of a \emph{single} heat-conducting fluid can be described in terms of the interacting components of an `enlarged' two-fluid system consisting of particles and entropy. This view was motivated by the thermodynamic limit of non-equilibrium relativistic kinetic theory, where we obtained a well defined entropy flux whose main effect is to redistribute the internal energy of the system towards the state of thermal equilibrium\footnote{See the definition at the beginning of section 2.2.}. In this manner, the internal dynamics of a dissipative fluid are given by a force-balance relation between a `material' and a `thermal' fluid\footnote{See the discussion in section 5.2.2. In particular, the variational definition of the forces \eqref{cv.forces}, and the force balance expressed by \eqref{cv.div} and \eqref{cv.fbal}.}. Thus, the dynamics of thermal equilibrium is given by
     \begin{subequations}
     \begin{align}
     \label{mult.divn}
     n^a_{\ ;a}     & = 0,\\
     \label{mult.divs}
     s^a_{\ ;a}     & = 0, \\
     T^{\ b}_{a\ ;b}  & = f^\n_a + f^\s_a = 0.
     \end{align}
     \end{subequations}
Here $T^{ab}$ is the canonically defined two-fluid energy momentum tensor [see equation \eqref{cemtensor}], $n^a$ and $s^a$ are the conserved particle number density and entropy density currents and, using the conservation laws \eqref{mult.divn} and \eqref{mult.divs}, the forces $f^\n_a$ and $f^\s_a$ are given by [see equations \eqref{cv.forces}]
    \begin{subequations}
    \label{ts.forceseq}
     \begin{align}
     \label{ts.01}
     f_a^\n &= 2 \mu_{[a;b]}n^b,\\
     \label{ts.02}
     f_a^\s &= 2 \theta_{[a;b]}s^b.
     \end{align}
     \end{subequations}

To obtain the linearised dynamics of the system, we consider only those perturbations which have no effect on the background geometry, that is, those for which $\delta g_{ab} = 0$. Thus, considering the set of linear perturbations
     \begin{subequations}
     \begin{align}
     n^a       \longrightarrow n^a + \delta n^a,        & \qquad    f^\n_a    \longrightarrow f^\n_a + \delta f^\n_a, \\
     s^a       \longrightarrow s^a + \delta s^a,        & \qquad    f^\s_a    \longrightarrow f^\s_a + \delta f^\s_a,
     \end{align}
     \end{subequations}
the linearised equations of motion are
	\begin{subequations}
     \begin{align}
     \label{ts.l1}
     \delta n^a_{\ ;a} = 0,    & \qquad   \delta f^\n_a = n^b\delta \mu_{[a;b]}, \\
     \label{ts.l2}
     \delta s^a_{\ ;a} = 0,    & \qquad   \delta f^\s_a = s^b\delta \theta_{[a;b]},
     \end{align}
     \end{subequations}
which must satisfy the energy conservation constraint expressed by     
     \beq
     \label{ts.fb}
     \delta f^\s_a + \delta f^\n_a = 0.
     \eeq

Let us make a remark about the dynamics of thermal equilibrium. As discussed at the end of section 5.2.2, in thermal equilibrium the two matter-spaces used to define each of the currents coincide. That is, for a compact region of spacetime $\Omega$, the one-parameter families of diffeomorphisms generated by each of the particle species' worldlines, are the same. Therefore, there is a unique normalised timelike vector field whose integral curves, up to reparametrization, represents the two different flows\footnote{See the discussion in section 5.1.}. In this case, where the dynamics are completely constrained, the system becomes completely integrable, {\it i.e.}
    \beq
     \label{ts.vort}
     \mu_{[a;b]} = \theta_{[a;b]} = 0.
     \eeq
Thus, each of the forces \eqref{ts.forceseq} vanishes independently. Moreover, if the perturbations are \emph{adiabatic}, we need to require, in addition of \eqref{ts.fb}, that the perturbed components of the force balance \eqref{ts.fb} individually satisfy  
     \begin{subequations}
     \begin{align}
     \label{mult.pertforce}
     n^b\delta \mu_{[a;b]} & = 0, \\
     s^b\delta \theta_{[a;b]} & = 0.
     \end{align} 
     \end{subequations}
If this where not the case, the non-zero value of the perturbed force acting on each displacement of the individual  currents would generate a relative flow between them, thus producing heat. 
     
In order to express the linearised equations of motion, let us recall the definition of the conjugate momenta $\mu_a$ and $\theta_a$ given in the previous chapter [equations \eqref{pin} and \eqref{pis}]
	\begin{align}
	\mu_a= & g_{ab}(\Bn n^b + \Ans s^b),\nn\\
	\theta_a = & g_{ab}(\Bs s^b + \Ans n^b),\nn
	\end{align}
where the coefficients $\Bn$, $\Bs$ and $\Ans$ are given by the partial derivatives of the master function $\Lambda(n,s,j)$ [equation \eqref{var.coefs}]
	\[
	\mathcal{B}^\n\equiv -2 \frac{\partial \Lambda}{\partial n^2}, \quad \mathcal{B}^\s\equiv -2 \frac{\partial \Lambda}{\partial s^2} \quad \text{and} \quad \mathcal{A}^{\n\s}\equiv-\frac{\partial \Lambda}{\partial j^2}.
	\]
Therefore, the perturbations of the  momenta which leave the background metric unchanged  take the form
     \begin{subequations}
     \label{mult.pertmom}
     \begin{align}
     \label{ts.p1}
     \delta \mu_a &= g_{ab} (n^b \delta \Bn + \Bn \delta n^b + s^b \delta \Ans + \Ans \delta s^b),\\
     \label{ts.p2}
     \delta \theta_a &= g_{ab} (s^b \delta \Bs + \Bs \delta s^b + n^b \delta \Ans + \Ans \delta n^b).
     \end{align}
     \end{subequations}
where the variations of the coefficients \eqref{var.coefs} are given by
     \beq
     \label{ts.perts}
     \delta \Bn  =  -2 \frac{\partial}{\partial n^2}\delta \Lambda,\quad
     \delta \Bs     =  -2 \frac{\partial}{\partial s^2}\delta \Lambda \quad \text{and} \quad
     \delta \Ans    = -\frac{\partial}{\partial j^2}\delta \Lambda.
     \eeq
Following the work by Andersson and Comer \cite{lrrnils}, we introduce the `abbreviations'  for the quantities 
     \begin{subequations}
     \begin{align}
     \label{ts.bn}
     \Bn_{ab}  &= \Bn g_{ab} - 2 \frac{\partial \Bn}{\partial n^2}g_{ac}g_{bd}n^c n^d,\\
     \Bs_{ab}  &= \Bs g_{ab} - 2 \frac{\partial \Bs}{\partial s^2}g_{ac}g_{bd}s^c s^d,\\
     \Ann_{ab} &= -g_{ac}g_{bd}\left(\frac{\partial\Bn}{\partial j^2}[n^c s^d + n^d s^c] + \frac{\partial\Ans}{\partial j^2}s^c s^d \right),\\
     \Ass_{ab} &= -g_{ac}g_{bd}\left(\frac{\partial\Bs}{\partial j^2}[s^c n^d + s^d n^c] + \frac{\partial\Ans}{\partial j^2}n^c n^d \right),\\
     \Ans_{ab} &= \Ans g_{ab} - g_{ac}g_{bd}\left(\frac{\partial \Bn}{\partial j^2}n^c n^d +\frac{\partial\Bs}{\partial j^2}s^c s^d + \frac{\partial \Ans}{\partial j^2}s^c n^d\right),\\
     \label{ts.xn}
     \Xns_{ab} &= -2 \frac{\partial\Bn}{\partial s^2} g_{ac}g_{bd}n^c s^d,\\
     \Xsn_{ab} &= -2 \frac{\partial\Bs}{\partial n^2} g_{ac}g_{bd}s^c n^d,
     \end{align}
     \end{subequations}
in order to separate the perturbed momenta in terms of the variations of the currents. Thus, we have
     \begin{subequations}
     \label{mult.pertmom02}
     \begin{align}
     \label{ts.p3}
     \delta \mu_a  &= (\Bn_{ab} + \Ann_{ab}) \delta n^b + (\Xns_{ab} + \Ans_{ab})\delta s^b,\\
     \label{ts.p4}
     \delta \theta_a  &= (\Bs_{ab} + \Ass_{ab}) \delta s^b + (\Xsn_{ab} + \Asn_{ab})\delta n^b.
     \end{align}
     \end{subequations}     

This is a very significant result. In the previous chapter, we defined entrainment as the misalignment between  one of the fluxes and its conjugate momentum due to the presence of the other fluid. Such an effect is a consequence  of the dependence of the master function $\Lambda$ on the relative flux between the currents, characterised by the coefficient $\Ans$ above. Here, equations \eqref{mult.pertmom02} tell us that, even in the case that the master function is independent of the relative flux, the perturbed momenta have non-vanishing components in the direction of both perturbed currents.

To understand better this more complicated setting, we step back for a moment to analyse wave propagation in a  perfect  single fluid.


\section{Sound waves in a single fluid}
\label{TS.sound}

To set  the stage for our forthcoming  analysis, we consider first the propagation of sound waves on a single perfect fluid given by the conserved particle current $n^a$. In this section, we perform a local analysis of linear perturbations of the fluid propagating on a generic background geometry given by the spacetime metric $g_{ab}$.  

Let us make the following plane-wave ansatz for the particle number density current 
     \beq
     \label{ts.sw}
     \delta n^a = A^a \exp(-i k_b x^b).
     \eeq
Here we assume that both, the amplitude $A^a$\glo{$A^a$}{Amplitude for a plane wave} and the wave four-vector $k^a$\glo{$k^a$}{Wave four vector},  have vanishing covariant derivatives
     \beq
     A^a_{\ ;b} = 0 \quad \text{and} \quad k_{a;b} = 0.
     \eeq

For a single perfect fluid, the perturbed momentum \eqref{ts.p3} reduces to
     \beq
     \delta \mu_{a} = \Bn_{ab} \delta n^b.
     \eeq
Following the convention set in the previous chapter, we work in the Eckart frame. Thus,  we can decompose the amplitude and the wave vector into their co-moving and transverse components
	\begin{subequations}
     \begin{align}
     \label{ts.wavev}
     k^a_\n & = k_\n(\sigma_\n u^a + \hat k_\n^a),\\
     A^a_\n & = A^\n_{\parallel} u^a + A^a_{\n\perp}.
     \end{align}
     \end{subequations}
Here, we have made the usual definitions
     \beq
     \label{ts.wavevv}
     k_\n\sigma_\n  = - k_a u^a \quad \text{and} \quad     k_\n \hat k^\n_a = h^b_{\ a}k_b .
     \eeq
where $\hat k^\n_a$ is a unit three-vector pointing in the spatial direction of the wave propagation. Similarly, the components of the  amplitude are expressed by
     \beq
     A^\n_{\parallel} = -A^a_\n u_a \quad \text{and}    A^a_{\n\perp}    = h^a_{\ b} A^b_\n,
     \eeq
and $\sigma_\n$ measures the phase velocity of the waves as seen in the Eckart frame.

The perturbed conservation law, equation \eqref{ts.l1}, becomes
     \begin{align}
     \delta n^a_{\ ;a} &= [A^a\exp(-\i k_bx^b)]_{;a}\nn\\
                       &= [A^a_{\ ;a} + i A^a(-k_{b;a}x^b -\delta^b_{\ a}k_b)]\exp(-i k_bx^b)\nn\\
                       &= i A^ak_a \exp(-\i k_dx^d) = 0
     \end{align}
which implies that the waves are transverse in the space-time sense
     \beq
     \label{ts.ort}
     k_a A^a = 0.
     \eeq

It is a straightforward exercise to show that the contraction $A^c\Bn_{c[a;b]}$ is identically zero. Therefore,   \eqref{mult.pertforce} becomes
     \begin{align}
     \label{ts.s1}
     \delta \mu_{[a;b]}n^b &= n^b[(\Bn_{ac}\delta n^c)_{;b} - (\Bn_{bc}\delta n^c)_{;a}]\nn\\
                              &= i n^bA^ck_{[b}\Bn_{a]c} \exp(-i k_dx^d) = 0,
     \end{align}
and thus we express the linearised force by
     \beq
     \label{ts.p5}
     n^bk_{[b}\Bn_{a]c}A^c =0.
     \eeq
     
 The dispersion relation can be easily obtained by contracting equation \eqref{ts.p5} with $k^b$. Assuming that $n_cA^c_\n \neq 0$, we may define the speed of sound by\glo{$c_\n$}{Speed of sound of matter waves}
     \beq
     \label{ts.sp}
     c_\n^2 = 1 + \frac{\partial\log \Bn}{\partial \log n}.
     \eeq
Thus, we can rewrite the coefficient $\Bn_{ab}$ in terms of the sound speed $c_\n$ as
     \begin{align}
     \Bn_{ab}  & = \Bn g_{ab} - 2 \frac{\partial \Bn}{\partial n^2}g_{ac}g_{bd}n^c n^d\nn\\
               & = \Bn \left[ g_{ab} + u_a u_b - u_a u_b -\frac{2n^2}{\Bn}\frac{\partial \Bn}{\partial n^2}u_a u_b\right]\nn\\
               & = \Bn \left[h_{ab} - u_a u_b \left(1 + \frac{\partial \log \Bn}{\partial \log n} \right) \right]\nn\\
               & = \Bn (h_{ab} - c_\n^2 u_a u_b).
     \end{align}
Substituting this result in the first term of the  anti-symmetrised bracket of the reduced perturbation of the force, equation \eqref{ts.p5}, we obtain
     \beq
     k^a n^b k_a \Bn_{bc} A^c_\n = \left(k_ak^a\right)n^b\Bn\left(h_{bc}-c_\n^2 u_b u_c\right)A^c_\n,                                 
     \eeq
whilst the second term is
     \beq
     k^a n^b k_b\Bn_{ac}A^c = \left(n^bk_b\right)k^a\left[\Bn\left(h_{ac}-c^2_\n u_a u_c\right)\right]A^c.                          
     \eeq
Therefore,  the single fluid dispersion relation takes the simple form
     \beq
     k^a n^b k_{[a}\Bn_{b]c}A^c_\n = 0.
     \eeq
Note that this is equivalent to\glo{$\sigma_\n$}{Phase velocity of matter waves}
	\beq
	\sigma_\n^2 - c_\n^2 = 0,
	\eeq
where, from the first equation in \eqref{ts.wavevv}, we have 
	\beq
	\sigma_\n = -\frac{1}{k_\n}k^a_\n u_a.
	\eeq

Now we can see that $\sigma_n$ corresponds to the phase velocity of the waves as measured in the Eckart frame. Now, we just need to show that $c_\n$ is effectively the speed of sound. To see that this is indeed the case, it will suffice to introduce the pressure and recall that, for the single fluid case, the energy density is  related to the master function simply by
	\beq
	\rho=-\Lambda.
	\eeq
In this case, the energy density is a function of the particle number density alone. Thus, the Gibbs relation is just 
     \beq
     \d \rho    = \mu \d n. 
     \eeq
The  pressure $\Psi$ is defined by the standard Euler relation $\rho + \Psi = \mu n$, and in this very simple case we can write it as
     \beq
     \Psi = -\rho + n \frac{\d \rho}{\d n} = -\rho + \Bn n^2.
     \eeq
Therefore,
     \begin{align}
     \label{ts.sp2}
     \frac{\d \Psi}{\d \rho}   &= -1 + n^2 \frac{\d \Bn}{\d \rho} + 2n\Bn \frac{\d n}{\d \rho}\nn\\
                               &=  1 + n^2 \frac{\Bn}{\d \rho}\frac{\Bn}{\Bn}\frac{\d n}{\d n}\nn\\
                               &= 1 + \frac{\d \log \Bn}{\d \log n},
     \end{align}
and we can see that, indeed, $c_\n$  is the speed of sound. 

Let us recapitulate what we have achieved with this exercise. We studied the perturbed equations of motion of a single perfect fluid through a plane wave ansatz. We found the dispersion relation and  defined the speed of sound in terms of variations of the master function for the matter model of our choice. This allows us to rule out non-physical models, for instance, if the perturbations end up propagating faster than the speed of light. In this sense, the relativistic analysis constrains our choices for master functions, and hence, of equations of state. Thus, equation \eqref{ts.sp2} serves to analyse when the matter model we use will give non-physical results. Therefore, let us discuss a bit further the issue of local causality.

\subsubsection{On the local causality of the selected matter model}

Consider the wave vector $k^a$ of the plane wave ansatz. From the above dispersion relation we see that
     \beq
     \label{ts.wvect}
     k_a k^a = k_\n^2(1-c_\n^2).
     \eeq
Thus, for causal wave propagation, the speed of sound should be less than the speed of light, in natural units
	\beq
	\label{mult.cn}
	c_\n^2 \leq 1.
	\eeq 
Note that the amplitude $A^a$ of the perturbed current defines the character (timelike, spacelike or null) of the perturbation. In order to have causal disturbances, $A^a$ must be a timelike vector field. Thus, from the transversality of the waves, equation \eqref{ts.ort}, $k^a$ is a spacelike vector.

Let us also note that, when the force equation \eqref{ts.p4} is evaluated in terms of the solution to the dispersion relation,  we find that, 
     \beq
     A_{\n\perp}^a = \sigma_n A_\parallel^\n \hat k^a_\n.
     \eeq
Thus, the waves are transverse in the ordinary three dimensional sense.

Equation \eqref{mult.cn} is the first encounter we have with a restriction imposed by causality. This implies a constraint on the master function 
     \beq
     \frac{\partial \log\Bn}{\partial \log n} + 1 \leq 1.
     \eeq
In addition, in order to avoid absolute instabilities, the speed of sound should also be  bound from below, {\it i.e.}
	\beq
	c_\n^2\geq 0.
	\eeq
This restriction prevents cases in which  $\sigma_\n^2\leq 0$. This can be written in terms of the master function by the inequality
     \beq
     \frac{\partial\log\Bn}{\partial \log n} + 1 \geq 0.
     \eeq
Combining these two results, we obtain a criterion that must be satisfied by any proposed matter model that guarantees absolute stability and local causality of the perturbed system in terms of the defining master function alone. That is
     \beq
     \label{ts.caus00}
     -1 \leq \frac{\partial\log\Bn}{\partial \log n}\leq 0.
     \eeq

In the following section, we will extend this analysis to obtain a stability and causality criterion for the two-fluid system we are concerned with.
     
\section{Sound waves in the matter and entropy system}

Having discussed plane wave propagation for a perfect fluid, we use the same technique for the conducting model developed in chapter 5. The aim of this section is to work out the dispersion relation for wave propagation in the particle-entropy two-fluid system. To make the discussion more general, we relax the thermal equilibrium assumption and work out the perturbations in the case where there exists a relative flow between the fluids. In this sense, the particular case of thermal equilibrium (aligned flows), will be discussed separately in section 6.3.2. 

Using the plane wave ansatz [equation \eqref{ts.sw}], we write the perturbations of the two conserved currents as
     \begin{subequations}
     \begin{align}
     \label{ts.ptn}
     \delta n^a     &= A^a_\n \exp(-i k_b x^b),\\
     \label{ts.pts}
     \delta s^a     &= A^a_\s \exp(-i k_b x^b).
     \end{align}
     \end{subequations}
Let us note that in these expressions we do not attach a constituent index to the wave vector $k^a$, since waves in a system are such that the constant wave vector is the same for all fluids. Thus, from the component decomposition of the wave vector, equation \eqref{ts.wavev}, we have
     \beq
     \label{ts.wvect2}
     k^a = k_\n(\sigma_\n u^a + \hat k^a_\n) = k_\s (\sigma_\s u^a_\s + \hat k^a_\s).
     \eeq
Also, since the dispersion relation is  a scalar equation, it will be useful to express the inner product $\hat w^a \hat k_a^\n$ (where $\hat w^a = w^a/w$) in terms of the angle $\beta_{\n\s}$ between the two vectors
     \beq
     \hat w^a \hat k^\n_a = \cos \beta_{\n\s}.
     \eeq

Note that there is significant difference from the previous single fluid case. For a generic multifluid system, we have to take into account the relative flow $w^a$ between the fluids introduced in section \ref{ThI}. There, we expressed the entropy flux in the Eckart frame by the relation \eqref{thermal01}, that is,
     \beq
     \label{ts.svel}
     s^a= s u^a_\s = s\gamma (u^a + w^a).
     \eeq
Having introduced the entropy current raises a subtlety. The quantities $\sigma_\n$, and $\hat k^a_\n$ are what would be measured in the matter frame. We could choose the frame defined by the entropy flux instead, and make our measurements with respect to the entropy flow. Although, from the mathematical point of view there are well-defined transformations between the two descriptions, working in the entropy frame would produce unnecessary complications due to the fact that thermometers and accelerometers are naturally carried by the particles in the fluid; not by the phonons of the system.  Moreover, the relative flow of the particles with respect to the entropy frame, which we may call $w^a_\s$, will clearly have the same magnitude as $w^a$  and is just
     \beq
     w^a_\s = -\gamma (w^2 u^a + w^a).
     \eeq

Denoting the normalised four velocity associated with the entropy flux by $u^a_\s$, it is useful to note that we can write the two four-velocities in terms of their relative flow as
     \begin{align}
     u^a       &= -w^{-2}(w^a + \gamma^{-1}w^a_\s)\\
     u^a_\s    &= -w^{-2}(w^a_\s + \gamma^{-1}w^a).
     \end{align}
Thus, by contracting each of the four-velocities with the wave vector $k_a$, we obtain a matrix equation of the form
     \beq
     \left[\begin{array}{cc}
     w \sigma_\n - \cos \beta_{\n\s}    & -\gamma^{-1} \cos \beta_{\s\n}\\
     -\gamma^{-1} \cos \beta_{\n\s}     & w \sigma_\s - \cos \beta_{\s\n}
     \end{array}
     \right]\left[
     \begin{array}{c}
     k_\n \\
     k_\s
     \end{array}
     \right] = \left[\begin{array}{c}
     0 \\
     0
     \end{array}\right].
     \eeq
The determinant of the $2 \times 2$ matrix should vanish, and this leads to\glo{$\sigma_\s$}{Phase velocity of entropy waves}\glo{$c_\s$}{Speed of sound of entropy waves}
     \beq
     \label{ts.strans}
     \sigma_\s = \cos \beta_{\s\n}\frac{\sigma_\n - w\cos \beta_{\n\s}}{w\sigma_\n - \cos \beta_{\n\s}}.
     \eeq
Thus, we have shown that  if $\sigma_\n^2\leq 1$ then $\sigma^2_{\s}\leq 1$. This is a natural conclusion since causality is a frame independent requirement.




 It was shown in the Newtonian case \cite{AnComPrix}, that two chemically coupled fluids may become two-stream unstable. This result was later generalised in \cite{larslam} for the relativistic case. In what follows, we reproduce the same analysis.
 
 
Let us begin by noting that the perturbed conservation laws, equations \eqref{ts.l1} and \eqref{ts.l2}, give the same transversality condition as in \eqref{ts.ort}, that is,
     \begin{align}
     k_a A^a_\n = k_a A^a_\s = 0.
     \end{align}
However, the perturbed forces are  more complicated than in the previous case.  The covariant derivative of the perturbed momentum \eqref{ts.p3} is
     \begin{align}
     \delta \mu_{a;c}      &= \delta n^b_{\ ;c}(\Bn_{ab} + \Ann_{ab}) + \delta s^b_{\ ;c}(\Xns_{ab} + \Ans_{ab})\nn\\
                              &= i k_c (A^b_\n \Bn_{ab} + A^b_\n \Ann_{ab} A^b_\s \Xns_{ab} + A^b_\s\Ans_{ab})\exp(-i k_d x^d).
     \end{align}

Thus, we  encounter our first problem. The perturbed equations of motion satisfy a similar force balance as in the previous chapter, but that does not directly imply the vanishing of the individual perturbed forces.  In a generic multi-fluid system, the discussion becomes significantly more elaborate at a very early stage. In our case, however, we can use the fact that, in the strict equilibrium sense, we can use the integrability of the individual forces to obtain
     \begin{align}
     \label{ts.dfn1}
     \delta f^\n_a =n^c\delta \mu_{[a;c]} &= n^c\{(k_{[c}\Bn_{a]b} + k_{[c}\Ann_{a]b})A^b_\n + (k_{[c}\Xns_{a]b} + k_{[c}\Ans_{a]b})A^b_\s\}   \nn\\
                              &= \mathcal{K}^\n_{ab} A^b_\n + \mathcal{K}^{\n\s}_{ab}A^b_\s = 0,
     \end{align}
where we have introduced the `n-dispersion' tensors\glo{$\mathcal{K}$}{Dispersion tensor}
     \begin{align}
     \mathcal{K}^\n_{ab}      &= n^c(k_{[c}\Bn_{a]b} + k_{[c}\Ann_{a]b}),\\
     \mathcal{K}^{\n\s}_{ab}  &= n^c(k_{[c}\Xns_{a]b} + k_{[c}\Ans_{a]b}).
     \end{align}
Similarly, the perturbed entropy force is
     \beq
     \label{ts.dfs1}
     \delta f^\s_a = \mathcal{K}^\s_{ab}A^b_\s + \mathcal{K}^{\s\n}_{ab}A^b_\n = 0
     \eeq 
where the `s-dispersion' tensors are
     \begin{align}
     \mathcal{K}^\s_{ab}      &= s^c(k_{[c}\Bs_{a]b} + k_{[c}\Ass_{a]b})\\
     \mathcal{K}^{\s\n}_{ab}  &= s^c(k_{[c}\Xsn_{a]b} + k_{[c}\Asn_{a]b}).
     \end{align}

Following the argument in section \ref{TS.sound}, we contract both equations \eqref{ts.dfn1} and \eqref{ts.dfs1} with the wave vector $k^a$. This will result in a $2 \times 2$ matrix problem which we address by systematically increasing the degree of interaction between the two fluids. In order to do so, it is convenient to isolate further the different contributions that appear in the dispersion matrices. 

The bulk contribution in the $\mathcal{K}^\n_{ab}$ and $\mathcal{K}^\s_{ab}$ matrices can be reduced to \cite{larslam}
     \begin{align}
     \label{ts.aux1}
     \mathscr{B}_{ab}^\n & = n^c k_{[c}\Bn_{a]b} = -\frac{1}{2}\Bn n (k_\n \sigma_\n h_{ab} + c_\n^2 k_a u_b),\\
     \label{ts.aux01}
     \mathscr{B}_{ab}^\s & = s^c k_{[c}\Bs_{a]b} = -\frac{1}{2}\Bs s (k_\n \sigma_\s h_{ab}^\s + c_\s^2 k_a u_b^\s),     
     \end{align}
while the entrainment piece becomes
     \begin{align}
     \mathscr{A}_{ab}^\n = n^c k_{[c}\Ann_{a]b}  =& \frac{1}{2} \gamma n {\Big\{} \Bn_{,\n\s}[(k_\n \sigma_\n w_a - 2k_a)u_b - k_a w_b]\nn\\
                                                &+ \gamma \frac{s}{n} \Ans_{,\n\s}(k_\n \sigma_\n w_a - k_a)(u_b+w_b){\Big\}},\\
     \mathscr{A}_{ab}^\s = s^c k_{[c}\Ass_{a]b}  =& \frac{1}{2} \gamma s {\Big\{} \Bs_{,\s\n}[(k_\s \sigma_\s w_a^\s - 2k_a)u_b^\s - k_a w_b^\s]\nn\\
                                                 &+ \gamma \frac{n}{s} \Asn_{,\s\n}(k_\s \sigma_\s w_a^\s - k_a)(u_b^\s + w_b^\s){\Big\}}.
     \end{align}
We have also introduced here the orthogonal projector  $h_{ab}^\s$ associated with  the entropy four-velocity $u^a_\s$ and  the speed of sound for the entropy perturbations $c_\s^2$ [{\it c.f.} equation \eqref{ts.sp}]
     \beq
     c_\s^2 = 1 + \frac{\partial \log \Bs}{\partial \log s}.
     \eeq 
We have introduced the shorthand notation for derivatives of the coefficients $\Bn$, $\Bs$ and $\Ans$ 
     \begin{align}
     \Bn_{,\n\s}    &\equiv \frac{1}{\Bn\Bs}\left(2ns \frac{\partial \Bn}{\partial j^2} \right),\\
     \Bs_{,\s\n}    &\equiv \frac{1}{\Bs\Bn}\left(2sn \frac{\partial \Bs}{\partial j^2} \right),\\
     \Ans_{,\n\s}   &\equiv n s \frac{\partial \Ans}{\partial j^2},
     \end{align}
and  for the cross-coupling terms in $\mathcal{K}^{\n\s}_{ab}$ 
     \begin{align}
     \label{ts.cc1}
     \mathscr{X}^{\n\s}_{ab} &= n^c k_{[c}\Xns_{a]b} = -\frac{1}{2}\mathcal{C}\gamma n \sqrt{\Bn\Bs}k_a(u_b + w_b),\\
     \label{ts.cc2}
     \mathscr{X}^{\n\s}_{ab} &= s^c k_{[c}\Xsn_{a]b} = -\frac{1}{2}\mathcal{C}\gamma s \sqrt{\Bn\Bs}k_a(u_b^\s + w_b^\s).    
     \end{align}
Here,
     \beq
     \mathcal{C} \equiv \frac{1}{\Bn\Bs}\left(2 n s \frac{\partial \Bn}{\partial s^2} \right) = \frac{1}{\Bn\Bs}\left(2 n s \frac{\partial \Bs}{\partial n^2} \right).
     \eeq
Finally, we see that the entrainment part has significantly more presence
     \begin{align}
     \mathscr{A}^{\n\s}_{ab} &= n^c k_{[c}\Ans_{a]b}\nn\\
                             &= \frac{n}{2}\Big{\{}-\Ans k_\n \sigma_\n h_{ab} - \left[\Ans + \frac{s}{n}\mathscr{B}^\n_{,\n\s}+\gamma \left(\gamma \frac{s}{n}\Bs_{,\n\s} + \Ans_{,\n\s} \right)\right]k_a u_b\nn\\
                             & +\gamma k_\n\sigma_\n \left(\gamma\frac{s}{n}\Bs_{,\n\s} + \Ans_{,\n\s} \right)w_a u_b \Big{ \}},\\
                             \label{ts.aux2}
     \mathscr{A}^{\s\n}_{ab} &= s^c k_{[c}\Asn_{a]b}\nn\\
                             &= \frac{s}{2}\Big{\{}-\Asn k_\s \sigma_\s h_{ab}^\s - \left[\Asn + \frac{n}{s}\mathscr{B}^\s_{,\s\n}+\gamma \left(\gamma \frac{n}{s}\Bn_{,\s\n} + \Asn_{,\s\n} \right)\right]k_a u_b^\s\nn\\
                             & +\gamma k_\s\sigma_\s \left(\gamma\frac{n}{s}\Bn_{,\n\s} + \Asn_{,\n\s} \right)w_a^\s u_b^\s \Big{ \}}.                        
     \end{align}
     
As stated in \cite{larslam}, a multi-fluid system must have non-zero bulk properties, the other terms can be absent depending on the equation of state. We now proceed with our sound  propagation analysis by  gradually increasing the coupling  between the matter and entropy fluids.


\subsection{The completely decoupled case}

In this very simple case, the two fluids are completely independent of one another.  The forces vanish independently and the system is absolutely integrable. Thus, although the currents may not be aligned, the dynamics are not affected by any relative flow.  In this particular case, the only non-zero contributions to the dispersion matrices $\mathcal{K}$ are equations \eqref{ts.aux1} and \eqref{ts.aux01} and we can proceed as in section 6.2. Thus, the contraction of equation \eqref{ts.dfn1} and \eqref{ts.dfs1} with the wave vector 
     \begin{align}
     k^a \delta f^\n_a &= 0,\\
     k^a \delta f^\s_a &=0,
     \end{align}
produces the simple $2 \times 2$ matrix problem
     \beq
     \left[
     \begin{array}{cc}
     \Bn(\sigma_\n^2 - c^2_\n)     &               0          \\
               0                   & \Bs(\sigma_\s^2 - c_\s^2) 
     \end{array} \right]
     \left[ 
     \begin{array}{c}
     u_a A^a_\n \\
     u_a^\s A^a_\s
     \end{array}\right] =
     \left[
     \begin{array}{c}
     0 \\
     0
     \end{array}\right].
     \eeq
Therefore, the dispersion relation is simply given by
     \beq
     (\sigma_\n^2 - c_\n^2)(\sigma_\s^2 - c_\s^2) = 0.
     \eeq
     
As we have seen before, $\sigma_\n^2$ corresponds to the squared phase three-velocity  as measured in the Eckart frame. Similarly, $\sigma_\s^2$ is the squared phase velocity  of the `entropy waves' as measured in the frame defined by $u^a_\s$, whose transformation to the Eckart frame is given by \eqref{ts.strans}.  Thus we see that, for the uncoupled system, we must require that the equation of state satisfies \eqref{ts.caus00} for both, particles and entropy.

This simple case has served the purpose of defining the quantities $c_\n$ and $c_\s$. As we have argued in the single fluid case, these quantities are identified with the decoupled sound speeds. It will be useful to keep this in mind in the following more elaborate discussion.

\subsection{Cross-constituent coupling}

The next situation corresponds to that where the system is sensitive to the relative flow $w^a$ but not to entrainment. That is, in addition of \eqref{ts.aux1} and \eqref{ts.aux01}, we have contributions due to the cross-coupling tensors \eqref{ts.cc1} and \eqref{ts.cc2}. Again, we reduce the problem to a $2 \times 2$  matrix by contracting \eqref{ts.dfn1} and \eqref{ts.dfs1} with the wave vector $k^a$. The cross-coupling produces the expected  off-diagonal terms
     \beq
     \left[
     \begin{array}{cc}
     \Bn(\sigma_\n^2 - c_\n^2)     &    -\sqrt{\Bn\Bs}\mathcal{C}     \\
     -\sqrt{\Bn\Bs}\mathcal{C}     &    \Bs(\sigma_\s^2 - c_\s^2)
     \end{array}
     \right]\left[
     \begin{array}{c}
     u_a A^a_\n     \\
     u_a^\s A^a_\s
     \end{array}
     \right]=\left[
     \begin{array}{c}
     0    \\
     0
     \end{array}
     \right],
     \eeq
where $\mathcal{C}^2$ is the cross constituent coupling defined by
     \beq
     \label{ts.cc}
     \mathcal{C}^2 = \frac{1}{\Bn\Bs}\left( 2n s \frac{\partial \Bn}{\partial s^2}\right)^2.
     \eeq

The dispersion relation is again the determinant of the $2 \times 2$ matrix and in this case gives
     \beq
     \label{ts.discc}
     (\sigma_\n^2 - c_\n^2)(\sigma_\s^2 - c_\s^2)=\mathcal{C}^2.
     \eeq
     
We can see from the definition of $\mathcal{C}^2$, \eqref{ts.cc}, that in order for it to be less than zero, either $\Bn$ or $\Bs$ should be negative. For ordinary matter, or entropy, this is not generally the case. Hence we do not deal with such possibilities in this work.

Equation \eqref{ts.discc} shows the  richer phenomenology of the coupling between matter and entropy, or of a two-fluid system in general. This allows for two-stream unstable flows, as discussed below (see also \cite{larslam} and \cite{AnComPrix}). Once more we are going to take small steps to analyse this situation.

\subsubsection{Aligned flow: the stability of thermal equilibrium}

We can now use  the angular relationship between $\sigma_\n$ and $\sigma_\s$ given by \eqref{ts.strans} to write a dispersion relations solely in terms of $\sigma_\n$. Addressing the question of absolute stability ($\sigma_\n^2\geq 0$), we set the relative flow $w^a=0$ in \eqref{ts.strans}, for which we would have $\sigma_\s = \sigma_\n$ and \eqref{ts.discc} becomes the quadratic expression in $\sigma_\n^2$\glo{$\mathcal{C}$}{Chemical coupling coefficient}
     \beq
     (\sigma_\n^2)^2 - (c_\n^2 + c_\s^2)\sigma_\n^2 +(c_\n^2 c_\s^2 - \mathcal{C}^2) = 0,
     \eeq
whose roots are
     \begin{align}
     \label{ts.root1}
     \sigma_\n^2    &= \frac{1}{2}\left(c_\n^2 + c_\s^2 \pm \sqrt{(c_\n^2 + c_\s^2)^2 - 4(c_\n^2 c_\s^2 - \mathcal{C}^2)} \right)\nn\\
                    &= \frac{1}{2}\left(c_\n^2 + c_\s^2 \pm \sqrt{(c_\n^2 - c_\s^2)^2 + 4\mathcal{C}^2} \right)
     \end{align}
     
In order to avoid complex $\sigma^2_\n$, the square root in \eqref{ts.root1} has to be real, which follows from the fact that $\mathcal{C}^2\geq 0$. Now, to have absolute stability it is sufficient to require that the absolute value of the negative square root in \eqref{ts.root1} be less than the first term, that is 
     \beq
     (c_\n^2 + c^2_\s)^2 \geq (c_\n^2 - c_\s^2) + 4 \mathcal{C}^2,
     \eeq
that is equivalent to the requirement 
     \beq
     \label{ts.astabcc}
     \mathcal{C}^2 \leq c_\n^2 c_\s^2.
     \eeq

This can be written in terms of the bulk quantities we expect to be positive, $\Bn \Bs > 0$, and so absolute stability constrains the master function to satisfy
     \beq
     \frac{\partial \log\Bn}{\partial\log s}\frac{\partial \log\Bs}{\partial \log n} \leq \left(1+\frac{\partial\log\Bn}{\partial \log n} \right)\left( 1 + \frac{\partial\log\Bs}{\partial \log s}\right).
     \eeq

The causality constraint requires $\sigma_\n^2\leq 1$. For this we notice that if the absolute stability condition \eqref{ts.astabcc} is satisfied, and $\mathcal{C}^2\geq 0$, we need only to guarantee that the positive root of \eqref{ts.root1} is causal, since the negative one is always smaller
     \begin{align}
     \label{ts.caus}
     \frac{1}{2}(c_\n^2 + c_\s^2) + \sqrt{(c_\n^2 - c_\s^2)^2 + 4\mathcal{C}^2} &\leq 1\nn\\
                                    \sqrt{(c_\n^2 - c_\s^2)^2 + 4\mathcal{C}^2} &\leq (1-c_\n^2)+(1-c_\s^2)\nn\\
                                    (c_\n^2 - c_\s^2)^2 + 4\mathcal{C}^2        &\leq [(1-c_\n^2)+(1-c_\s^2)]^2\nn\\
                                    \mathcal{C}^2                               &\leq (1-c_\n^2)(1-c_\s^2),   
     \end{align}
which in terms of the master function becomes
     \beq
     \label{mult.thermaleqstab}
     \frac{\partial \log\Bn}{\partial\log s}\frac{\partial \log\Bs}{\partial \log n} \leq \frac{\partial\log\Bn}{\partial \log n}\frac{\partial\log\Bs}{\partial \log s}.
     \eeq

Note that causality and absolute stability set upper bounds for $\mathcal{C}^2$. In general it can be shown \cite{larslam} that if $c_\n^2 + c_\s^2 \leq 1$ then any absolutely stable master function is also causal. Conversely, if $c_\n^2 + c_\s^2\geq 1$, any causal master function is absolutely stable.

Equation \eqref{mult.thermaleqstab} is one of our most important results since it represents the stability of thermal equilibrium. Thus, a matter model whose master function (equation of state) fails to satisfy the inequality \eqref{mult.thermaleqstab} can be regarded as non-physical.  

In order to obtain a comparison with the Newtonian problem, we consider the case of `small' relative flow. Although this would not strictly correspond to an adiabatic perturbation, it will allow us to show the similarities with a better known problem. In this case, we will be in position to examine if a dynamical two-stream instability is present by solving \eqref{ts.discc} for some relative flow and then take the the limit in which the flows are aligned. In what follows we will assume that the  master function  is absolutely stable and causal.

 \subsubsection{Small relative flow: `almost' adiabatic perturbations}
 
Assuming that $w$ and $\sigma_\n$ are both much smaller than the speed of light ($c=1$) in \eqref{ts.strans}, the dispersion relation \eqref{ts.discc} becomes
     \beq
     (\sigma_\n^2 - c_\n^2)[(\sigma_\n - w\cos \beta_{\n\s})^2 - c_\s^2] = \mathcal{C}^2.
     \eeq
     
Introducing the variables
     \beq
     \label{ts.rescal}
      x = \frac{\sigma_\n}{c_\s}, \quad y=\frac{w\cos \beta_{\n\s}}{c_\s}, \quad b^2 =\left(\frac{c_\n}{c_\s} \right)^2, \quad a^2 = \frac{\mathcal{C}^2}{c_\s^4},
      \eeq
we get
     \beq
     \frac{x^2 - b^2}{a^2}[(x-y^2) - 1] = 1.
     \eeq
     
Written in this form, the problem is identical to the Newtonian plane wave propagation discussed by Andersson, Comer and Prix \cite{AnComPrix} where they show the regime in which the two-stream instability operates. It is also shown in \cite{larslam} that the particular example used by Andersson, Comer and Prix obeys the causality and absolute stability criteria  derived above.  First, let us  note that, due to the presence of a velocity scale given by the speed of light, we cannot completely scale out the velocities. Thus, the relativistic analysis of stability will, in general, contain an extra parameter compared to the Newtonian case. Following \cite{larslam}, we take that parameter to be the speed of sound for the entropy, $c_\s$, which without loss of generality, we can take larger than the $c_\n$.  Using these parameter the absolute stability criterion \eqref{ts.astabcc} becomes
     \beq
     \frac{a^2}{b^2}\leq 1,
     \eeq 
which is satisfied in the model discussed above.



We now turn to the relativistic dispersion relation \eqref{ts.root1}. Written as an equation for $\sigma_\n$, it constitutes a non-trivial quartic. If we use the same re-scaling \eqref{ts.rescal}, then \eqref{ts.root1} becomes
     \beq
     \label{ts.quartic}
     \frac{x^2-b^2}{a^2}\left[\frac{(x-y)^2}{\gamma^{-2}(1-c_\s^2 x^2) + c_\s^2(x-y)^2} -1 \right] = 1.
     \eeq
     
\begin{figure}
  \includegraphics[height=0.6\textwidth,width=0.9\textwidth,clip]
   {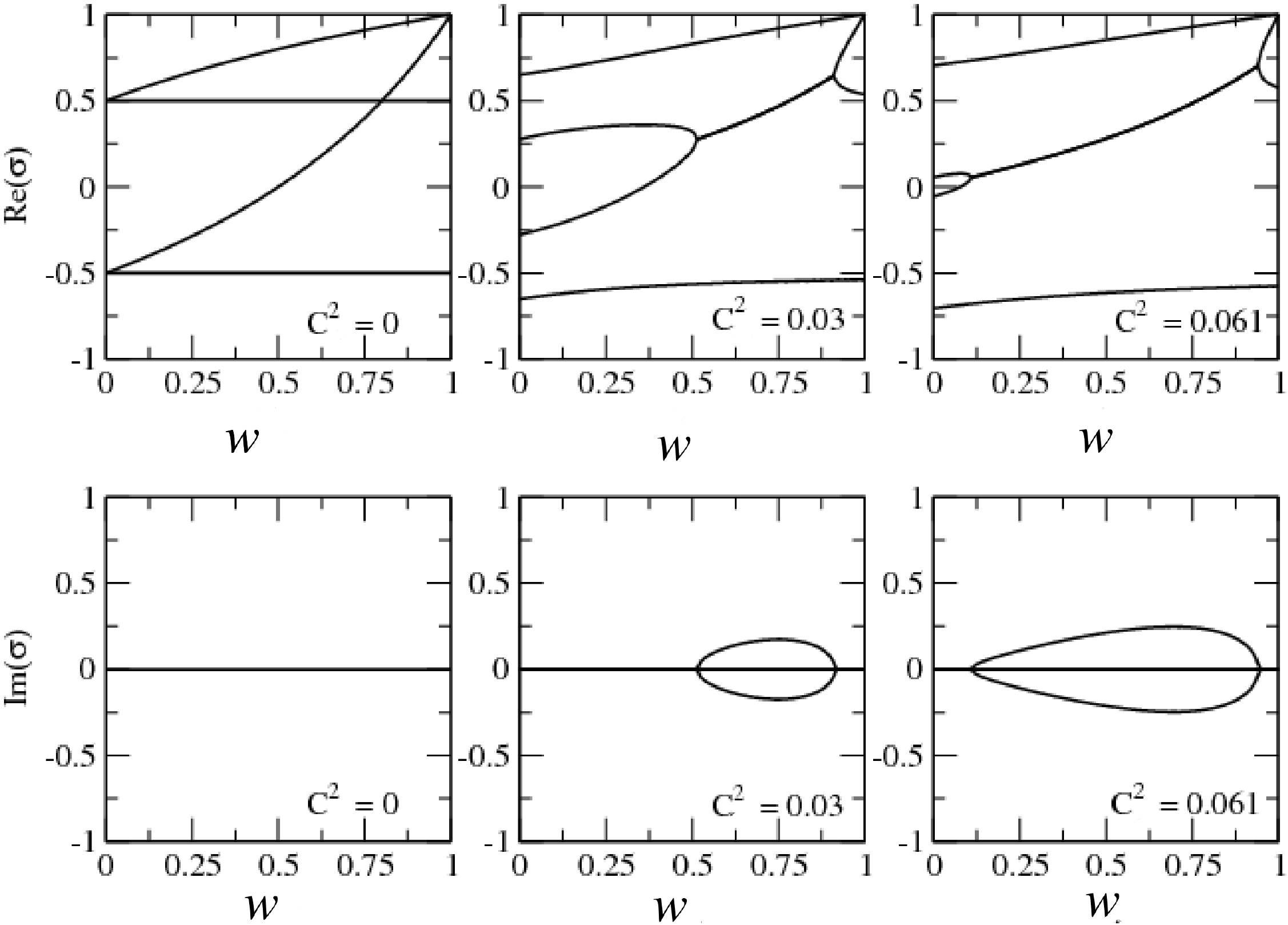}
   \caption{Plots of the real (top) and imaginary (bottom) parts of the mode
   frequencies $\sigma_\n$ as functions of $w$, for $c_\s = 0.5$, $b^2
   = 1.0$, and $\beta_{\n \s} = 0$. \ The merger of two frequencies, and
   subsequent non-zero imaginary parts, signal the presence of a two-stream
   instability.} \label{sigcccplt0}
\end{figure}

 Although the general solution of \eqref{ts.quartic} is available, it is quite a complicated expression and offer very little insight for our analysis. Thus, we tackle the problem numerically for a set of parameter values that maintain absolute stability and causality. Figure \ref{sigcccplt0} provides plots of the real and imaginary parts for the solutions of \eqref{ts.quartic} in the case when the particle and entropy flow are \emph{aligned}. The solutions for $\sigma_\n$ are taken to be functions of the relative flow parameter $y$ and the coupling $\mathcal{C}^2$, with the choice $c_\s^2=0.5$, $b^2=1$ and $\theta=0$. Non-zero values for ${\rm Im}(\sigma_\n)$ in the figures indicates the presence of an unstable mode. The appearance of unstable modes is reflected in the real parts whenever two frequencies merge. This behaviour is typical for this kind of dynamical instability \cite{AnComPrix,larslam}.  Note that, although in the central figure the instability sets at a  `high' relative flow ($w\sim 0.6$), unstable modes can be produced for arbitrary small values of the coupling $\mathcal{C}$, as can be seen from the third column. 


\subsection{The role of entrainment}     
     
We finally turn our attention to the full problem. Up to this point we have introduced three equation of state  parameters, {\it i.e.} $c_\n$, $c_\s$ and $\mathcal{C}$, that are obtained as second derivatives of the master function $\Lambda$. Given that we have established  the two-stream instability for the general cross-constituent coupling and arbitrary background flow, the main reason for discussing entrainment is because it is an essential ingredient for the correct interpretation of relaxation times in the Cattaneo-like equations derived in Chapter 5. We will simplify the entrainment case by assuming that the system is close to equilibrium, that is the relative velocity is such that $w\ll 1$,  and that the flows are aligned. This does not mean that the sound wave speeds or phase velocities have to be small, nor do the flows $u^a$ nor $u^a_\s$ have to be similarly restricted.

If we keep deviations from thermal equilibrium to $\mathcal{O}(w^2)$, the master function can be approximated as \cite{nils1,nils2}
     \beq
     \Lambda = \lambda_0(n^2,s^2) + \lambda_1(n^2,s^2)(j^2 - \sqrt{n^2 s^2}),
     \eeq
which immediately implies
     \begin{align}
     \Bn       &= -2 \left[\frac{\partial \lambda_0}{\partial n^2} + \frac{\partial\lambda_1}{\partial n^2}(j^2 - \sqrt{n^2 s^2}) - \lambda_1 \frac{s}{2n}\right]\\
     \Bn_{,ns} &= -2 n s \frac{\partial \lambda_1}{\partial n^2}\\
     \Bs_{,ns} &= -2 n s \frac{\partial \lambda_1}{\partial s^2}
     \end{align}
and 
     \beq
     \Ans=-\lambda_1, \qquad \Ans_{,ns} = 0.
     \eeq

In the spirit of understanding a complicated problem, we shall make a further simplifying approximation by assuming that  $\lambda_1$ is constant, so that
     \beq
     \Bn_{,ns}= \Bs_{,ns} = 0.
     \eeq
 
This leaves the problem with the simple entrainment parameter $\lambda_1$. With this approximation we find 
     \beq
     \label{ts.entm}
     \left[
          \begin{array}{cc}
               \Bn (\sigma_\n^2 - c_\n^2)    &    -\mathcal{C}\sqrt{\Bs\Bn} - \Ans \frac{\epsilon_{\n\s k_\n}}{\epsilon_{\s\n} k_\s}(\sigma_\n^2-1)     \\
               -\mathcal{C}\sqrt{\Bs\Bn} - \Ans \frac{\epsilon_{\s\n k_\s}}{\epsilon_{\n\s} k_\n}(\sigma_\s^2-1)     &    \Bs(\sigma_\s^2 - c_\s^2)    
          \end{array}
      \right]\left[ 
          \begin{array}{c}
               u_a A_\n^a     \\
               u_a^\s A^a_\s
          \end{array}
      \right]=\left[
          \begin{array}{c}
               0    \\
               0
          \end{array}
      \right].
     \eeq

It follows from \eqref{ts.wvect} and \eqref{ts.wvect2} that
     \beq
     \frac{k_\n}{k_\s} = \sqrt{\frac{1-\sigma_\s^2}{1-\sigma^2_\n}},
     \eeq
and hence the dispersion relations becomes\glo{$\mathcal{C}^\natural$}{Entrainment coupling}
     \beq
     \label{ts.dispen}
     (\sigma_\n^2-c_\n^2)(\sigma_\s^2-c_\s^2)-\left[\mathcal{C}+\mathcal{C}^\natural \sqrt{(1-\sigma_\n^2)(1-\sigma_\s^2)}\right]^2=0,
     \eeq
where 
     \beq
     \mathcal{C}^\natural = \epsilon_{\n\s}\epsilon_{\s\n}\frac{\Ans}{\sqrt{\Bn\Bs}},
     \eeq
and because the inverse of \eqref{ts.entm} must exist, $|\mathcal{C}^\natural|\neq 1$.

We stress that, unlike the simpler cases, the dispersion relation \eqref{ts.dispen} does not hold for arbitrary propagation direction with respect to the relative velocity. This mirrors the fact that entrainment enters the master function in a fundamentally different way: At first-order in the relative velocity squared. In the dispersion relation, however, entrainment contributes even in the limit of zero relative velocity, as we have seen in the previous section, because there are still two sets of interacting sound waves.

In the limit of aligned flows and small deviations from thermal equilibrium ($w\ll 1$), we see from \eqref{ts.strans} that $\sigma_\n^2 = \sigma_\s^2$, and so
     \beq
     (\sigma_\n^2 - c_\n^2)(\sigma_\n^2 - c_\s^2) - \left[\mathcal{C} + \mathcal{C}^\natural (1-\sigma_\n^2) \right]^2 = 0.
     \eeq

Even though when it is difficult to think of a system of fluids which are entrained and yet not chemically coupled, it is particularly instructive to consider the entrainment case alone, {\it i.e.} set $\mathcal{C}= 0$. The corresponding dispersion relation is a quadratic for $\sigma_\n^2$, and has solutions
     \beq
     \label{ts.sigentr}
     \sigma_\n^2 = \frac{c_\n^2 + c_\s^2 - 2 {\mathcal{C}^\natural}^2 \pm \left[ (c_\n^2 - c_\s^2)^2 + 4 {\mathcal{C}^\natural}^2(1-c_\n^2)(1-c_\s^2) \right]^{\frac{1}{2}}}{2(1-{\mathcal{C}^\natural}^2)}.
     \eeq 
     
For speeds of sound satisfying the causality requirement, $c_\n, c_\s \leq 1$ the discriminant is obviously positive and hence the $\sigma_\n^2$ are real. In order to analyse absolute stability and causality we need to consider the ranges $0 \leq {\mathcal{C}^\natural}^2 < 1$ and $1 < {\mathcal{C}^\natural}^2$ separately. We will look at ${\mathcal{C}^\natural}^2>1$ first.
 
We rewrite \eqref{ts.sigentr} so that the denominator is positive
     \beq
     \sigma_\n^2 = \frac{2 {\mathcal{C}^\natural}^2 - c_\n^2 - c_\s^2  \pm \left[ (c_\n^2 - c_\s^2)^2 + 4 {\mathcal{C}^\natural}^2(1-c_\n^2)(1-c_\s^2) \right]^{\frac{1}{2}}}{2({\mathcal{C}^\natural}^2 -1 )}.
     \eeq

Since we are imposing $c_\n$, $c_\s \leq 1$ and ${\mathcal{C}^\natural}^2 > 1$, the terms outside the square root in the numerator are positive. therefore the positive solution is absolutely stable. But, we can also show that it cannot be causal. As for the minus solution, it can be shown that it is absolute stable  only if $\mathcal{C}^\natural <1$ \cite{larslam}, which is not the case we are considering at the moment. Therefore, ${\mathcal{C}^\natural}^2>1$ does not lead to both absolutely stability and causality, and is therefore ruled out.

In a manner similar to the cross coupling case, the region $0 \leq {\mathcal{C}^\natural}^2<1$ is a different story, because the terms outside the radical in the numerator of \eqref{ts.sigentr} are not of any definite sign. This affects the  absolute stability analysis more than the determination of causality. In fact the causality requirement is sufficiently straightforward in this range for $\mathcal{C}^\natural.$ 

In order to assess  absolute stability, it is useful to introduce \cite{larslam}
     \beq
     \tau = \left[(c_\n^2 - c_\s^2)^2 + 4 {\mathcal{C}^\natural}^2 (1 - c_\n^2)(1 - c_\s^2)\right]^\frac{1}{2}.
     \eeq
     
This allows the numerator \eqref{ts.sigentr} to be written in such a way that the absolute stability condition becomes
     \beq
     \label{ts.tau}
     \tau^2 - \left[\pm 2 (1 - c_\n^2)(1 - c_\s^2)  \right] + \left[ c_\n^2 (1 - c_\s^2) + c_\s^2 (1 - c_\n^2)\right]\left[ 2 - (c_\n^2 + c_\s^2) \right]\leq 0,
     \eeq
where the $\pm$ corresponds to that of \eqref{ts.sigentr}. The final step is to factorise \eqref{ts.tau} to obtain
     \beq
     \left\{ \tau \pm \left[ 2 - (c_\n^2 + c_\s^2)\right]\right\}\left\{ \tau \mp \left[ c_\n^2 (1- c_\s^2 + c_\s^2 (1 - c_\n^2)) \right]\right\} \leq 0,
     \eeq
where  if the `$+$' is taken from the first factor, then the `$-$' must be taken in the second, and vice versa. In either case, the factor that has the $+$ is positive definite, and so the other factor must be less than zero. When the first factor takes the $-$ the inequality leads to ${\mathcal{C}^\natural}^2>1$. The other choice leads to the more restrictive condition ${\mathcal{C}^\natural}^2\leq c_\n^2 c_\s^2$.

\begin{figure}
  \includegraphics[height=0.6\textwidth,width=0.9\textwidth,clip]
   {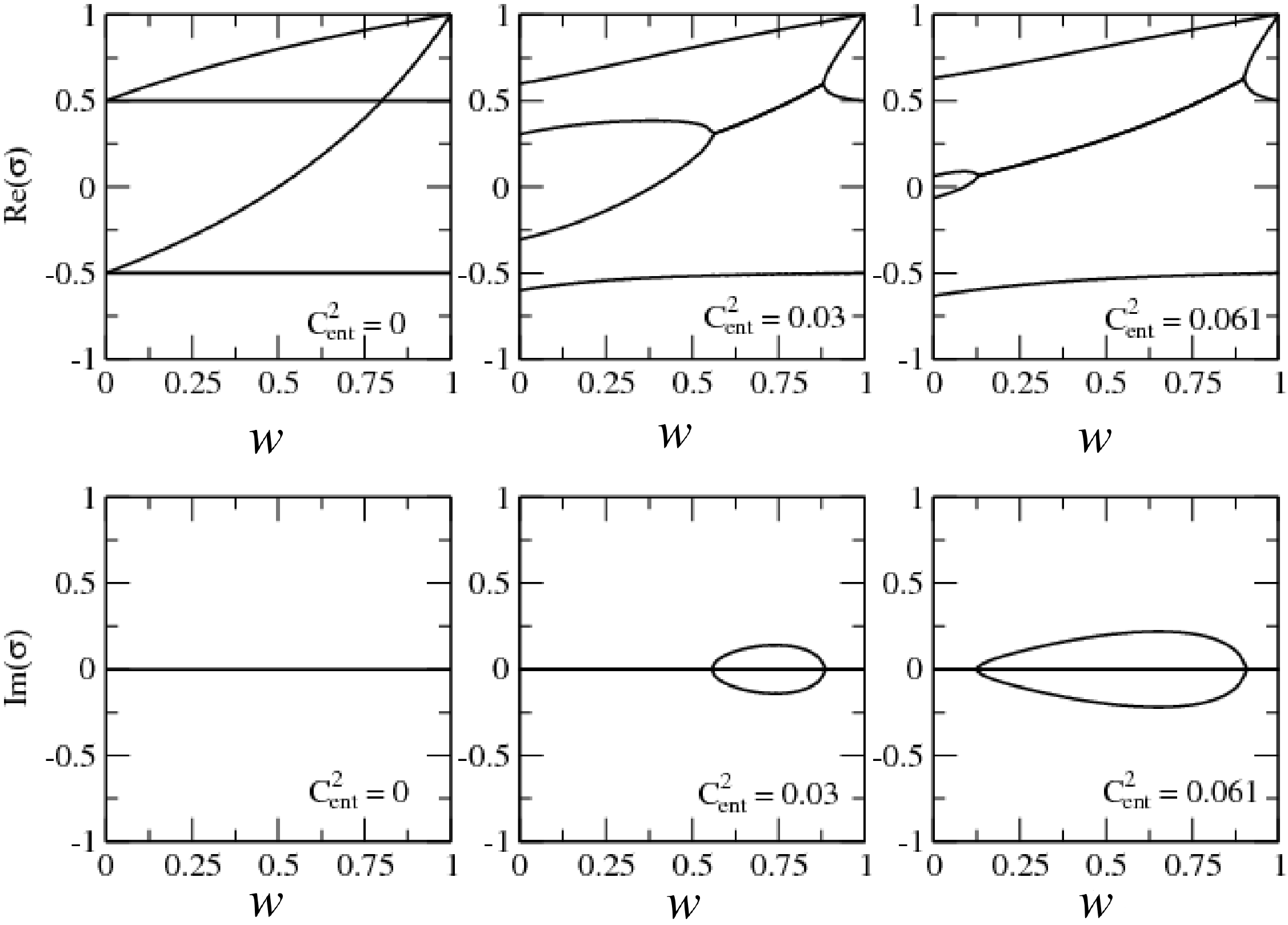}
   \caption{Plots of the real and imaginary parts of the mode frequencies
   $\sigma_\n$ as functions of $y$ and ${\mathcal{C}^\natural}^2$, for $c_s = 0.5$, and
   $b^2 = 1.0$. \ The lines of intersection and subsequent merger of two
   frequencies signal the presence of a two-stream instability.}
   \label{entplt}
\end{figure}

To summarise, we have shown that when ${\mathcal{C}^\natural}^2>1$, there is either no absolute stability or causality, which makes this range non-physical. Meanwhile, for $0 \leq {\mathcal{C}^\natural}^2 \leq c_\n^2 c_\s^2$ this system is causal and absolutely stable. In terms of our previous definitions, this translates to
     \beq
     (\Ans)^2 \leq \Bn \Bs \left(1 + \frac{\partial\log \Bn}{\partial \log n } \right)\left(1 + \frac{\partial \log \Bs}{\partial \log s} \right),
     \eeq
as a constraint on the master function, when the relative speed between the two fluids is sufficiently small. A numerical analysis shows a similar behaviour to the one shown in Figure \ref{entplt}.

\subsubsection{What have we learned?}

This chapter built on the relativistic two-stream instability analysis developed in \cite{AnComPrix,larslam}. The central results of this chapter can be summarised by the following two points:

	\begin{enumerate}
	\item Although it was not the aim of this chapter to present a two-stream instability analysis for a system of `generic' fluids, let us note that there is still work to be done in this direction. This is important because, although we have made some progress in our understanding of the role of entrainment in linear perturbations, we have not been able to produce a complete analysis of the fully coupled two-fluid system. Such analysis should become relevant in answering specific questions regarding the onset of instability for particular matter models, where the two-fluids under consideration are strongly coupled. 
	\item  In our present thermodynamical discussion we note that in all three cases - decoupled, chemically coupled and entrained - equilibrium is  always stable against adiabatic perturbations. This is, perhaps, not surprising. However, notice that for some matter models, if the couplings are strong enough, there might be unstable modes in an arbitrarily small neighbourhoods of  equilibrium, {\it i.e.} in the regime of \emph{almost} adiabatic perturbations. In this sense, the emergence of a two-stream instability for the matter-entropy system is a concrete prediction of our multifluid programme for relativistic dissipation.
	\end{enumerate}

\chapter[Non-adiabatic perturbations]{Non-adiabatic perturbations of equilibrium}
     
In the previous chapter, we have found the criteria that matter models should satisfy in order to have a stable state of thermal equilibrium. However, thus far we have only considered adiabatic perturbations, {\it i.e.} those which do not produce heat. Here, we will take a first step towards testing the stability of thermal equilibrium against deviations `down hill' the summit of maximum entropy; those where heat has a role to play.

We have shown in Chapter 5 that the heat conduction problem has two degrees of freedom. However, we have only focused on the thermal one, leaving aside the one related to the material degree of freedom. Taking both into account, let us consider the case where the original state is that of thermal equilibrium, so that  the two conservation laws for particles and entropy, equations  \eqref{mult.divn} and \eqref{mult.divs}, hold. As discussed in the previous chapter, by adiabatic perturbations we mean those which produce heat. Here, we consider the case of `small' non-adiabatic deviations from thermal equilibrium, that is, those which are linear in the heat flux. Thus, we  linearise the momentum equations associated with the transport phenomena given by the Cattaneo-like relations [{\it c.f.} section 5.4] to obtain
	\begin{subequations}
	\begin{align}
	\alpha \dot q_a + \dot \alpha q^a + h^b_{\ a} \mu_{;b} + \mu \dot u_a           & = 0, \\
	\tau \dot q_a   + q_a             +\kappa h^b_{\ a} T_{;b} + \kappa T \dot u_a	& = 0,
	\end{align}
	\end{subequations} 
where the coefficient $\alpha$ is directly related to the entrainment and the equilibrium temperature $T$ by
	\beq
	\label{dec.alpha}
	\alpha = \frac{\Ans}{T} = \frac{\kappa - \tau s}{\kappa n},
	\eeq
and we have used the definitions \eqref{can.beta} and \eqref{can.reltime} to  eliminate $\Ans$ in favour of the material properties $\kappa$ and $\tau$.

Note that here we use the local equilibrium hypothesis since  $s^*$ and $\theta^*$ differ from their equilibrium values only at second order. Moreover, the first order pressure is obtained from the standard equilibrium Gibbs relation
	\beq
	p_{;a} = \mu_{;a} n + T_{;a}s.
	\eeq
Therefore, in the case of linear deviations from thermal equilibrium, the fundamental Euler relation involves only equilibrium variables, that is
	\beq
	\label{dec.euler}
	\rho + p = \mu n + T s.
	\eeq
	
The key point to note is that the first-order model remains cognizant of its higher order origins. Specifically, $\alpha$ and, therefore, $\tau$ depend on the entrainment of the two-fluid system. Thus, although the deviations from equilibrium are linear, the dependence of the master function on the relative flux between particles and entropy is a second order effect\footnote{See for instance the original discussion by Carter presented in the last section of chapter 4.} which is a crucial ingredient to address the problem of non-adiabatic perturbations.

In order to analyse the linearised dynamics of the heat problem, let us consider perturbations away from a uniform background state. We denote such perturbations by $\delta$. Since we start at equilibrium, we have  $q^a = \dot u^a = \dot \alpha =  0$. Therefore, the perturbed momentum equations can be written as
	\begin{subequations}
	\label{dec.pertmom}
	\begin{align}
	\alpha \delta \dot q_a + h^{b}_{\ a} \delta \left(\mu_{;b}\right) + \mu \delta \dot u_a   & =  0,\\
	\tau \delta \dot q_a + \delta q_a + \kappa h^{b}_{\ a}\delta\left(T_{;b} \right) + \kappa T \delta \dot u_a & =  0.
	\end{align}
	\end{subequations}

Now, we can divide our  analysis into two parts. First, we will study the transverse mode to obtain the requirements that need to be fulfilled in order to have a stable equilibrium state against non-adiabatic perturbations. Secondly, we will look at the dispersion relation for the longitudinal mode, containing further information about the stability of the theory and where we can impose local causality as described in the previous sections.

\section{Transverse modes: first stability criterion}

Noting that the perturbed momenta \eqref{dec.pertmom} only have spatial components with respect to the Eckart frame, we can take the spatial curl in coordinates associated with the background flow. Throughout this section, we will denote spatial indices in the traditional manner by the Latin letters $i$, $j$ and $k$ (not to be confused with spacetime indices $a$, $b$, $c$, $d$ ...). Thus, acting with the curl operator $\epsilon^{ijk} \nabla_k$ on the perturbed momentum equations we obtain
	\begin{subequations}
	\label{dec.pertmomt}
	\begin{align}
	\mu \dot U^i + \alpha \dot Q^i & = 0,\\
	\tau \dot Q^i  + Q^i + \kappa T \dot U^i & = 0,
	\end{align}
	\end{subequations}
where we have defined the spatial three vectors $U^i$ and $Q^i$ by
	\beq
	U^i = \epsilon^{ijk}\nabla_j u_k \quad \text{and} \quad Q^i = \epsilon^{ijk} \nabla_j q_k,
	\eeq
and $\epsilon^{ijk}$ is the totally-antisymmetric three-volume form whilst $\nabla_j$ is the spatial gradient operator.

Assuming that the perturbations have a dependence on time of the form $e^{i\omega t}$, we arrive at the dispersion relation for the transverse perturbations
	\beq
	\label{dec.transdisp}
	i\omega\left[\mu\left(1 + i\omega\tau\right) - i \omega \alpha\kappa T \right] = 0.
	\eeq
	
Obviously, the case $\omega =0$ is a solution. With the aid of the fundamental Euler relation, equation \eqref{dec.euler}, it is not difficult to show that the other solution of \eqref{dec.transdisp} can be written as
	\beq
	\omega = i \frac{n \mu}{p + \rho}\left(\tau - \frac{\kappa T}{p + \rho} \right)^{-1}.
	\eeq
This is quite a remarkable result. It shows that the relaxation time $\tau$ is essential in order for the system to be stable. We need  the imaginary part of the frequency $\omega$ to be positive. Thus, given that the chemical potential,  the particle number density and absolute equilibrium temperature are positive quantities, and assuming that the weak energy condition\footnote{For matter described by the energy momentum tensor $T_{ab}$, the weak energy condition asserts:  that for every future directed timelike vector field $X$, the energy density $\rho = X^a X^b T_{ab}\geq 0$. In particular, for a perfect fluid, it implies $\rho + p \geq 0$.} holds, we have 
	\beq
	\frac{\mu n}{\rho + p}\geq 0 \quad \text{and} \quad \frac{\kappa T}{\rho + p} > 0.
	\eeq
Therefore, the stability criterion of the system is given in terms of the relaxation time through 
	\beq
	\label{first.taucrit}
	\tau > \frac{\kappa T}{\rho + p}.
	\eeq
Thus, we arrive at a similar result to the one obtained by Hiscock and Olson \cite{hiscolson} regarding the stability of first-order theories. In particular, the Eckart model - for which the relaxation time is identically zero - is inherently unstable.
	

\section{Longitudinal modes: causality and  the onset of instability}
	
For the case of longitudinal perturbations, we take the three-divergence of the perturbed momenta, equations \eqref{dec.pertmomt}, to get
	\begin{subequations}
	\begin{align}
	\mu \frac{\partial}{\partial t}\nabla_i \delta u^i + \nabla^2\delta \mu + \alpha\frac{\partial}{\partial t} \nabla_i \delta q^i & =0,\\
	\tau\frac{\partial}{\partial t}\nabla_i \delta q^i + \nabla_i \delta q^i + \kappa \delta^2 T + \kappa T \nabla_i \delta u^i & =0.
	\end{align}
	\end{subequations}
	
Here, we also need to consider the divergence of the perturbed conservation laws. These take the form
	\begin{subequations}
	\begin{align}
	\nabla_i \delta u^i & = - \frac{1}{n}\frac{\partial}{\partial t}\delta n,\\
	\nabla_i \delta q^i & = - nT\frac{\partial}{\partial t}\delta \hat s,
	\end{align}
	\end{subequations}	
where we have defined the specific entropy as $\hat s = s/n$. This problem is clearly more involved
than the transverse one. To make progress we need to make some decisions. First, we need to choose an appropriate set of variables to work with. Secondly, we need to provide an equation of state for matter. In the following we will opt to work
with the perturbed densities $\delta n$ and $\delta \hat s$. In this case, we have the set of thermodynamic relations
	\begin{subequations}
	\begin{align}
	\delta \mu  & = \left.\frac{\partial \mu}{\partial n}\right|_{\hat s} \delta n + \left.\frac{\partial \mu}{\partial \hat s}\right|_n \delta \hat s,\\
	\delta T    & = \left.\frac{\partial T}{\partial n}\right|_{\hat s} \delta n  + \left.\frac{\partial T}{\partial \hat s}\right|_n \delta \hat s.
   	\end{align}
	\end{subequations}
	
It is also useful to keep in mind that  the temperature can be interpreted as the `entropy chemical potential', {\it i.e.} it is defined by 
	\beq
	T = \left.\frac{\partial \rho}{\partial s}\right|_n,
	\eeq
where $\rho= \rho(n,s)$ represents the equation of state. This immediately implies the other thermodynamic relation
	\beq
	\left.\frac{\partial T}{\partial n}\right|_s = \left.\frac{\partial \mu}{\partial s}\right|_n
	\eeq
This relation allows us to reduce the number of unspecified thermodynamic quantities, which is
the key reason for the choice to work with the perturbed densities. It is also worth noting that
	\beq
	\left.\frac{\partial \mu}{\partial \hat s}\right|_{n} = n\left.\frac{\partial \mu}{\partial s}\right|_{n}.
	\eeq
	
Again, we will consider the plane wave solutions such that the perturbations behave as $\exp(i \omega t + k_jx^j)$. In a similar way as in the previous sections, we introduce the phase velocity of the waves $\sigma = \omega/k$, where $k^2=k_ik^i$. Thus, the above relations lead to the coupled equations
	\begin{subequations}
	\label{sym}
	\begin{align}
	\frac{\mu}{n} \left[\sigma^2 - \frac{n}{\mu}\frac{\partial\mu}{\partial n} \right]\delta n & + n\alpha T \left[\sigma^2 - \frac{1}{\alpha T}\frac{\partial T}{\partial n} \right]\delta \bar{s}  = 0\\
	\label{sym02}
	\kappa \frac{T}{n}\left[\sigma^2 - \frac{n}{T}\frac{\partial T}{\partial n} \right]\delta n & + n\tau T \left[\sigma^2 - i \frac{1}{k\tau}\sigma - \frac{\kappa}{\tau T}\frac{\partial T}{\partial s} \right] \delta \bar{s}  =0 .
	\end{align}	
	\end{subequations}
		
The dispersion relation we will obtain from \eqref{sym} will be a quartic on $\sigma$. Furthermore, we immediately notice the appearance of a complex root. To facilitate the interpretation of the roots, we complete the square in the last bracket of \eqref{sym02}
	\beq
	\sigma^2 - i \frac{1}{k\tau}\sigma - \frac{\kappa}{\tau T}\frac{\partial T}{\partial s}  =  \left(\sigma -i\frac{1}{2\tau k}\right)^2 - \left( \frac{\kappa}{\tau T}\frac{\partial T}{\partial s} - \frac{1}{4 k^2 \tau^2}\right).
	\eeq
	
Let us introduce the speed of sound, specific heat and the coefficient $b$ given by
		\beq
	c_n^2 = \frac{n}{\mu}\frac{\partial \mu}{\partial n}, \quad\frac{1}{\cv} = \frac{1}{T}\frac{\partial T}{\partial s}  \quad  \text{and}\quad b=\frac{1}{2\tau k},
	\eeq
together with the auxiliary `velocities'
	\begin{subequations}
	\label{def.aux1}
	\begin{align}
	c_1^2 &= \frac{n}{T}\frac{\partial T}{\partial n},\\
	c_2^2& =\frac{1}{\alpha T}\frac{\partial T}{\partial n},\\
	c_3^2 &= c_n^2,\\
	c_4^2& =\frac{\kappa}{\tau \cv} - \frac{1}{4 k^2 \tau^2}.
	\end{align}
	\end{subequations} 
	
With the above definitions, the system of equations \eqref{sym} takes the more manageable form
		\begin{align}
	\frac{\mu}{n}\left[\sigma^2 - c_3^2 \right] \delta n & + n\alpha T\left[\sigma^2 -c_2^2 \right] \delta \bar{s} = 0\\
	\kappa\frac{T}{n} \left[\sigma^2 - c_1^2 \right] \delta n & + n\tau T\left[ \left(\sigma - i b \right)^2 - c_4^2 \right] \delta \bar{s} =0,
	\end{align}
where we can read off the dispersion relation as 
	\beq
	\label{disp}
	 A\left[\left(\sigma^2 - c_1^2\right)\left(\sigma^2 - c_2^2\right) \right] - \left[\sigma^2 - c_3^2 \right]\left[\left(\sigma - i b \right)^2 - c_4^2 \right]  = 0.
	\eeq	
Here, we have introduced the additional abbreviation
	\beq
	A = \frac{\alpha \kappa T}{\mu\tau}. 
	\eeq
Expanding \eqref{disp} gives us the quartic polynomial	
	\beq
	\label{quartic}
	(1-A)\sigma^4 + i  B \sigma^3 +  C \sigma^2 + i  D \sigma +  E = 0,
	\eeq
where each of the coefficients are given by
	\begin{subequations}
	\begin{align}
	 B &= 2b,\\
	 C &= b^2 + c_4^2 + c_3^2 - A\left(c_1^2 + c_2^2\right),\\
	 D &= -2bc_3^2,\\
	 E &= Ac_1^2 c_2^2 -c_3^2b^2 - c_3^2 c_4^2.
	\end{align}
	\end{subequations}
This expression is obviously too complicated for us to be able to make definite statements about
the solution, without making further assumptions. The most direct strategy would be to consider
an explicit example equation of state, work out the relevant thermodynamics quantities, and then
solve \eqref{quartic}. This would allow us to establish whether the considered model
is stable and causal. This route is, however, not particularly attractive given the need to introduce
an explicit model. Nevertheless, we still have hope to make some progress by attempting the  alternative involution method to find some stability criterion in the following manner. Suppose $r_i$, $i=0,1,2,3$ are the roots of \eqref{quartic}. If we now transform this set of solutions by an involution into
	\begin{subequations}
	\begin{align}
	s_0 & = \frac{1}{2}\left(r_0 + r_1 + r_2 + r_3 \right),\\
	s_1 & = \frac{1}{2}\left(r_0 - r_1 + r_2 - r_3 \right),\\
	s_2 & = \frac{1}{2}\left(r_0 + r_1 - r_2 - r_3 \right),\\
	s_3 & = \frac{1}{2}\left(r_0 - r_1 - r_2 + r_3 \right),
	\end{align}
	\end{subequations}
where we know from \eqref{quartic} that \cite{galois}
	\beq
	s_0 = i\frac{B}{2(1-A)}.
	\eeq 
Thus, we can evaluate $s_0$  directly in terms of known thermodynamic quantities 
	\beq
	s_0 = \frac{1}{2} \sum_{k=0}^3 r_k = - i\frac{B}{2(1-A)} = - i \frac{b}{A-1} = -\frac{i}{2\tau k} \left[\frac{\alpha k T}{\mu \tau} - 1\right]^{-1},
	\eeq
which, by substituting the definition of $\alpha$ given by equation \eqref{dec.alpha}, reduces to
		\beq
	s_0 = -\frac{i}{2 k}n\mu\left[(\kappa - \tau s)T - \tau n \mu \right]^{-1}
	\eeq
	
Writing $s_0$ as the sum of the real and imaginary parts of the roots of \eqref{quartic}
	\beq
	s_0 = \frac{1}{2}\left[\sum_{k=0}^3 {\rm Re}(r_k) + i \sum_{k=0}^3 {\rm Im}(r_k)\right]
	\eeq	
it follows that
	\beq
	\sum_{k=0}^3 {\rm Re}(r_k) = 0
	\eeq
Therefore, there would be no hope for the system to be stable if the sum of the imaginary parts is negative,
	\beq
	\sum_{k=0}^3 {\rm Im}(r_k) < 0.
	\eeq
However, this condition  requires that $(\kappa - \tau s )T - \tau n \mu \geq 0$, that is,
	\beq
	\label{first.taucrit2}
	\frac{\kappa T}{\rho + p} \geq \tau
	\eeq	
for arbitrary small values of the temperature. Thus, the system is absolutely unstable only in the case in which the relaxation time vanishes. This is not a new result but it shows that our longitudinal analysis is consistent with the transverse one [see equation \eqref{first.taucrit}].

If we want to continue to consider the problem in (at least to some extent)
generality, then we need to resort to approximations. As we will see, this is a very instructive
route. Thus, let us consider the case for which $s_0$ vanishes. This corresponds to the limit of very large wave number. In this limit, we can see that \eqref{quartic} reduces to the bi-quadratic
	\beq
	\label{bicuad}
	(1-A)\sigma^4 + C \sigma^2 + E = 0
	\eeq 
whose roots are given by
	\beq
	\sigma^2_\pm = \frac{1}{2(1-A)}\left\{-C \pm \left[C^2 - 4(1-A)E\right]^{1/2} \right\}.
	\eeq
Let us examine this expression carefully. First, we  evaluate the coefficient
	\beq
	1-A = 1 -\frac{\alpha \kappa T}{\mu \tau}   = \frac{1}{\tau \mu n}\left[\left(\rho + p \right)\tau - \kappa T \right] > 0.
	\eeq
Thus, to guarantee the stability of the system ($\sigma^2_\pm\geq 0$), the following conditions must hold
	\begin{subequations}
	\begin{align}
	\label{c.const}
	C & = \frac{\kappa}{\tau \cv} + c_n^2 - \frac{\kappa}{\mu \tau}\frac{\partial T}{\partial n}\left(2-\frac{\tau s}{\kappa} \right) > 0,\\
	\label{e.const}
	E  & = \frac{\kappa}{\tau}\left[\frac{1}{T}\frac{n}{\mu}\left(\frac{\partial T}{\partial n} \right)^2 - \frac{c_n^2}{\cv}\right] \leq 0. 
	\end{align}
	\end{subequations}
This last condition leads immediately to
	\beq
	c_n^2 \geq \frac{n \cv}{\mu T}\left(\frac{\partial T}{\partial n} \right)^2,
	\eeq
	
It is worth noting that this last condition is readily satisfied by all separable equations of state of form
$\rho = f (n) + g(s)$. 

Our analysis leads to two further conditions that must be satisfied. Condition \eqref{c.const} leads to
	\beq
	\tau > \kappa \left[\frac{\mu}{\cv} + 2 \frac{\partial T}{\partial n} \right]\left[s \frac{\partial T}{\partial n}+\mu c_n^2\right]^{-1}.
	\eeq
which, provides us with a further lower bound for the relaxation time. Thus, we have a second criterion to establish local stability in a given matter model whose equation of state is separable.

Finally, we can approximate the finite wavelength regime through an expansion in powers of $\epsilon = k^{-1}$. Thus, consider the expression
	\beq
	\label{dec.sig01}
	\sigma = \sigma_\pm + \frac{1}{k}\sigma_1 = \sigma_\pm + \epsilon \sigma_1,
	\eeq
where $\sigma_\pm$ solves \eqref{bicuad}. We can evaluate \eqref{dec.sig01} up to the fourth power and neglect terms of order $\mathcal{O}(\epsilon^2)$. Thus, we have
	\begin{subequations}
	\label{powers}
	\begin{align}
	\sigma^4 &= (\sigma_\pm + \epsilon \sigma_1)^4 = \sigma_\pm^4 + 4\epsilon \sigma_\pm^3 \sigma_1 + O(\epsilon^2),\\
	\sigma^3 &= (\sigma_\pm + \epsilon \sigma_1)^3 = \sigma_\pm^3 + 3\epsilon \sigma_\pm^2 \sigma_1 + O(\epsilon^2),\\
	\sigma^2 &= (\sigma_\pm + \epsilon \sigma_1)^2 = \sigma_\pm^2 + 2\epsilon \sigma_\pm \sigma_1 + O(\epsilon^2).
	\end{align}
	\end{subequations}

Substituting \eqref{powers} into \eqref{quartic} while keeping only linear terms in $\epsilon$ gives
	\beq
	4(1-A)\sigma_\pm^2 \sigma_1 + \frac{i}{\tau}\sigma_\pm^2 + 2 C\sigma_1 - \frac{i}{\tau}c_n^2 =0,
	\eeq
which leads immediately to
	\beq
	\sigma_1 = \frac{i}{2\tau}\frac{c_n^2 - \sigma_\pm^2}{2(1-A)\sigma_\pm^2 + C}.
	\eeq
Since all quantities in this expression are constrained to be real and stability requires ${\rm Im}(\sigma_1)\geq 0$, we have two cases:

	\begin{enumerate}
		\item $2(1-A)\sigma_{+}^2 + C = +\left[B^2 - 4(1-A)E\right]^{1/2}\geq 0 \rightarrow
			c_n^2 < \sigma_{+}^2$.
		\item  $2(1-A)\sigma_{-}^2 + C = -\left[B^2 - 4(1-A)E\right]^{1/2}\leq 0 \rightarrow c_n^2 > \sigma_{-}^2$.	
	\end{enumerate}
Therefore, we can write the conditions for the stability of our \emph{complete} first-order theory as
	\beq
	\label{first.last}
	\sigma_{-}^2 \leq c_n^2 \leq \sigma_{+}^2,
	\eeq
which is consistent with the notion that mode-mergers signal the onset of instability.

\section{What have we learned?}

In this section, we constructed a consistent first-order expansion away from thermal equilibrium. We  argued in chapter 4 that first order theories are `pathological' due to instabilities and causality violations. Therefore, two remarks are in order:
	\begin{enumerate}
		 \item On the one hand, the presence of instabilities is not sufficient to rule out a particular matter model. There are physical instabilities, as the two-stream case, which provide essential information about the dynamics of the system. Moreover, in situations far away from thermal equilibrium, thermal instabilities are responsible for structure formation, {\it e.g.} Benard cells, Taylor instability, Kelvin-Helmholtz, etc... 
		 \item On the other hand, causality violations may lead to unstable behaviour. However, even when  no instabilities are present, in any relativistic matter model this is a clear violation of our mathematical construction. Therefore, causality is a requirement to consider a matter model as physical.  
	\end{enumerate} 
	
Here, we have shown that, in contrast with the conclusion of Hiscock and Lindblom \cite{hiscock01}, a first-order expansion away from thermal equilibrium is possible provided we do not exclude  second-order terms prematurely. This is manifest in our first stability criterion for transverse modes, equation \eqref{first.taucrit}, where we can see that the Eckart model not only fails to be stable, but also causal. That is because from the construction, the relaxation time - which is a second-order effect - was absent from the dynamics of the model. Moreover, the stability criterion \eqref{first.taucrit} gives a non-trivial lower bound for the relaxation time of transverse [and longitudinal waves, see \eqref{first.taucrit2}] modes and, hence, an upper bound for the speed of propagation for thermal disturbances.

To tackle the problem of stability and causality in an analytic manner for arbitrary wave numbers is a very difficult task. However, we can obtain approximate criteria from an expansion away from the `geometrical optics' regime, where the dispersion relation reduces to a bi-quadratic providing the upper and lower bounds for the phase velocity. In this sense, we obtained the onset of instability as a merger of modes in our expansion away from thermal equilibrium. Notice that our only true criterion to rule out the instabilities as non-physical is well established. That is, if the onset of instability occurs beyond the causality regime, every unstable mode will necessarily be non-physical. However, if the instability region is within the causality domain, we cannot rule out the instability on the grounds of local causality. Moreover, \eqref{first.last} gives us the time-scale for the unstable behaviour, providing us with the relevant information to observe their evolution given a particular matter model, {\it e.g.} superfluid states of matter inside a neutron star or, from the non-adiabatic pressure perturbations in a cosmological setting \cite{adam2}.

This is just the first step towards a better understanding of the dynamical basis of dissipative systems in a general relativistic setting. There are still many issues to address. Of immediate interest is to apply our stability analysis to specific matter models to test and predict the dynamics of heat transport in relativistic systems.

			\chapter{Closing remarks}

The central aim of the research presented in this thesis was to elaborate on the geometric formalism pioneered by Carter  with the aim of  understanding heat flow in a large range of physical scales and situations. We formulated a theory of relativistic heat conduction
through a completely generally covariant constrained variational principle for a multi-fluid system. In
such an approach, we dealt only with the matter sector of the Einstein-Hilbert action for a system consisting of
two coupled fluids - particles and entropy - whose interaction is a crucial ingredient to guarantee stability and
causality in situations away from thermodynamic equilibrium.

We started our discussion by introducing the basic empirical principles upon which we construct general relativity - the equivalence principle and general covariance - and the way they lead to a dynamical theory of geometry and matter. We also discussed that the laws of thermodynamics are a collection of self-consistent empirical facts which must be added as additional hypotheses to our fundamental theory of motion. In this sense, these empirical laws rule out entire classes of phenomena predicted by the dynamics but whose occurrence in nature may never be observed, {\it e.g.} the flow of heat from colder to warmer bodies. This constitutes one of the central topics of fundamental debate on the apparent time-asymmetry of nature.  In this work we have not entered this debate but it would have been remiss not to acknowledge the conceptual difficulties inherent in the inclusion of the second law of thermodynamics into any dynamical framework. 

Variational techniques are powerful and elegant mathematical tools. We have formulated a theory of relativistic heat conduction through a completely \emph{generally covariant} constrained variational principle for a multi-fluid system. In such an approach, we dealt only with the matter sector of the Einstein-Hilbert action for a system consisting of two coupled fluids - particles and entropy - whose interaction is a crucial ingredient to guarantee stability and causality in situations away from thermodynamic equilibrium. 

The motivation for treating entropy as another `fluid' coupled to the matter current - rather than considering it merely as a component -  stemmed from the hydrodynamic limit of relativistic kinetic theory. The emergence of an entropy density current,  generally not aligned with the matter flow, is generic to every theory of relativistic dissipation. The novel feature in our programme is that we incorporate the entropy flux into the action principle, allowing us to truly analyse the dynamics of the coupled matter and entropy system, where the \emph{entrainment} effect plays a crucial role.

Compared to the standard single-fluid analysis, a multi-fluid system has more degrees of freedom; we need to account for relative flows between them. This is an essential requirement for the two-stream instability. It is a generic phenomenon that does not require particular fine-tuning to be triggered, nor is limited to any specific physical system. The only requirements are that there must be a relative (background) flow and a coupling between the
fluids. Such instabilities are known to exist for a variety of configurations. For example, in shearing motion at
an interface, this corresponds to the well known Kelvin--Helmholtz instability.  A key result from our work was a covariant generalization of the two-stream instability analysis to relativistic situations. Here, we have shown that the equilibrium configurations of our variational construction, as expected, are absolutely stable. Moreover, we argued that, for some equations of state, the onset of instability is arbitrarily close to the equilibrium regime. It is worth noting that this instability for thermodynamic systems is a prediction of our programme for relativistic heat conduction.  The dynamical role of this instability has not been explored for particular matter models beyond the linear regime. 

Finally, we took our first step in non-adiabatic equilibrium perturbations at the \emph{linear} regime. We obtain the criteria for the stability of this type of perturbations. It follows from our analysis that the Eckart model is intrinsically unstable because it prematurely  ignores higher order terms in deviations from equilibrium. In this sense, our results  constitute  a consistent `first-order' theory of relativistic dissipation.

\section{Further developments}

There are a number of future directions which our formalism will allow us to explore. To list a few, we have:

%

	\subsection{Microscopic foundations of relativistic thermal dynamics} 
	
	There are many directions that can be explored regarding the connection between the microscopic canonical dynamics of a relativistic system and its averaged thermal dynamics on spacetime. Here, we enlist three steps of a particular strategy:  
	
	\begin{enumerate}
	
	\item \emph{Kinetic limit of the two-fluid formalism.}
	
	 In order to improve our understanding of the variational model in a physical setting, and the role
of the various non-equilibrium terms, we need to consider our heat model in the context of relativistic
kinetic theory. The aim should be to develop a consistent model for the entrainment and the thermal relaxation of a dissipative fluid.
A detailed  exploration of the constrains on the relativistic single-particle phase-space imposed by the two-fluid nature of the  macroscopic system is needed.  This would help us  to understand the connection between the variational formalism and the geometry of the  canonical formulation  for the microscopic dynamics of a dissipative system. Let us note that the standard approach to relativistic kinetic theory  is based on two fundamental assumptions:
the notions of a dilute gas and the (time-asymmetric) molecular chaos hypothesis. It is relevant to ask whether these concepts need to be extended in order to match more complex macro-scale physics. After all, the variational approach is
readily extended to more complicated multi-component systems ({\it e.g}. with varying composition). It also
accounts for non-linear phenomena that would require the inclusion of higher moments  of the distribution function at the microscopic level.

Thus, our formalism could allow us to  address a range of conceptual issues concerning the entropy of a relativistic system. Most importantly we may be able to develop a link between the two-fluid phenomenology and statistical physics. An understanding of this
issue is important for a number of relevant applications. The key point is that, from a technical point-
of-view, the variational model can account for a number of `distinct' entropy components {\it e.g.} phonons, rotons, etc... This may
be an advantage in order to describe specific systems, like relativistic superfluids at finite temperatures.
However, it is not clear how these models are linked to statistical physics. To explore this problem, one can extend existing models for phonons in finite temperature superfluids  (and the associated phonon
hydrodynamics) to more complex setting with several `distinct' entropy-carrying components.

	\item{The matter-space dynamics and the geometry of dissipation.}

	The notion of individual matter spaces associated with the various fluid components played a central
role in the variational construction of chapter 5. In this case, the matter-space construct essentially boils
down to the variation of flow lines associated to the individual fluid elements. This problem is relatively simple
because the matter space has no real physical `structure'. However, there are situations where it is necessary to
generalize this idea. A particular case concerns elastic matter \cite{carterquin,karlars}, where the elasticity is encoded in
the matter-space geometry. This problem is fairly well understood. What is less clear at the present time is how we should handle evolving system, {\it e.g.} when elastic matter undergoes plastic flow or when there are composition changes due to reactions in a
multi-component system. This problem is central to the description of irreversible phenomena. After all,
the second law requires us to allow for entropy production. In order to do this, as in the heat-flow model
described above, we need to relax the variational assumptions.

A natural step forward in this direction would be to develop the variational formalism further by considering `evolving' matter spaces. This idea is technically challenging, but we see no reason why one should not be able to make progress on it.
We will need to add more structure to the description (as in the elastic case) and likely account for interaction between individual matter spaces (particularly if there are reactions involved), but the mathematics
should be sufficiently flexible to deal with these issues. The main challenge may be conceptual.
At the deeper level we need to develop a link between the matter-space and the coarse-graining of
phase-space in statistical physics. It is, in fact, relevant to ask at what level the statistics should be considered. Is it at the spacetime level, or is it in the lower-dimensional matter-space? In principle,
both answers seem viable but the latter would be an attractive (possibly quite revolutionary) solution. In
analogy with the description of elastic matter one may envisage a model based on the notion of an evolving `thermal geometry' (in matter-space) directly linked to the entropy change between hypersurfaces in
spacetime. The model also requires a dynamical map between the matter space and spacetime, in order to link
the changes in the local geometric structure of the matter configuration (described in terms of normal
coordinates) to the macroscopic evolution of the system.

	\item{The thermodynamics of the gravitational degrees of freedom}

	This direction is, perhaps, one of the most challenging from a foundational point-of-view. It stems from the fact that there seems to be an intimate link between gravity and thermodynamics. This follows immediately from the
equivalence of energy and mass, which implies that heat must affect the gravitational field. Nevertheless,
from a conceptual point of view it is not understood to what extent the gravitational field is `hot'. Basically, we do not yet have a clear definition of the entropy associated with an evolving gravitational field.

The variational model for heat accounts for the coupling to (and evolution of) the gravitational field
via the Einstein field equations. However, so far our main focus has been on the matter sector of the
problem. In order to explore the role of the gravitational field  one can consider the coupled system
of equations in a number of model scenarios (plane waves and simplistic matter models). The aim is to
established how the second law of thermodynamics feeds into the evolution of the gravitational field, and
(conversely) how variations in the gravitational field affect the entropy and the heat flow. An interesting
conceptual issue concerns the link between observers and the increase of gravitational entropy.

	\end{enumerate}

	\subsection{The `near' and `far' from equilibrium entropy conditions}

	We have argued in chapter 2 that, although the evolution towards equilibrium of a thermodynamic system is governed by the second law, there is no criterion to `select the path' towards equilibrium, {\it i.e.} there is no fundamental principle of extremal entropy production. It would be interesting to explore how systems
thermalise once the entropy production is constrained. The variational model provides us with a useful context
for such work. However, we would need to develop  a clearer understanding of the entropy production associated with
a given initial configuration.

To make progress in this direction, one can consider two extreme cases; the near and far from
equilibrium regimes. In the former case, we can rely on an expansion away from thermal equilibrium (as in the
moment expansion of kinetic theory), while in the latter we need to account for the non-linearities of
the variational model. Non-linear systems are, in fact, very interesting because the available models for
relativistic heat differ in the quadratic (and higher order) terms. Thus, we need to explore to what extent
these differences lead to distinguishable differences in the evolution. To this end, one can  build
on experience of non-linear evolutions of relativistic material systems in general relativity, accounting for
a suitably constrained thermodynamics.
	\begin{enumerate}

		\item {\it Near-equilibrium regime.} In chapter 2, we gave an example of the principle of minimum entropy production for systems close to thermal equilibrium. However, this notion is closely related to the Onsager's symmetry principle of the Linear Irreversible Thermodynamics programme. It is, of course, not at all obvious that usual
derivation can be extended to the relativistic problem. However, if the argument fails then we will have
learned something fundamental about non-equilibrium physics. 

An understanding of the near-equilibrium behaviour gained from a perturbative approach would obviously be incomplete. Moreover, as we have shown at the end of chapter 6, even at linear order in a perturbation expansion we need information about the full non-linear regime. Nevertheless, such analysis gives us the causality criterion and the onset of thermodynamic instability. In the case of physical unstable modes, these criteria would be serve to set limit cases for the evolutions in the non-linear regime.

	\item{\it Far-from-equilibrium regime.} To fully understand the role of the second law and its association with pattern and structure formation we need a non-linear framework which must account for (as in numerical relativity) the coupling between matter and spacetime. In order to explore the non-linear sector, we may consider the maximum
entropy principle. This idea is based on the notion that when estimating a probability distribution,
it is natural to select the distribution that leaves the largest remaining uncertainty (maximum entropy)
consistent with the system constraints. The advantage of this approach is that it minimises the information required (which indeed was Jaynes' motivation for the approach) in the calculation. The maximum
entropy principle is intimately linked to information theory, identifying the entropy in that context with
the statistical physics. This connection is relevant for the gravity problem, especially as it may provide
insight into the issues like information loss associated with black-hole horizons. 
  
  To sum up, one could extend the variational model in such a way that the evolution ensures maximum entropy
production. This can be done using Lagrange multipliers, but in order to ensure computational efficiency
we need to learn from other contexts where this principle has been implemented ({\it e.g.} for biological
systems). Successful implementation of this idea is likely to lead to a real breakthrough in non-linear
relativity simulations of heat.

	\end{enumerate}

 	\subsection{Classical origins of the arrow of time}

 	The research described in this thesis is entirely within the realm of classical physics. Yet, it is clear that any real understanding of the nature of time will have to account for quantum aspects. While we
make progress at the classical level, we should prepare the ground for future explorations of the quantum
arena. In absence of a theory for quantum gravity this is obviously a huge step, but some basic principles seem clear. For example, if we want to make direct contact (and build upon) Rovelli's notion of thermal
time (associated with the evolution of pre-symplectic systems), then we need to develop a Hamiltonian
description for the relativistic thermo/hydrodynamics. This is known to be a challenging problem, but
Dirac's procedure for developing a Hamiltonian system from a given Lagrangian is (at least in principle)
clearly laid out. However, the steps involved are far from straightforward in practice. This is particularly
true in the case of a constrained variational model, as in the present case.

\section{Conclusion}

The work presented in this thesis represents a few fundamental steps towards a deeper understanding of a relativistic theory of irreversible processes. Thus, we have learned that the dynamical contribution of heat plays a crucial role in the evolution of a material system in spacetime. Moreover, we have seen that, to correctly account for such dynamics, we need to include the  contribution of heat directly into the variational principle through the entropy current. In this manner, the equations of motion for an internal observer (one which is moving with the material particles) include a thermodynamic description where the inertial properties of heat are an essential feature for the system's evolution towards equilibrium. To show this, we have made a linear perturbation analysis for both the adiabatic and non-adiabatic cases. In the former, we generalise the two-stream instability analysis for linear perturbations of equilibrium to a relativistic setting. The consequences of this analysis are yet to be explored. In the latter case, we obtained a consistent first order perturbative analysis of deviations from thermal equilibrium. This is a novel and unexpected result which provides new light on the old second-order paradigm. However, let us emphasise that such a consistent first-order model can only stem from an underlying second order dynamical theory which contains the complete thermodynamic information for the propagation of linear waves. Finally, we have presented a number of exciting directions in which this programme can be extended.  With some luck, one of them may give us a hint in a quest that, just like entropy itself, \emph{grows} as we walk ahead.

		

\cleardoublepage

\begin{figure}
	\begin{center}
		\includegraphics[width=.95\columnwidth,angle=180]{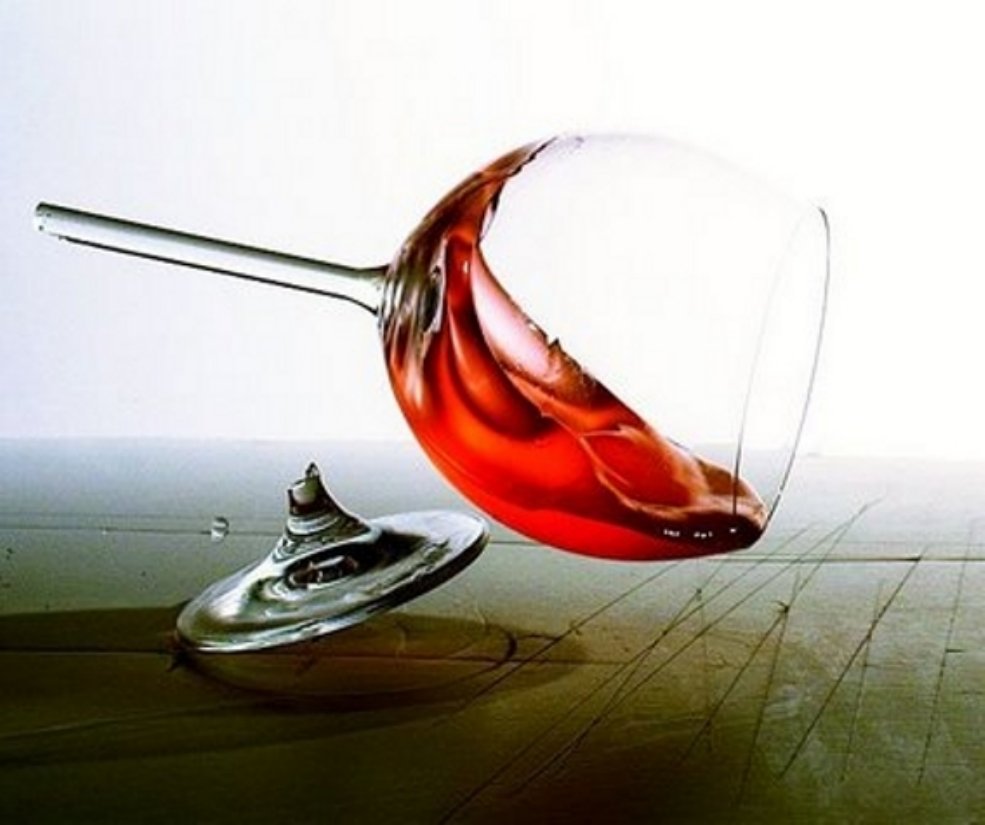}
	\end{center}
	\caption{Gravity and entropy production: What can our intuition tell us about the \emph{future} of this \emph{initial} configuration? [{\small This image is property of Steve Strawn, splutphoto.com}]}
	\label{graf.02}
\end{figure}

		\appendix
			\chapter{Regular and Anomalous Heat Conduction}
\begin{singlespace}
\begin{quote}
This appendix contains an extract of a set of notes by Brandon Carter \cite{cartermiss}. We have included this transcript here in order to make more accessible a reference which, otherwise, would be a bit less than impossible to get\footnote{The author is thankful to Greg Comer for this communication.}. The material below is a collection of the original ideas which motivated the later developments presented in this work. 
\end{quote}
\end{singlespace}

The purpose of this work is to offer a fresh approach to an old problem by providing an
appropriate relativistic formulation of classical thermodynamics. In particular, of the law of
heat conduction. The essential feature of this approach is the prominence given to the canonical 4-momentum
covector associated with the entropy flux. The main result of this work is to provide a more satisfactory alternative to the traditional Eckart heat conduction equation.

Eckart's equation [{\it c.f.} Eckart's hypothesis, equation \eqref{heat01}] differs from the non-relativistic Fourier's equation only by the second term on the left hand side involving the acceleration $\dot u^a$. Many authors have found that \eqref{heat01} is
unsatisfactory because it leads to a causality paradox (heat conduction faster than light), and 
it has been suggested, notably by Cattaneo \cite{cattaneo} and Kranys \cite{kranys} that this should be remedied by replacing the LHS of Eckart's equation by the {\it ad-hoc} expression $q^a + \tau h^a_{\ b}\dot q^b$, where the coefficient $\tau$ of the additional term is a new phenomenological parameter, to be interpreted as representing a microscopic relaxation time.

In the present approach, based purely on the most intrinsically plausible thermodynamic
axioms, we find that an extra term having essentially the same effect as one of the Kranys
type turns up automatically on the LHS of \eqref{heat01}. Part of our purpose is to point out that it is
not surprising that the original form of \eqref{heat01} gave unsatisfactory results. The reason for this, is that the traditional textbook derivation of Eckart's conduction equation is based on a rather incoherent set
of simplifying assumptions. The weak point in the traditional derivation of \eqref{heat01} is the (usually
implicit rather than explicit) assumption that inertia associated with the heat current is zero. {\it A priori}, this might not seem unreasonable in a course-tuned treatment aimed at maximum
simplicity, but its utter inappropriateness is made evident {\it post-facto} by the fact that the Eckart
acceleration term in  is naturally interpretable ({\it c.f.} Ehlers \cite{ehlers}) as being due to an effective
inertia which amounts quantitatively (in units for which Boltzmann’s constant is unity) to the
rest mass associated with the energy  per unit entropy. It follows that instead of ignoring the
inertia associated with the heat current, a much more reasonable ansatz would be to suppose
consistently from the outset that there is an effective mass per unit entropy given precisely by
the temperature $\theta$. When this condition is satisfied, we shall describe the heat flux as \emph{regular}. In
the regular case, we shall show that an appropriately modified version of the standard textbook
derivation of \eqref{heat01} leads unambiguously to the appearance of an extra term $\mathcal{L}_{\rm u}(q^a/s)$ on the LHS,
where $s$ is the rest frame entropy density, and $\mathcal{L}_{\rm u}$ denotes Lie differentiation with respect to the
fundamental flow vector $u^a$.

Although it is certainly not fair to suggest that this regularity condition is `ad-hoc' (as
was implied by Israel  with reference to an abbreviated oral communication of this work) it
would nevertheless be entirely justifiable to describe it as `naive', since it is certainly not a
logical necessity that inertia appearing in different contexts should always be strictly the same.
In defence of the regularity condition all that one can say is the following.

Firstly, it will be accurately valid in the limit when thermal energy is entirely in the form
of a perfect gas of black body equilibrium radiation with negligible coupling to the background
medium, and hence it will be a good approximation in many problems involving radiative transfer in radiation pressure dominated stellar or cosmological media. Secondly, in many other
contexts the difference between the various conceivable correction terms will be important only
for variations of such high frequency that a straightforward thermodynamic treatment would fail
anyway, so that one is in practice free to choose the most mathematically satisfactory hypothesis, {\it i.e.}
 that of regularity. Nevertheless, in addition to these two kinds of physical situation ({\it i.e.} those
where regularity holds accurately, and those where there is no practical loss of accuracy in imposing it for convenience) there are also situations where anomalous behaviour ({\it i.e.} deviations
from regularity) can give important measurable effects, the most obvious example being that of superfluidity. One may draw a close analogy with the case of regular and anomalous magnetic moments: for the electron, the `naive' Dirac's
theory applied very well. However, for more strongly interacting particles, such as the proton, a significant anomaly occurs. In a somewhat analogous manner, the strong correlations between atoms of superfluid helium make possible the occurrence an effective inertia per unit entropy that is enormously greater than the regular value (which is of course negligibly small in such a highly
non-relativistic context).

		\cleardoublepage
		\addcontentsline{toc}{chapter}{Glossary}
		
		\cleardoublepage
		\addcontentsline{toc}{chapter}{Bibliography}
		\bibliographystyle{plain}
		\bibliography{myrefs}

\end{document}